\documentclass[prb,aps,twocolumn,superscriptaddress,showpacs]{revtex4-1}

\makeatletter
\renewcommand*{\@fnsymbol}[1]{\ensuremath{\ifcase#1\or \dagger\or \dagger\or \ddagger\or
		\mathsection\or \mathparagraph\or \|\or **\or \dagger\dagger
		\or \ddagger\ddagger \else\@ctrerr\fi}}
\makeatother
\usepackage{color}
\usepackage{graphicx}
\usepackage{epstopdf}
\usepackage[colorlinks=true,linkcolor=black, citecolor=blue, urlcolor=blue, 
unicode=true,breaklinks=true]{hyperref}

\usepackage{float}
\usepackage{lipsum}
\usepackage{amsmath} 
\usepackage{amssymb}	
\usepackage{color}
\usepackage{hyperref} 
\usepackage{anysize}
\usepackage{hhline,multirow,float}
\usepackage{makecell}
\usepackage{array}
\usepackage{dcolumn}
\newcolumntype{?}{!{\vrule width 1pt}}
\usepackage{color}
\usepackage{placeins}
\usepackage{mathtools}
\usepackage{xr}
\definecolor{model1}{RGB}{0,0,0}
\definecolor{model2}{RGB}{255,0,0}
\definecolor{model3}{RGB}{135,81,81}
\definecolor{model4}{RGB}{0,176,80}
\definecolor{model5}{RGB}{255,165,0}
\definecolor{model6}{RGB}{128,128,0}
\definecolor{model7}{RGB}{0,0,255}

\usepackage{multirow}
\usepackage[table]{xcolor}
\usepackage{capt-of}
\usepackage{wasysym}

\marginsize{1.5 cm}{1.5 cm}{0.5 cm}{0.5 cm}

\usepackage{etoolbox} 
\usepackage{lipsum} 

\makeatletter
\appto{\appendix}{%
	\@ifstar{\def\theequation@prefix{A.}}%
	{}%
}
\makeatother

\begin{document}
\newcommand\bbone{\ensuremath{\mathbbm{1}}}
\newcommand{\ul}{\underline}
\newcommand{\bp}{{\bf p}}
\newcommand{\vl}{v_{_L}}
\newcommand{\vc}{\mathbf}
\newcommand{\be}{\begin{equation}}
\newcommand{\ee}{\end{equation}}
\newcommand{\bk}{{{\bf{k}}}}
\newcommand{\bK}{{{\bf{K}}}}
\newcommand{\cE}{{{\cal E}}}
\newcommand{\bQ}{{{\bf{Q}}}}
\newcommand{\br}{{{\bf{r}}}}
\newcommand{\bg}{{{\bf{g}}}}
\newcommand{\bG}{{{\bf{G}}}}
\newcommand{\hbr}{{\hat{\bf{r}}}}
\newcommand{\bR}{{{\bf{R}}}}
\newcommand{\bq}{{\bf{q}}}
\newcommand{\hx}{{\hat{x}}}
\newcommand{\hy}{{\hat{y}}}
\newcommand{\hd}{{\hat{\delta}}}
\newcommand{\bea}{\begin{eqnarray}}
\newcommand{\eea}{\end{eqnarray}}
\newcommand{\ra}{\rangle}
\newcommand{\la}{\langle}
\renewcommand{\tt}{{\tilde{t}}}
\newcommand{\upa}{\uparrow}
\newcommand{\dna}{\downarrow}
\newcommand{\bS}{{\bf S}}
\newcommand{\vS}{\vec{S}}
\newcommand{\dg}{{\dagger}}
\newcommand{\pdg}{{\phantom\dagger}}
\newcommand{\tphi}{{\tilde\phi}}
\newcommand{\cf}{{\cal F}}
\newcommand{\ca}{{\cal A}}
\renewcommand{\ni}{\noindent}
\newcommand{\ct}{{\cal T}}
\newcommand{\brf}{\bar{F}}
\newcommand{\brg}{\bar{G}}
\newcommand{\jeff}{j_{\rm eff}}
\newcommand{\cvo}{$\alpha$-CoV$_{3}$O$_{8}$}

\newcommand{\appropto}{\mathrel{\vcenter{
  \offinterlineskip\halign{\hfil$##$\cr
    \propto\cr\noalign{\kern2pt}\sim\cr\noalign{\kern-2pt}}}}}

\title{Spin-Orbit Excitons in CoO}

\author{P.~M.~Sarte}
\affiliation{California NanoSystems Institute, University of California, Santa Barbara, California 93106-6105, USA}
\affiliation{Materials Department, University of California, Santa Barbara, California 93106-5050, USA} 
\affiliation{School of Chemistry, University of Edinburgh, Edinburgh EH9 3FJ, United Kingdom}
\affiliation{Centre for Science at Extreme Conditions, University of Edinburgh, Edinburgh EH9 3FD, United Kingdom}
\author{M.~Songvilay}
\affiliation{Centre for Science at Extreme Conditions, University of Edinburgh, Edinburgh EH9 3FD, United Kingdom}
\affiliation{School of Physics and Astronomy, University of Edinburgh, Edinburgh EH9 3FD, United Kingdom}
\author{E.~Pachoud}
\affiliation{School of Chemistry, University of Edinburgh, Edinburgh EH9 3FJ, United Kingdom}
\affiliation{Centre for Science at Extreme Conditions, University of Edinburgh, Edinburgh EH9 3FD, United Kingdom}
\author{R.~A.~Ewings}
\affiliation{ISIS Pulsed Neutron and Muon Source, STFC Rutherford Appleton Laboratory, Harwell Campus, Didcot, Oxon, OX11 OQX, United Kingdom}
\author{C. D.~Frost}
\affiliation{ISIS Pulsed Neutron and Muon Source, STFC Rutherford Appleton Laboratory, Harwell Campus, Didcot, Oxon, OX11 OQX, United Kingdom}
\author{D.~Prabhakaran} 
\affiliation{Department of Physics, Clarendon Laboratory, University of Oxford, Park Road, Oxford, OX1 3PU, United Kingdom}
\author{K.~H.~Hong}
\affiliation{School of Chemistry, University of Edinburgh, Edinburgh EH9 3FJ, United Kingdom}
\affiliation{Centre for Science at Extreme Conditions, University of Edinburgh, Edinburgh EH9 3FD, United Kingdom}
\author{A.~J.~Browne}
\affiliation{School of Chemistry, University of Edinburgh, Edinburgh EH9 3FJ, United Kingdom}
\affiliation{Centre for Science at Extreme Conditions, University of Edinburgh, Edinburgh EH9 3FD, United Kingdom}
\author{Z.~Yamani}
\affiliation{National Research Council, Chalk River, Ontario K0J 1JO, Canada }
\author{J.~P.~Attfield}
\affiliation{School of Chemistry, University of Edinburgh, Edinburgh EH9 3FJ, United Kingdom}
\affiliation{Centre for Science at Extreme Conditions, University of Edinburgh, Edinburgh EH9 3FD, United Kingdom}
\author{E.~E.~Rodriguez} 
\affiliation{Department of Chemistry and Biochemistry, University of Maryland, College Park, Maryland 20742, USA}
\author{S.~D.~Wilson} 
\affiliation{California NanoSystems Institute, University of California, Santa Barbara, California 93106-6105, USA}
\affiliation{Materials Department, University of California, Santa Barbara, California 93106-5050, USA} 
\author{C.~Stock}
\affiliation{Centre for Science at Extreme Conditions, University of Edinburgh, Edinburgh EH9 3FD, United Kingdom}
\affiliation{School of Physics and Astronomy, University of Edinburgh, Edinburgh EH9 3FD, United Kingdom}

\date{\today}

\begin{abstract}

CoO has an odd number of electrons in its unit cell, and therefore is expected to be metallic. Yet, CoO is strongly insulating owing to significant electronic correlations, thus classifying it as a Mott insulator.  We investigate the magnetic fluctuations in CoO using neutron spectroscopy.  The strong and spatially far-reaching exchange constants reported in [Sarte \emph{et al.} Phys. Rev. B \textbf{98} 024415 (2018)], combined with the single-ion spin-orbit coupling of similar magnitude [Cowley \emph{et al.} Phys. Rev. B \textbf{88}, 205117 (2013)] results in significant mixing between $j\rm{_{eff}}$ spin-orbit levels in the low temperature magnetically ordered phase.  The high degree of entanglement, combined with the structural domains originating from the Jahn-Teller structural distortion at $\sim$ 300~K, make the magnetic excitation spectrum highly structured in both energy and momentum.  We extend previous theoretical work on PrTl$_{3}$ [Buyers \emph{et al.} Phys. Rev. B \textbf{11}, 266 (1975)] to construct a mean-field and multi-level spin exciton model employing the aforementioned spin exchange and spin-orbit coupling parameters for coupled Co$^{2+}$ ions on a rocksalt lattice.   This parameterization, based on a tetragonally distorted type-II antiferromagnetic unit cell, captures both the sharp low energy excitations at the magnetic zone center, and the energy broadened peaks at the zone boundary.  However, the model fails to describe the momentum dependence of the excitations at high energy transfers,  where the neutron response decays faster with momentum than the Co$^{2+}$ form factor.  We discuss such a failure in terms of a possible breakdown of localized spin-orbit excitons at high energy transfers.
  
\end{abstract}

\maketitle

\section{Introduction}

Mott insulators are materials where conventional band theory fails, predicting metallic behavior owing to half-filled bands, with the origin of the insulating response indicative of strong electronic correlations~\cite{Anisimov91:44,Mott49:62,Phillips06:321,brandow76:10,brandow77:26}.  Mott insulators are parent materials for high-temperature cuprate superconductivity~\cite{Lee06:78,Kastner,zhao11:7}. Moreover, there have been some suggestions that these insulators may even be implicated as being parent to some iron-based superconductors~\cite{Si08:101,Zhu10:104,McCabe14:89,Stock16:28,wilson10:82}.  These Mott insulators display well-defined spin excitations, however rapidly breakdown~\cite{Plumb14:89,Headings10:105,Stock10:82,Stock07:75} on charge doping towards superconductivity.~\cite{Ho01:86}    More recently, Mott insulators with strong spin-orbit coupling have been of particular interest in the search for unconventional topological states~\cite{moore10:464,schaffer16:79,Witczak14:5}.  These studies have focussed on 4$d$ and 5$d$ transition metals with strong spin-orbit coupling resulting in $j\rm{_{eff}}=\frac{1}{2}$ ground states, and new Kitaev bond directional phases~\cite{jakeli09:102,winter17:29,takagi19:xx}. However, much of the single-ion physics that results in these $j\rm{_{eff}}=\frac{1}{2}$ ground states is present in Co$^{2+}$-based compounds that also display strong spin-orbit coupling~\cite{Sano18:97,Liu18:97}.  

In this context, it is timely to investigate the classic Mott insulator CoO, where significant spin-orbit coupling is present and comparable to the magnetic exchange.   In this study, we investigate the mixed spin-orbit transitions in CoO through their parameterization with a multi-level spin-orbit exciton model extending previous theoretical work on PrTl$_{3}$~\cite{buyers75:11}.  While this model reproduces the experimental data at low-energy transfers, we show that its failure  at high energy transfers is accompanied by a possible breakdown of these excitations.

For the past several decades, CoO has been one of the most extensively studied Mott insulators.  The 3$d$ metal monoxide was among the first orbitally-ordered materials to be investigated with neutron diffraction~\cite{Shull51:102}.  Its primitive unit cell consists of one 3$d^{7}$ Co$^{2+}$ and one 2$p^{6}$ O$^{2-}$ with 2 electrons in the n=2 shell, which corresponds to 15 valence electrons. With an odd number of valence electrons, conventional band theory~\cite{Norman89:40,Terakura84:30} would predict CoO to be metallic. However, CoO is a very strong insulator~\cite{Engel09:103} with a room temperature resistivity of 10$^{8}$~$\Omega \cdot \rm{cm}$, and an optical band gap of 2.5~eV~\cite{Elp91:44,gillen13:25,gvishi72:33}, with evidence for metallic behavior being found only under extremely high pressures on the order of 100 GPa~\cite{Cohen97:275}. 

\begin{figure}[htb!]
	\centering
	\includegraphics[width=1\linewidth]{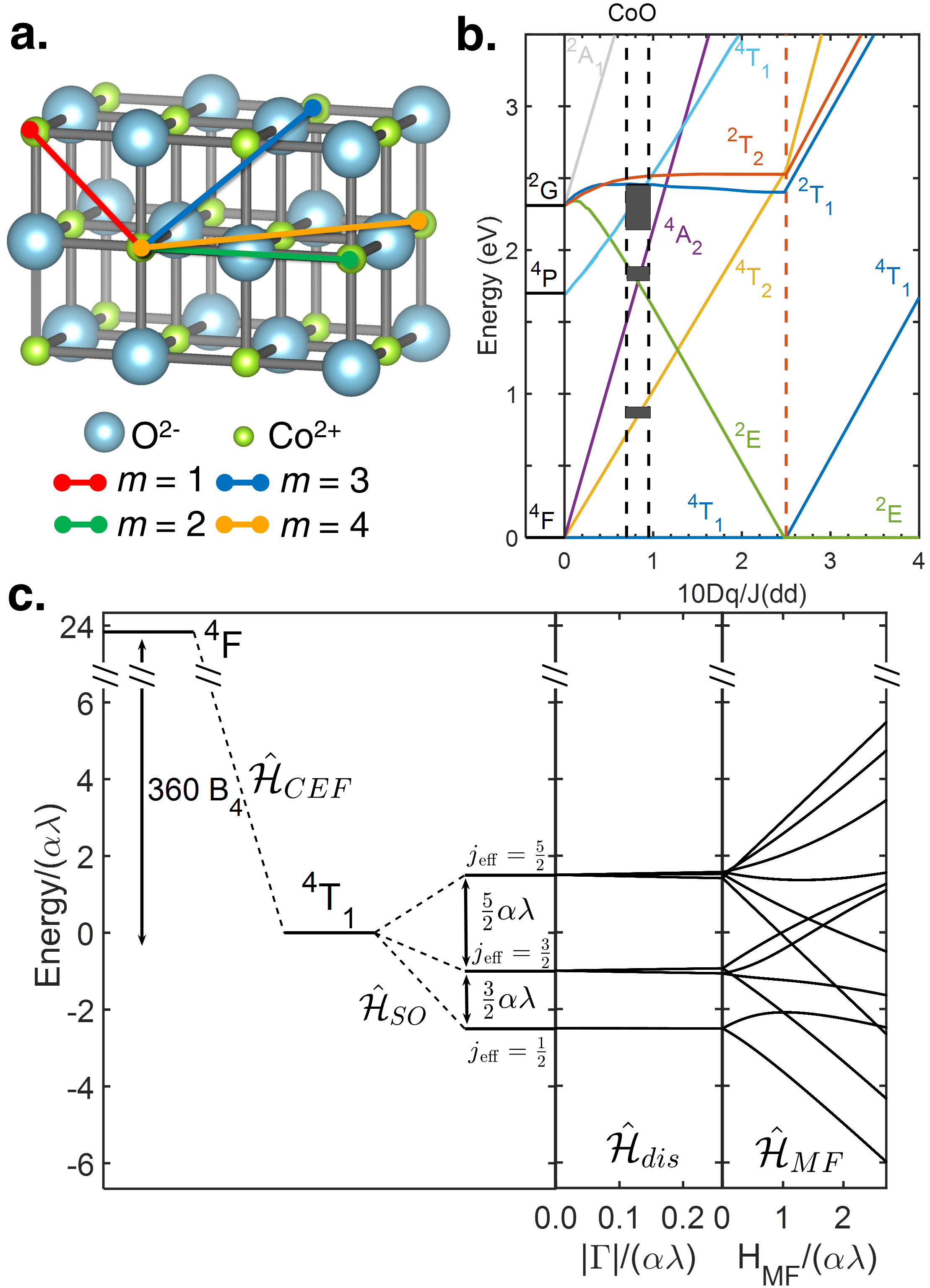}
	\caption{(a) First four coordination shells of the high temperature
		CoO rocksalt structure. (b) Tanabe-Sugano diagram for $d^{7}$ Co$^{2+}$ in octahedral coordination calculated by Cowley \emph{et al.}~\cite{cowley13:88}. Shaded rectangles correspond to experimentally measured excitations for cubic CoO at room temperature with heights and the width corresponding to experimental errors in energy and the statistical error of the refined value for $10Dq/J(dd)$, respectively. The dashed red line at $10Dq/J(dd)\sim2.5$ denotes the spin crossover from (left) high-spin $S=\frac{3}{2}$, $^{4}T_{1}$ to (right) low-spin $S=\frac{1}{2}$, $^{2}E$. (c) Calculated normalized energy variation as a function of the tetragonal distortion ($\hat{\mathcal{H}}_{dis}$) and the magnetic molecular field ($\hat{\mathcal{H}}_{MF}$) perturbations to the $j\rm{_{eff}}$ manifolds from the ground state crystal field triplet $^{4}T_{1}$ of Co$^{2+}$ in octahedral coordination. Both the energy eigenvalues and individual parameters are presented to scale.}
	\label{fig:fig1}
\end{figure}

Possessing a cubic $Fm\bar{3}m$ structure~\cite{jauch01:64,Satoshi79:55,tombs50:165,Ok68:168,liu06:110} at room temperature (Fig.~\ref{fig:fig1}(a)), CoO assumes long-range antiferromagnetic order at $T\rm{_{N}}\sim290$~K~\cite{greenwald53:6}, in contrast to the long-range ferromagnetism predicted by general band coupling models that assume a dominant direct exchange~\cite{deng10:96,Walsh08:100}. Despite being the subject of many neutron diffraction studies, its magnetic structure has proven to be particularly contentious, with both collinear~\cite{Roth58:110,Ressouche06:385} and non-collinear~\cite{vanLaar65:138,Tomiyasu04:70} models describing diffraction patterns equally well~\cite{fiebig02:93,vanLaar66:141,Kern03:15,Shishidou98:67,timm16:91}. 

As illustrated in Fig.~\ref{fig:fig1}(b), the $^{4}T_{1}$ crystal field ground state for CoO corresponds to the $d^{7}$ Co$^{2+}$ assuming a high spin ($S=\frac{3}{2}$) configuration, yielding an orbital triplet with one hole in the $t_{2g}$ orbital manifold. The resulting orbital degeneracy, coupled with both a Jahn-Teller driven~\cite{jauch01:64} unit cell distortion and various far-reaching large exchange interactions, yields a complex magnetic excitation spectrum that results from the strong entanglement of multiple spin-orbit levels (Fig.~\ref{fig:fig1}(c)).  The resulting multi-parameter spin-orbital Hamiltonian incorporating both exchange and spin-orbit coupling of similar magnitude, further complicated by the complex magnetic ordering and structural distortions, has made the understanding of the magnetic excitations in this material particularly difficult~\cite{sakurai68:167}. 

By employing both the spin-orbit coupling constant $\lambda$ and the magnetic exchange constants $J$ that were experimentally determined in our previous work~\cite{cowley13:88,sarte18:98} on the magnetically diluted monoxide Mg$_{0.97}$Co$_{0.03}$O, we will show that the low energy magnetic excitation spectrum of CoO measured in the N\'{e}el regime by inelastic neutron spectroscopy is reproduced by a mean-field multi-level spin-orbit exciton model based on Green's functions~\cite{buyers75:11}.  Our parameterization successfully captures the fine structure of well-defined low energy spin excitations present at the magnetic zone center and also the broadening in both momentum and energy at the zone boundaries. In contrast, the model fails to reproduce the high energy response consisting of high velocity excitations that decay with momentum faster than the Co$^{2+}$ form factor.  We suggest that this failure of the model provides evidence for a breakdown of localized spin-orbit excitations at high energy transfers, possibly replaced by delocalized or itinerant-like fluctuations, despite CoO being a strong Mott insulator. 

This paper is divided into three general sections.  In the first section, we describe the theoretical framework that we apply to parameterize the neutron scattering response in CoO.  We first outline the single-ion response defining the crystal field Hamiltonian, and then discuss the coupled equations-of-motion that were used to numerically derive the neutron response.  In the second section, we first present the experimental data as measured with neutron spectroscopy, followed by a direct comparison to our multi-level spin-orbit exciton model.  To conclude, we discuss the high energy excitations and the poor agreement with the Co$^{2+}$ form factor and speculate as to their origin.

\section{Theory}

We first discuss the theoretical framework used to describe the localized magnetic response in CoO. The neutron magnetic cross section is proportional to the magnetic dynamic structure factor $S(\bf{Q},\omega)$ defined by

\begin{equation}
S({\bf{Q}},\omega)=g_{L}^{2}f^2({\bf{Q}})\sum_{\alpha \beta} (\delta_{\alpha \beta}-\hat{Q}_{\alpha}\hat{Q}_{\beta}) S^{\alpha \beta}({\bf{Q}},\omega),
\label{dynamic_ss}
\end{equation}

\noindent where $g_{L}$ is the Land\'{e} $g$-factor, $f({\bf{Q}})$ is the magnetic form factor,~\cite{Kern03:15,Kern04:350} and $S^{\alpha \beta}(\bf{Q},\omega)$ corresponds to the dynamic spin structure factor. Since the orbital angular momentum is quenched ($i.e.$ $\left\langle \hat{\mathbf{L}} \right\rangle = 0$) for $3d^{7}$ Co$^{2+}$, the orbital contribution to the scattering cross section is assumed to be weak, and therefore the spin operators provide the dominant contribution to the neutron scattering cross section~\cite{zal05:book}. This assumption allows $S^{\alpha \beta}(\bf{Q},\omega)$ to be defined in terms of expectation values of spin operators $\hat{S}^{\nu}(i,t)$ of index $\nu=$ $+$, $-$, or $z$, acting on a site $i$ at a time $t$. Such a definition of the dynamic structure factor is given by 

\begin{equation}
S^{\alpha \beta}({\bf{Q}},\omega)= {1 \over {2\pi}} \int dt \ e^{i\omega t} \langle \hat{S}^{\alpha}({\bf{Q}},t) \hat{S}^{\beta} (-{\bf{Q}},0) \rangle,
\nonumber
\end{equation}

\noindent whose relationship to the response function $G^{\alpha \beta}(\bf{Q},\omega)$ is given by the fluctuation-dissipation theorem as 

\begin{equation}
S^{\alpha \beta}({\bf{Q}},\omega)=-{1\over \pi} {1 \over {1-\exp(\omega/k{\rm{_{B}}}T)}} \Im{G^{\alpha \beta} (\bf{Q},\omega)}.
\label{eq:spin_structure}
\end{equation} 

Motivated by previous work on PrTl$_{3}$~\cite{buyers75:11,fulde70:241}, the theoretical portion of this paper begins by first writing down the equations-of-motion for the response function in terms of commutators involving the magnetic Hamiltonian $\hat{\mathcal{H}}$. We then investigate this magnetic Hamiltonian in CoO to define both the single-ion states and how these states are coupled from site to site on the rocksalt lattice.   Finally, we apply mean-field theory to decouple the equations-of-motion, thereby reducing the formula for the response functions to a set of coupled linear equations that can be computed numerically and directly compared with experiment.  In this approach, we use creation and annihilation operators of the single-ion states rather than the Holstein-Primakoff transformation for a single spin operator.  This approach allows for both the incorporation of spin-orbit level mixing, and the explicit inclusion of the single-ion terms in the Hamiltonian, such as spin-orbit coupling, rather than employing anisotropy terms that incorporate the orbital physics through perturbation theory~\cite{yosida}.

\FloatBarrier

\subsection{The Equation-of-Motion for the Response Function}
  
According to linear response theory, the response function measured with neutrons is proportional to the Fourier transform of the retarded Green's function that is given by

\begin{equation}
\begin{split}
G^{\alpha\beta}(ij, t)  & =  G(\hat{S}^{\alpha}(i,t),\hat{S}^{\beta}(j,0))  \\
 & =  -i\Theta(t)\langle[\hat{S}^{\alpha}(i,t),\hat{S}^{\beta}(j,0)]\rangle,
\end{split}
\end{equation}

\noindent where $\Theta(t)$ is the Heaviside function. As shown in Section~I of the \emph{Supplementary Information}~\cite{suppl},  by taking the first time derivative of $G^{\alpha\beta}(ij, t)$, applying the Heisenberg equation-of-motion, and Fourier transforming from time to energy, one arrives at the following equation-of-motion

\begin{equation}
\omega G(\hat{A},\hat{B},\omega) = \langle [\hat{A},\hat{B}] \rangle + G([\hat{A},\hat{\mathcal{H}}],\hat{B},\omega),
\label{eq:10}
\end{equation}

\noindent where $\hat{A}$ and $\hat{B}$ denote generic spin operators. Eq.~\ref{eq:10} indicates that deriving a model for the neutron scattering response functions relies both on the understanding of the Hamiltonian $\hat{\mathcal{H}}$ and its commutator with the spin operators. We now investigate the individual contributions to magnetic Hamiltonian in CoO.

\subsection{The Total Magnetic Hamiltonian $\hat{\mathcal{H}}$} \label{sec:mag_ham}

The total magnetic Hamiltonian consisting of crystal field ($CF$) contributions and coupling between Co$^{2+}$ on sites $i$ and $j$ can be written as

\begin{equation}
\begin{split}
\hat{\mathcal{H}} &= \hat{\mathcal{H}}_{CF} + \sum\limits_{ij}J(ij)\hat{\mathbf{S}}(i) \cdot \hat{\mathbf{S}}(j).
\nonumber
\end{split} 
\end{equation}

\noindent By defining a molecular field Hamiltonian

\begin{equation}
\hat{\mathcal{H}}_{MF}(i) = \sum\limits_{i}H_{MF}(i)\hat{S}_{z}(i),
\nonumber
\end{equation}

\noindent where

\begin{equation}
H_{MF}(i) = 2\sum\limits_{i>j}J(ij)\langle \hat{S}_{z}(j)\rangle,
\label{eq:molecular_field_2} 
\end{equation}

\noindent $\hat{\mathcal{H}}$ can be written as a sum of a single-ion ($\hat{\mathcal{H}}_{1}$) and an inter-ion ($\hat{\mathcal{H}}_{2}$) term given by

\begin{equation}
\hat{\mathcal{H}}_{1} = \sum\limits_{i}\hat{\mathcal{H}}_{CF}(i) + \sum\limits_{i}\hat{S}_{z}(i)\left(2\sum\limits_{i>j}J(ij)\langle \hat{S}_{z}(j)\rangle\right),
\label{single_ion}
\end{equation}

\noindent and

\begin{equation}
\begin{split}
\hat{\mathcal{H}}_{2} &= \sum\limits_{ij}J(ij)\hat{S}_{z}(i)[\hat{S}_{z}(j)-2\langle \hat{S}_{z}(j)\rangle] \\
&+ \frac{1}{2}\sum\limits_{ij}J(ij)[\hat{S}_{+}(i)\hat{S}_{-}(j) + \hat{S}_{-}(i)\hat{S}_{+}(j)],
\end{split}
\label{Inter_ion}
\end{equation}

\noindent where $\langle \hat{S}_{z}(j)\rangle$ denotes a thermal average given by 

\begin{equation}
\langle \hat{S}_{\alpha} \rangle = \sum\limits_{n}f_{n}\langle n|\hat{S}_{\alpha}|n\rangle \equiv \sum\limits_{n}\hat{S}_{\alpha nn}f_{n},
\label{eq:thermal_avg}
\end{equation}

\noindent with $\hat{S}_{\alpha nn}= \langle n|\hat{S}_{\alpha}|n\rangle$, and $f_{n}$ is the Boltzmann thermal population factor.  The inclusion of a factor of 2 in Eq.~\ref{single_ion}  follows the convention that was established in Ref.~\onlinecite{sakurai68:167} to explicitly account for the double counting in the sum over sites. 

The procedure we follow to derive the neutron response consists of two parts.  First, we diagonalize the single-ion component $\hat{\mathcal{H}}_{1}$ for a given molecular field such that

\begin{equation}
\hat{\mathcal{H}}_{1}=\sum_{n} \sum_{i}\omega_{n} C^{\dagger}_{n}(i)C_{n}(i), 
\label{ladder}
\end{equation}

\noindent where $C(i)$ and $C^{\dagger}(i)$ are ladder operators satisfying the commutation relations $[C_{n}(i),C^{\dagger}_{m}(j)]=\delta_{ij}\delta_{nm}$, and $\omega_{n}$ are the energy eigenvalues.  The second step consists of using these states to apply mean-field theory on the inter-ion $\hat{\mathcal{H}}_{2}$ term to then compute the neutron response using the equation-of-motion given in Eq.~\ref{eq:10}.  We will now discuss the eigenstates of the single-ion component of the Hamiltonian $\hat{\mathcal{H}}$, followed by the inter-ion component. This section concludes with the application of mean-field theory on the inter-ion component, allowing for the derivation of an expression for the Green's function that can be calculated numerically and compared directly to experiment. 

\subsubsection{Single-Ion Hamiltonian $\hat{\mathcal{H}}_{1}$}

As schematically illustrated in  Fig.~\ref{fig:fig1}(c), the single-ion component of the Hamiltonian consists of four components 

\begin{equation}
\begin{split}
\hat{\mathcal{H}}_{1} &= \hat{\mathcal{H}}_{CF}+ \hat{\mathcal{H}}_{MF}=\\
&(\hat{\mathcal{H}}_{CEF}+\hat{\mathcal{H}}_{SO} + \hat{\mathcal{H}}_{dis})+ \hat{\mathcal{H}}_{MF},
\nonumber
\end{split} 
\end{equation}

\noindent corresponding to the contributions from the crystalline electric field $\hat{\mathcal{H}}_{CEF}$, spin-orbit $\hat{\mathcal{H}}_{SO}$, structural distortion $\hat{\mathcal{H}}_{dis}$, and mean molecular field $\hat{\mathcal{H}}_{MF}$. Hyperfine nuclear transitions are neglected since previous measurements~\cite{Chatterji09:79} have indicated that these are on the order of $\sim \mu$eV, and thus beyond the experimental resolution of our study.  We now discuss each term of the Hamiltonian $\hat{\mathcal{H}}_{1}$.

\textit{The Crystalline Electric Field}, $\hat{\mathcal{H}}_{CEF}$: As illustrated in the Tanabe-Sugano~\cite{Tanabe54:9,Tanabe54:9_2,Liehr63:67} diagram in Fig.~\ref{fig:fig1}(b), for the case of CoO, the $3d^{7}$ Co$^{2+}$ is octahedrally coordinated by the weak field O$^{2-}$ ligand resulting in a crystal field splitting $10Dq$ that is weaker than the energy differences between the free-ion terms~\cite{Pratt59:116}. Consequently, the crystalline electric field contribution $\hat{\mathcal{H}}_{CEF}$ can be treated as a perturbation to the free-ion basis states that are defined by Hund's rules incorporating the effects of electron-electron Coulomb repulsion and the Pauli Exclusion Principle. A combination of Hund's first rule of maximum multiplicity and second rule requiring the total orbital angular momentum $L$ be maximized yields a total spin of $S=\frac{3}{2}$ and orbital angular momentum $L=3$, for the $3d^{7}$ Co$^{2+}$, corresponding to an orbital ground state term symbol of $^{4}F$.  

The crystalline electric field $\hat{\mathcal{H}}_{CEF}$ contribution corresponding to the octahedral coordination of the $^{4}F$ free-ion ground state by O$^{2-}$ ligands can be expressed in terms of the Stevens operators $\hat{\mathcal{O}}^{0}_{4}$ and $\hat{\mathcal{O}}^{4}_{4}$, and the numerical coefficient $B_{4}<0$~\cite{Abragam:book,Hutchings64:16} as 

\begin{equation}
\hat{\mathcal{H}}_{CEF} = B_{4}\left(\hat{\mathcal{O}}^{0}_{4} + 5\hat{\mathcal{O}}^{4}_{4}\right). 
\nonumber
\end{equation}     

\noindent Since the spin-orbit coupling is expected to be considerably weaker than the crystal field contribution for the $3d$ Co$^{2+}$, the complete set of commuting observables are $\hat{L}^{2}$, $\hat{L}_{z}$, $\hat{S}^{2}$ and $\hat{S}_{z}$ with corresponding good quantum numbers $L$, $m_{L}$, $s$ and $m_{s}$ in the Russell-Saunders $L$-$S$ coupling scheme; thus, by the Wigner-Eckart theorem, both Stevens operators in $\hat{\mathcal{H}}_{CEF}$ can be defined in the $|L,m_{L}\rangle$ basis, as summarized in Section~II of the \emph{Supplementary Information}~\cite{suppl}. 

The diagonalization of $\hat{\mathcal{H}}_{CEF}$ results in an orbital triplet ground state ($^{4}T_{1}$), an excited orbital triplet ($^{4}T_{2}$), and an orbital singlet $^{4}A_{2}$, where $\Delta(^{4}T_{1}\rightarrow^{4}T_{2})=480B_{4}$ and  $\Delta(^{4}T_{2} \rightarrow^{4}A_{2})=600B_{4}$.   The Stevens factor $B_{4}$ is related to the crystal field splitting by $10Dq=400 B_{4}$ (Fig. \ref{fig:fig1}(b)), where $10Dq$ was previously measured to be $\sim$1~eV~\cite{cowley13:88,schooneveld12:116,magnuson02:65,Chiuz08:78,Sakellaris12:116,Gorsch94:49}.  Since the $^{4}T_{1}$ crystal field ground state and $^{4}T_{2}$ first excited state are separated by $\sim$ 1~eV, it is a valid approximation that the $^{4}T_{1}$ ground state will exclusively determine the magnetic properties of CoO~\cite{Larson07:99,Haverkort07:99}. 

\textit{Spin-Orbit Coupling}, $\hat{\mathcal{H}}_{SO}$: The second perturbation to the $^{4}F$ free-ion ground state is spin-orbit coupling given by

\begin{equation}
\hat{\mathcal{H}}_{SO}=\lambda \hat{\mathbf{L}} \cdot \hat{\mathbf{S}}, 
\label{eq:SO_1}
\end{equation} 

\noindent where $\lambda$ is the spin-orbit coupling constant.  A common approach is to exclusively consider the $^{4}T_{1}$ ground state, requiring a projection from the original $|L=3,m_{L}\rangle$ basis onto a smaller basis $|l=1,m_{l}\rangle$ that defines the subspace that is spanned by the crystal field ground state.  As discussed by Abragam \& Bleaney~\cite{Abragam:book}, this particular projection can be performed using representation theory.  Here, we outline an alternate method based on the matrix representation of angular momentum operators~\cite{sarte18:98_2} that was inspired by the work on $4d$ and $5d$ transition metal oxides by Stamokostas and Fiete~\cite{stamokostas18:97}.

The matrix approach begins by first determining the set of eigenvectors $|\phi_{CEF}\rangle$ of the crystalline electric field Hamiltonian $\hat{\mathcal{H}}_{CEF}$ in the $|L=3,m_{L}\rangle$ basis.  Since $|\phi_{CEF}\rangle$ is also a basis, a transformation matrix $\mathcal{C}$ can be constructed that rotates from the $|L=3,m_{L}\rangle$ to the $|\phi_{CEF}\rangle$ basis.  The matrix $\mathcal{C}$ consists of columns corresponding to eigenvectors of $\hat{\mathcal{H}}_{CEF}$ in the $|L=3,m_{L}\rangle$ basis arranged in order of increasing energy eigenvalues. In the case of degenerate eigenvalues, a small perturbative Zeeman term of the form $\epsilon\hat{S}_{z}$, with $\epsilon$ being a small constant, was applied to remove the degeneracy and uniquely define the column order. For Co$^{2+}$ in octahedral coordination with $B_{4}$ set to $-1$, $\mathcal{C}$ is given by 

 \begin{equation}
\mathcal{C} = \left[ {\begin{array}{ccccccc}
	0 & 0  & -0.79 & 0.61 & 0& 0 & 0 \\
	0 & 0  & 0 & 0 & -0.71 & 0 & -0.71 \\
	0.61 & 0  & 0 & 0 & 0 & -0.79  & 0 \\
	0 & 1.00  & 0 & 0 & 0& 0 & 0 \\
	0 & 0  & -0.61 & -0.79 & 0& 0 & 0 \\
	0 & 0  & 0 & 0 & -0.71 & 0 & 0.71 \\
	0.79 & 0  & 0 & 0 & 0 & 0.61 & 0 \\
	\end{array} } \right].
\nonumber
\end{equation}     

\noindent Having obtained the transformation matrix, the rotation from the $|L=3,m_{L}\rangle$ to the $|\phi_{CEF}\rangle$ basis can then be accomplished by $\hat{\mathcal{O}}_{|\phi_{CEF}\rangle} = \mathcal{C}^{-1}\hat{\mathcal{O}}_{|L,m_{L}\rangle}\mathcal{C}$.

For the  $\hat{L}_{z}$ operator, this transformation yields

\begin{equation}
\mathcal{C}^{-1}\hat{L}_{z}\mathcal{C} = \left[ {\begin{array}{ccc|ccc|c}
	\mathbf{1.50} & \mathbf{0}  & \mathbf{0} & 0 & 0& -1.94 & 0 \\
\mathbf{0} & \mathbf{0}  & \mathbf{0} & 0 & 0& 0 & 0 \\
\mathbf{0} & \mathbf{0}  & \mathbf{-1.50} & -1.94 & 0& 0 & 0 \\
\hline
0 & 0  & -1.94 & 	\mathbf{-0.50} & \mathbf{0}& 	\mathbf{0} & 0 \\
0 & 0  & 0 & \mathbf{0} & 	\mathbf{0} & 	\mathbf{0} & 2.00 \\
-1.94 & 0  & 0 & 	\mathbf{0} & \mathbf{0}& \mathbf{0.50} & 0 \\
\hline
0 & 0  & 0 & 0 & 2.00& 0 & \mathbf{0}\\
	\end{array} } \right],
	\nonumber
\end{equation}

\noindent illustrating the ground state orbital triplet $^{4}T_{1}$, and the excited orbital triplet $^{4}T_{2}$ and singlet $^{4}A_{2}$ states. We note that the opposite sequence exists in the case of a tetrahedral environment with an orbital singlet ground state~\cite{Decaroli15:71}.  A comparison of the top 3 $\times$ 3 block matrix to the $\hat{L}_{z}$ operator in the $|l=1,m_{l}\rangle$ basis given by  

 \begin{equation}
 \hat{L}_{z} = \left[ {\begin{array}{ccc}
 	-1 & 0  & 0  \\
 	0 & 0  & 0  \\
 	0 & 0  & 1 \\
 	\end{array} } \right]
 \nonumber
 \end{equation} 

\noindent confirms that the block matrix is equivalent to the $\hat{L}_{z}$ operator in the $|l=1,m_{l}\rangle$ basis, with a projection factor $\alpha=-\frac{3}{2}$, in agreement with previous approaches based on representation theory.  Therefore, in the low temperature/energy limit, the spin-orbit Hamiltonian (Eq.~\ref{eq:SO_1}) can be rewritten as 

\begin{equation}
\hat{\mathcal{H}}_{SO} = \alpha\lambda\hat{\mathbf{l}}\cdot\hat{\mathbf{S}}, 
\nonumber
\end{equation} 

\noindent corresponding to a new Hamiltonian consisting of new orbital angular momentum operators that act on the projected $|l=1,m_{l}\rangle$ basis.

By assigning an effective angular momentum operator to the subspace spanned by the $|l=1,m_{l}\rangle$ basis, it is implied these new operators must follow the same commutation relations for general angular momentum operators.  To check this fundamental requirement is satisfied, we have transformed the three $\hat{L}_{x, y, z}$ operators, each of which is a $7 \times 7$ matrix, to the $|\phi_{CEF}\rangle$ basis with the procedure outline above. We then extracted the top $ 3 \times 3 $ block matrices of the projected matrices $\mathcal{C}^{-1}\hat{L}_{x,y,z}\mathcal{C}$ to define $\hat{l}_{x, y, z}$, and confirmed that these matrices do follow the commutation relations of angular momentum given by $\hat{\bf{l}} \times \hat{\bf{l}}=i\hat{\bf{l}}$.  We note that the presence of a thermally isolated low energy triplet does not guarantee that these commutation relations are followed. An example of such a failure has been recently discussed in the context of low energy doublets in the heavy fermion CeRhSi$_{3}$~\cite{pasztorova19:99}.

Having projected $\hat{\mathbf{L}}$ onto a fictitious operator $\hat{\mathbf{l}}$ to reflect the triplet orbital degeneracy of the $^{4}T_{1}$ ground state of $\hat{\mathcal{H}}_{CEF}$, we now derive the eigenstates of the perturbative $\hat{\mathcal{H}}_{SO}$ term.  The basis is now the 12 $|l=1, m_{l}; s={3\over 2}, m_{s} \rangle$ states, and based on both the Land\'{e} interval rule and the addition theorem of angular momentum, we expect this Hamiltonian to yield three levels defined by $j\rm{_{eff}}$=${1\over 2}$, ${3 \over 2}$, and ${5 \over 2}$.  Using the projection factor $\alpha=-\frac{3}{2}$, and the experimentally determined~\cite{cowley13:88} spin-orbit coupling constant $\lambda=-16$~meV, the diagonalization of the spin-orbit Hamiltonian $\hat{\mathcal{H}}_{SO}$ matrix yields 

\begin{widetext}
\begin{equation}
diag\left(\hat{\mathcal{H}}_{SO}\right) =	
			\setlength{\arraycolsep}{3pt}
				\def\arraystretch{1}
\left[\begin{array}{cccccccccccc}
		-60&0&0&0&0&0&0&0&0&0&0&0 \\
		0&-60&0&0&0&0&0&0&0&0&0&0 \\ 
		0&0&-24&0&0&0&0&0&0&0&0&0 \\
		0&0&0&-24&0&0&0&0&0&0&0&0 \\
		0&0&0&0&-24&0&0&0&0&0&0&0 \\
		0&0&0&0&0&-24&0&0&0&0&0&0 \\ 
		0&0&0&0&0&0&36&0&0&0&0&0 \\
		0&0&0&0&0&0&0&36&0&0&0&0 \\
		0&0&0&0&0&0&0&0&36&0&0&0 \\
		0&0&0&0&0&0&0&0&0&36&0&0 \\
		0&0&0&0&0&0&0&0&0&0&36&0 \\
		0&0&0&0&0&0&0&0&0&0&0&36 \\
		\end{array}	\right], 
		\nonumber
\end{equation}
\end{widetext}

\noindent confirming a doublet ground state with quartet and sextet excited manifolds (Fig.~\ref{fig:fig1}(c)) with $\Delta$(doublet$\rightarrow$quartet) = $-36$~meV $\equiv \frac{3}{2}\alpha\lambda$ and  $\Delta$(quartet$\rightarrow$sextet) = $-60$~meV $\equiv \frac{5}{2}\alpha\lambda$. Finally, we may confirm that each manifold corresponds to $j\rm{_{eff}}=\frac{1}{2}$, $\frac{3}{2}$, and $\frac{5}{2}$, respectively, by projecting the components of the effective total angular momentum operator $\hat{\mathbf{j}}=\hat{\mathbf{l}}+\hat{\mathbf{S}}$ onto the subspaces spanned by the spin-orbit manifolds of $\hat{\mathcal{H}}_{SO}$. As was the case for the projection onto the subspace spanned by the $^{4}T_{1}$ crystal field ground state, such a projection is accomplished by first defining a transformation $\mathcal{C}$ which rotates operators from the $| l,m_{l},s,m_{s}\rangle$ basis to the $|\phi_{SO}\rangle$ basis with the columns being the eigenvectors of $\hat{\mathcal{H}}_{SO}$ arranged in increasing energy. Rotating the $\hat{j}_{z}$ operator onto the $|\phi_{SO}\rangle$ basis yields 

\begin{widetext}
\begin{equation}
\mathcal{C}^{-1}\hat{j_{z}}\mathcal{C}=
					\setlength{\arraycolsep}{3pt}
						\def\arraystretch{1}
\left[\begin{array}{cc|cccc|cccccc}
\mathbf{-\frac{1}{2}}& 	\mathbf{0} &	0 &	0&	0& 	0&	0&	0&	0&	0&	0&	0 \\
\mathbf{0}&	\mathbf{\frac{1}{2}}&	0&	0&	0&	0&	0&	0&	0&	0&	0&	0 \\ \hline
0&	0&	\mathbf{-\frac{3}{2}}&	\mathbf{0}&	\mathbf{0}&	\mathbf{0}&	0&	0&	0&	0&	0&	0 \\
0&	0&	\mathbf{0}&	\mathbf{-\frac{1}{2}}&	\mathbf{0}&	\mathbf{0}&	0&	0&	0&	0&	0&	0 \\
0&	0&	\mathbf{0}&	\mathbf{0}&	\mathbf{\frac{1}{2}}&	\mathbf{0}&	0&	0&	0&	0&	0&	0 \\
0&	0&	\mathbf{0}&	\mathbf{0}&	\mathbf{0}&	\mathbf{\frac{3}{2}}&	0&	0&	0&	0&	0&	0 \\ \hline
0&	0&	0&	0&	0&	0&	\mathbf{-\frac{5}{2}}&	\mathbf{0}&	\mathbf{0}&	\mathbf{0}&	\mathbf{0}&	\mathbf{0} \\
0&	0&	0&	0&	0&	0& \mathbf{0}&	\mathbf{-\frac{3}{2}}&	\mathbf{0}&	\mathbf{0}&	\mathbf{0}&	\mathbf{0} \\
0&	0&	0&	0&	0&	0&	\mathbf{0}&	\mathbf{0}&	\mathbf{-\frac{1}{2}}&	\mathbf{0}&	\mathbf{0}&	\mathbf{0} \\
0&	0&	0&	0&	0&	0&	\mathbf{0}&	\mathbf{0}&	\mathbf{0}&	\mathbf{\frac{1}{2}} &	\mathbf{0}&	\mathbf{0} \\
0&	0&	0&	0&	0&	0&	\mathbf{0}&	\mathbf{0}&	\mathbf{0}&	\mathbf{0}&	\mathbf{\frac{3}{2}}&	\mathbf{0} \\
0&	0&	0&	0&	0&	0&	\mathbf{0}&	\mathbf{0}&	\mathbf{0}&	\mathbf{0}&	\mathbf{0}&	\mathbf{\frac{5}{2}}\\
		\end{array}	\right].
		\nonumber 
\end{equation}
\end{widetext}

\noindent A comparison of the block matrices of the projected $\hat{j}_{z}$ operator as given above to the $\hat{J}_{z}$ operator in the $|j{\rm{_{eff}}}=\frac{1}{2},m_{j}\rangle$, $|j{\rm{_{eff}}}=\frac{3}{2},m_{j}\rangle$, and $|j{\rm{_{eff}}}=\frac{5}{2},m_{j}\rangle$ bases confirms that the top 2 $\times$ 2, middle $4 \times 4$ and bottom $6 \times 6$ block matrices corresponds to $j\rm{_{eff}}=\frac{1}{2},\frac{3}{2}$, and $\frac{5}{2}$ manifolds, respectively.  By projecting the $\hat{j}_{x}$ and $\hat{j}_{y}$ operators, it can be shown that these block matrices also satisfy the canonical commutation relations of angular momentum. 

\textit{The Distortion Hamiltonian, $\hat{\mathcal{H}}_{dis}$:}  The next perturbation to the single-ion Hamiltonian corresponds to the structural deformation of the CoO unit cell that accompanies long-range antiferromagnetic order, resulting in the distortion of the crystalline electric field from ideal octahedral coordination~\cite{Schron12:86}.  While the exact symmetry of the low temperature phase has proven to be particularly contentious, we will consider a simple tetragonal distortion, corresponding to a uniaxial distortion along the $z$ axis.  Utilizing symmetry arguments,~\cite{walter84:45} the influences of such a distortion on the crystalline electric field is given by

\begin{figure*}[t!]
	\centering
	\includegraphics[width=\linewidth]{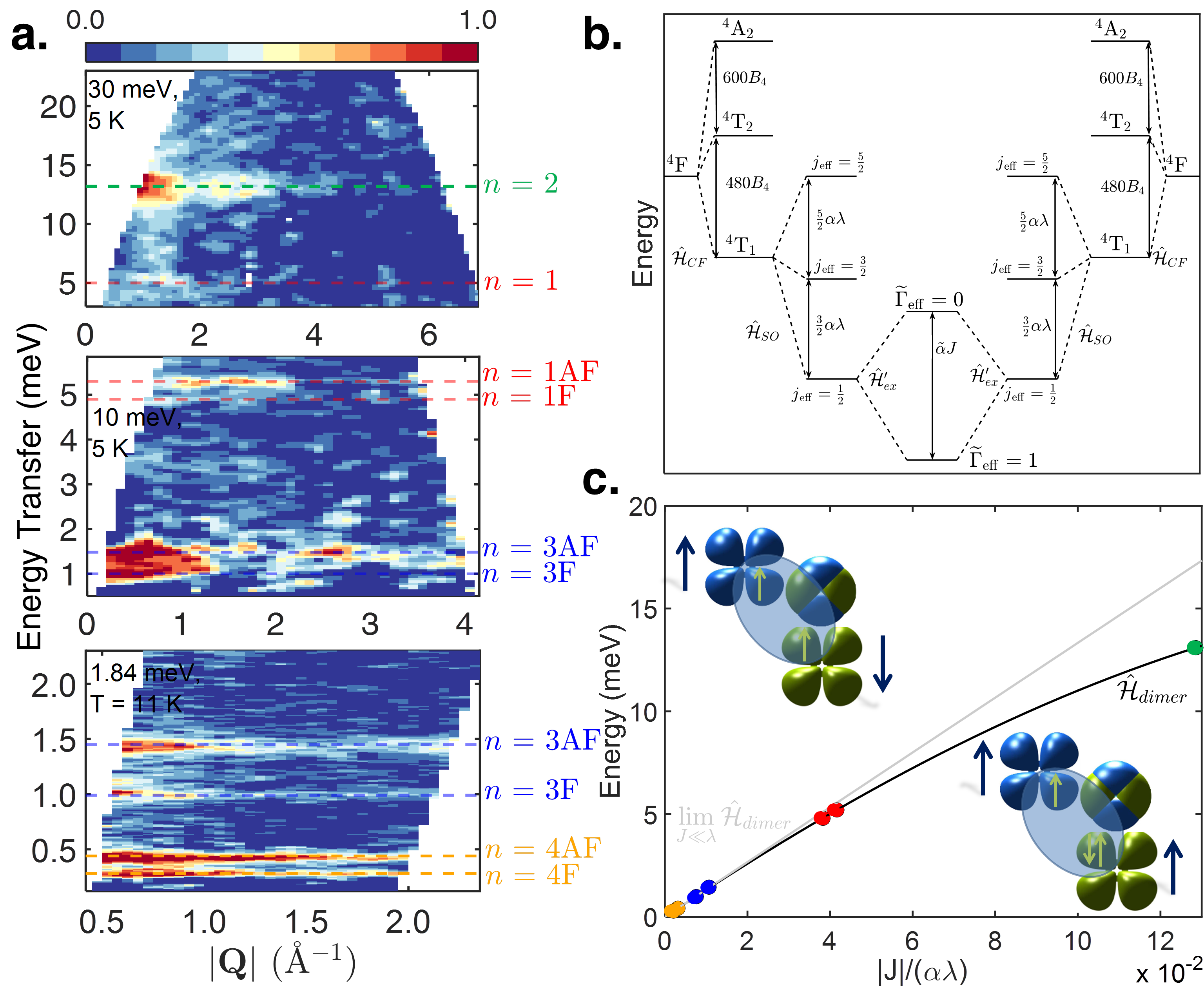}
	\caption{(a) Background (pure and nonmagnetic MgO) subtracted powder-averaged neutron-scattering intensity maps of Mg$_{0.97}$Co$_{0.03}$O measured on (top) MARI at 5~K with an $E_{i}$=30~meV, (middle) MARI at 5~K with an $E_{i}$=10~meV, and (bottom) IRIS at 11~K with an $E_{f}$ of 1.84~meV revealing seven
low-energy bands of dispersionless magnetic excitations. (b) Relevant energy scales for the effective pair Hamiltonian. (c) Calculated difference in the ground state manifold's energy eigenstates obtained from the diagonalization of the effective pair Hamiltonian (black line). The non-linearity is in constrast with the behavior predicted by the projection theorem  (gray line). (inset) The mechanism for antiferromagnetism (top) and weaker ferromagnetism (bottom) is a result of a combination of the 90$^{\rm{o}}$ Co$^{2+}$$-$O$^{2-}$$-$Co$^{2+}$ exchange pathway and the orbital degree of freedom in the $t_{2g}$ channel on each Co$^{2+}$, in agreement with the predictions of the Goodenough-Kanamori-Anderson rules. Yellow arrows denote local $t_{2g}$ spin configurations and teal arrows denote total spin configurations on each Co$^{2+}$.}
	\label{fig:fig3}
\end{figure*}

\begin{equation}
\hat{\mathcal{H}}_{dis}=B_{2}\hat{\mathcal{O}}^{0}_{2}=\Gamma  \left(\hat{l}^{2}_{z}-{2 \over 3} \right),
\nonumber
\end{equation}

\noindent with a distortion parameter $\Gamma$. 

\textit{The Molecular Field Hamiltonian, $\hat{\mathcal{H}}_{MF}$:}  The final term in the single-ion Hamiltonian corresponds to the effect of the molecular field that results from the magnetic order of Co$^{2+}$ moments in the rocksalt lattice. $\hat{\mathcal{H}}_{MF}$ behaves as a Zeeman-like term, resulting in a splitting of the nearly degenerate $j\rm{_{eff}}$ levels.  By considering a single dominant next-nearest neighbor $180^{\circ}$ Co$^{2+}$-O$^{2-}$-Co$^{2+}$  superexchange pathway with a magnetic exchange constant $J_{2}$, the corresponding mean molecular field is given by 

\begin{equation}
\hat{\mathcal{H}}_{MF}=\sum_{i}H_{MF}(i)\hat{S}_{z} =2z_{2}J_{2}\langle \hat{S_{z}}\rangle \hat{S}_{z},
\nonumber
\end{equation}

\noindent where $H_{MF}(i)$ is defined by Eq.~\ref{eq:molecular_field_2} above, and $z_{2}$ denotes the number of next-nearest neighbors.   As summarized by Fig.~\ref{fig:fig1}(c), the result of such a strong value of this exchange interaction is the significant entanglement between individual $j\rm{_{eff}}$ levels, in contrast with other Co$^{2+}$-based magnets such as CoV$_{2}$O$_{6}$,~\cite{wallington15:92} CoV$_{3}$O$_{8}$,~\cite{sarte18:98_2} and CoNb$_{2}$O$_{6}$~\cite{Cabrera14:90}, where the degree of mixing is much weaker. In these particular magnets, the spin-orbit split levels remain well-separated in energy, and therefore can be considered as $j{\rm_{eff}}=\frac{1}{2}$ magnets. The strong intertwining in CoO represents a limitation imposed on approaches based on conventional linear spin wave theory, and the reason why a multi-level spin-orbit exciton model needs to be considered.    

\subsubsection{Inter-Ion Hamiltonian $\hat{\mathcal{H}}_{2}$}

As summarized by Eq.~\ref{Inter_ion}, the inter-ion Hamiltonian is defined by the exchange parameters $J(ij)$ between sites $i$ and $j$.   In contrast to the parameters for the single-ion Hamiltonian $\hat{\mathcal{H}}_{1}$: $10Dq$, $\lambda$, and $\hat{\mathcal{H}}_{MF}$  discussed above, there does not exist a widely accepted set of experimentally determined exchange constants for CoO~\cite{feygenson11:83}.  Given the complexity of the mixed $j\rm{_{eff}}$ levels (Fig. \ref{fig:fig1}(c)), we have previously investigated the pair response in the dilute monoxide Mg$_{0.97}$Co$_{0.03}$O,~\cite{sarte18:98} where chemical dilution removes both the magnetic order-induced molecular field and the accompanying structural distortion that are originally present in CoO~\cite{Buyers84:30,furrer11:83,furrer13:85}.  A summary of the experimental results is presented in Fig.~\ref{fig:fig3}, taken from Ref. \onlinecite{sarte18:98}. Utilizing probabilistic arguments, it was shown that the series of well-defined low energy magnetic excitations (Fig.~\ref{fig:fig3}(a)) present in Mg$_{0.97}$Co$_{0.03}$O correspond to excitations of Co$^{2+}$ pairs. These pairs are described by the effective pair Hamiltonian given by 

\begin{equation}
\hat{\mathcal{H}}_{pair} = \alpha\lambda\hat{\mathbf{l}}_{1}\cdot\hat{\mathbf{S}}_{1} + \alpha\lambda\hat{\mathbf{l}}_{2}\cdot\hat{\mathbf{S}}_{2} + 2J_{1,2}\hat{\mathbf{S}}_{1}\cdot\hat{\mathbf{S}}_{2},
\label{pair_ham}
\end{equation} 

\noindent corresponding to a 144 $\times$ 144 matrix in terms of the two-particle basis of $|l_{1}=1,m_{l_{1}},s_{1}=\frac{3}{2},m_{s_{1}} \rangle \otimes |l_{2}=1,m_{l_{2}},s_{2}=\frac{3}{2},m_{s_{2}} \rangle$, where $l_{i}$, $m_{l_{i}}$, $s_{i}$, and $m_{s_{i}}$ denote the eigenvalues of the $\hat{\mathbf{l}}$, $\hat{l}_{z}$, $\hat{\mathbf{S}}$, and $\hat{S}_{z}$ operators, respectively, for the $i^{th}$ particle. As summarized schematically in Fig. \ref{fig:fig3}(b), the pair Hamiltonian $\hat{\mathcal{H}}_{pair}$ describes individual $j\rm{_{eff}}=\frac{1}{2}$ pair excitations as transitions between triplet ($\tilde{\Gamma}=1$) and singlet ($\tilde{\Gamma}=0$) levels separated by an energy $\Delta E = \tilde{\alpha}J$, where $\tilde{\alpha}$ behaves as an effective conversion factor between the energy transfer measured experimentally and the corresponding desired magnetic exchange constants.  The solution to $\hat{\mathcal{H}}_{pair}$ as a function of exchange energy $J_{1,2}$ is shown in Fig. \ref{fig:fig3}(c), with the solid black line and colored points corresponding to the exact solution to the above Hamiltonian and the measured energy positions, respectively. For comparison, the behavior predicted by the projection theorem of angular momentum is also presented. The deviation of the exact solution from the linear behavior predicted by the projection theorem is a consequence of the coupling of the ground state $j\rm{_{eff}}=\frac{1}{2}$ and higher energy $j\rm{_{eff}}={3\over 2}$ manifolds. Since the degree of coupling increases as $|J| \rightarrow |\lambda|$, the predicted values for exchange constants with larger magnitudes, specifically $J_{2}$, are particularly sensitive to the value of the spin-orbit coupling constant $\lambda$.  This point will be addressed below in the context of the analysis of the single crystal data. 

While the energy dependence affords estimates of the exchange constants, the relative distance $\mathbf{R}$ between the two Co$^{2+}$ spins that participate in the exchange interaction, and thus the relative coordination shell (Fig.~\ref{fig:fig3}(a)) can be determined for each magnetic excitation from their momentum dependence \emph{via} the first moment sum rule~\cite{hohenberg74:10,sarte18:98_2,heraldsen05:71}, as is summarized in Fig.~\ref{fig:fig3}.   Given the ground state for antiferromagnetically/ferromagnetically coupled Co$^{2+}$ ions is a triplet/singlet, the temperature dependence was used to establish the sign of the exchange constant~\cite{xu00:84}.  A final summary of the estimates of the exchange constants for the first four coordination shells of CoO is presented in Table \ref{tab:1} below.

\subsection{Mean-Field Theory for Multi-Level Spin-Orbit Excitons} \label{sec:MF_theory}

As discussed above, the modeling of the neutron response requires an understanding of both the Hamiltonian and its commutator with the spin operators.   In the previous section, we diagonalized the single-ion Hamiltonian $\hat{\mathcal{H}}_{1}$ such that $\hat{\mathcal{H}}_{1}|n\rangle = \omega_{n}|n\rangle$.  Since all terms of the inter-ion Hamiltonian $\hat{\mathcal{H}}_{2}$ are based on the components of the spin operator $\hat{\mathbf{S}}$, these operators can be rotated onto the basis states of the single-ion Hamiltonian by use of the ladder operators that were previously defined in Eq.~\ref{ladder} with such a coordinate rotation being given by    

\begin{equation} 
 	\hat{S}_{(\pm,z)}  = \sum\limits_{mn}\hat{S}_{(\pm,z)mn}C^{\dagger}_{m}C_{n} \label{eq:der_11}.
 \end{equation}
 
By writing the full Hamiltonian $\hat{\mathcal{H}}= \hat{\mathcal{H}}_{1} + \hat{\mathcal{H}}_{2}$ in terms of the ladder operators as defined in Eq.~\ref{eq:der_11}, and using the definition of the inter-level susceptibility $\hat{G}$ defined by

\begin{equation}
G^{\alpha\beta}(i,j,\omega) = \sum\limits_{mn}\hat{S}_{\alpha mn}\hat{G}^{\beta}(m,n,i,j,\omega), 
\label{eq:sum} 
\end{equation}

\noindent where the indices $\alpha,\beta$ are either $+$, $-$, or $z$, the second term on the RHS of the equation-of-motion of the Green's function (Eq.~\ref{eq:10}) reduces to three sets of commutators, termed  \textit{diagonal}, \textit{transverse}, and \textit{longitudinal}, with each involving spin operators rewritten in terms of ladder operators, as discussed in Section~III of \emph{Supplementary Information}~\cite{suppl}. Buyers \emph{et al.},~\cite{buyers75:11} demonstrated that by combining the \emph{random phase decoupling method}~\cite{wolff60:120,cooke73:7,yamada67:22,yamada66:21} (Section IV of the \emph{Supplementary Information}~\cite{suppl}) with the definitions of both the single-site response function    

\begin{equation}
g^{\alpha\beta}(E) = \sum\limits_{n}\left\{\frac{S_{\alpha 0n}S_{\beta n0}}{E +i \Delta  - E_{n0}} - \frac{S_{\alpha n0}S_{\beta 0n}}{E + i \Delta + E_{n0}} \right\},
\label{eq:small_g} 
\end{equation}

\noindent and the Fourier transform of the exchange interaction $J(\mathbf{Q})$  

\begin{equation}
J(\mathbf{Q}) = \sum\limits_{i \neq j}J_{ij}e^{i\mathbf{Q}\cdot\mathbf{d}_{ij}},
\label{eq:j_q}
\end{equation} 

\noindent the full neutron response  Fourier transformed into momentum $\mathbf{Q}$-space can be written as a set of coupled linear equations given by

\begin{widetext}
\begin{equation}
G^{\alpha\beta}(\mathbf{Q},E) = g^{\alpha\beta}(E) + g^{\alpha+}(E)J(\mathbf{Q})G^{-\beta} (\mathbf{Q},E) +  g^{\alpha-}(E)J(\mathbf{Q})G^{+\beta} (\mathbf{Q},\omega) + 2g^{\alpha z}(E)J(\mathbf{Q})G^{z\beta} (\mathbf{Q},E),
\label{eq:alphabeta} 
\end{equation}
\end{widetext}

\noindent where $\omega$ has been relabeled as $E=\hbar\omega$. Here, we have employed the $T\rightarrow0$~K single-site response function since the energy transfers under consideration in the current investigation ($\geq~20$~meV) are much larger than the sample temperature ($\sim$~0.5~meV). The denominator of $g^{\alpha\beta}$ consists of $E_{no}=\hbar \omega_{n}-\hbar \omega_{0}$ corresponding to the energy associated with a transition from the ground state $|0\rangle$ to the $|n\rangle$ eigenstate of the single-ion Hamiltonian $\hat{\mathcal{H}}_{1}$, while the presence of the positive infinitesimal $\Delta$ is to ensure analyticality, and was set to 50\% of the experimental resolution width (HWHM) on MERLIN~\cite{bewley06:385} that was calculated by PyChop~\cite{PyChop}. Coupling between the single-site response functions, and thus the dispersion of the total response functions $G^{\alpha\beta}$ is defined by $J(\mathbf{Q})$ which is parameterized by both $J_{ij}$ and $\mathbf{d}_{ij}$ denoting the exchange constant and displacement vector, respectively, between sites $i$ and $j$.  As a first approximation, our calculations have considered the simplest case where the exchange interaction is spatially isotropic.  We note that in general this is not case, owing to the anisotropy of the orbital configuration of Co$^{2+}$.

By considering all possible combinations of indices $\alpha,\beta$ in Eq.~\ref{eq:alphabeta} and noting that the non-zero single-site response functions for Co$^{2+}$ in such a highly symmetric environment are: $g^{+-}$, $g^{-+}$, and $g^{zz}$, only three non-zero Green's functions are obtained:

\begin{align}
	G^{+-}(\mathbf{Q},E)  &=  g^{+-}(E) + g^{+-}(E)J(\mathbf{Q})G^{+-} (\mathbf{Q},E) \nonumber \\
	G^{-+}(\mathbf{Q},E)  &= g^{-+}(E) + g^{-+}(E)J(\mathbf{Q})G^{-+} (\mathbf{Q},E)  \nonumber \\
	G^{zz}(\mathbf{Q},E)  &= g^{zz}(E) + 2g^{z z}(E)J(\mathbf{Q})G^{zz} (\mathbf{Q},E),
	\label{eq:Gab}
\end{align}

\noindent with both $G^{++}$ and G$^{--}$ being both zero, as required by definition of the retarded Green's function. 

The simplest model for the long-range antiferromagnetic order in CoO is a type-II collinear antiferromagnetic magnetic structure~\cite{Shull51:102}. Corresponding to (111) ferromagnetic sheets stacked antiferromagnetically along the [111] direction, this type of magnetic structure has been observed in CoO under pressure, despite the suppression of the structural distortion~\cite{Ding06:74}. Such a model implies that CoO can be reduced to two unique magnetic sublattices; thus, the site indices $i$ and $j$ assume labels of either 1 or 2, and Eq.~\ref{eq:Gab} becomes four coupled linear equations 

\begin{align} 
G^{+-}_{11}(\mathbf{Q},E) &= g^{+-}_{1}(E) + g^{+-}_{1}(E) J_{s}(\mathbf{Q})G^{+-}_{11}(\mathbf{Q},E) \nonumber \\
&+ g^{+-}_{1}(E)J_{d}(\mathbf{Q})G^{+-}_{21}(\mathbf{Q},E)   \nonumber \\ 
G^{+-}_{21}(\mathbf{Q},E) &= g^{+-}_{2}(E)J_{s}(\mathbf{Q})G^{+-}_{21}(\mathbf{Q},E) \nonumber \\
&+ g^{+-}_{2}(E)J_{d}(\mathbf{Q})G^{+-}_{11}(\mathbf{Q},E)    \nonumber \\ 
G^{+-}_{12}(\mathbf{Q},E) &= g^{+-}_{1}(E)J_{s}(\mathbf{Q})G^{+-}_{12}(\mathbf{Q},E) \nonumber \\
&+ g^{+-}_{1}(E)J_{d}(\mathbf{Q})G^{+-}_{22}(\mathbf{Q},E)    \nonumber \\ 
G^{+-}_{22}(\mathbf{Q},E) &= g^{+-}_{2}(E) + g^{+-}_{2}(E)J_{s}(\mathbf{Q})G^{+-}_{22}(\mathbf{Q},E) \nonumber \\
&+ g^{+-}_{2}(E)J_{d}(\mathbf{Q})G^{+-}_{12}(\mathbf{Q},E) \label{eq:ch6_4a}, 
\end{align}

\noindent and

\begin{align} 
G^{zz}_{11}(\mathbf{Q},E) &= g^{zz}_{1}(E) + 2g^{zz}_{1}(E) J_{s}(\mathbf{Q})G^{zz}_{11}(\mathbf{Q},E) \nonumber \\
&+ 2g^{zz}_{1}(E)J_{d}(\mathbf{Q})G^{zz}_{21}(\mathbf{Q},E)  \nonumber \\ 
G^{zz}_{21}(\mathbf{Q},E) &= 2g^{zz}_{2}(E)J_{s}(\mathbf{Q})G^{zz}_{21}(\mathbf{Q},E) \nonumber \\
&+ 2g^{zz}_{2}(E)J_{d}(\mathbf{Q})G^{zz}_{11}(\mathbf{Q},E)    \nonumber \\ 
G^{zz}_{12}(\mathbf{Q},E) &= 2g^{zz}_{1}(E)J_{s}(\mathbf{Q})G^{zz}_{12}(\mathbf{Q},E) \nonumber \\
&+ 2g^{zz}_{1}(E)J_{d}(\mathbf{Q})G^{zz}_{22}(\mathbf{Q},E)    \nonumber \\ 
G^{zz}_{22}(\mathbf{Q},E) &= g^{zz}_{2}(E) + 2g^{zz}_{2}(E)J_{s}(\mathbf{Q})G^{zz}_{22}(\mathbf{Q},E) \nonumber \\
&+ 2g^{zz}_{2}(E)J_{d}(\mathbf{Q})G^{zz}_{12}(\mathbf{Q},E) \label{eq:ch6_4b},
\end{align}

\noindent with $J_{s}$ and $J_{d}$ denoting $J(\mathbf{Q})$ on the same ($i=j$) and different ($i\neq j$) sublattices, respectively. Solving these four coupled equations yields 

\begin{widetext}
\begin{align}
\begin{split}
G^{+-}(\mathbf{Q},E) \equiv \sum\limits_{ij}G^{+-}_{ij}(\mathbf{Q},E)  = 
{{g^{+-}_{1}(E)+g^{+-}_{2}(E)+2g^{+-}_{1}(E)g^{+-}_{2}(E)[J_{d}(\mathbf{Q}) -J_{s}(\mathbf{Q})]} \over {[1-g^{+-}_{1}(E)J_{s}(\mathbf{Q})]\cdot [1  -g^{+-}_{2}(E)J_{s}(\mathbf{Q})]-g^{+-}_{1}(E)g^{+-}_{2}(E)[J_{d}(\mathbf{Q})]^{2}}},\\ 
\\
G^{zz}(\mathbf{Q},E) \equiv \sum\limits_{ij}G^{zz}_{ij}(\mathbf{Q},E)  = 
{{g^{zz}_{1}(E)+g^{zz}_{2}(E)+4g^{zz}_{1}(E)g^{zz}_{2}(E)[J_{d}(\mathbf{Q}) -J_{s}(\mathbf{Q})]} \over {[1-2g^{zz}_{1}(E)J_{s}(\mathbf{Q})]\cdot [1 -2g^{zz}_{2}(E)J_{s}(\mathbf{Q})]-4g^{zz}_{1}(E)g^{zz}_{2}(E)[J_{d}(\mathbf{Q})]^{2}}},
\end{split}
\nonumber
\end{align}

\end{widetext}

\noindent where $G^{-+}(\mathbf{Q},E)$ has the same form as $G^{+-}(\mathbf{Q},E)$ with indices $+$ $\longleftrightarrow$ $-$. The equations above are a function of the single-site response function $g^{\alpha\beta}$ and the Fourier transform of the exchange interaction $J(\mathbf{Q})$.

\begin{figure}[htb!]
	\includegraphics[width=1.0\linewidth]{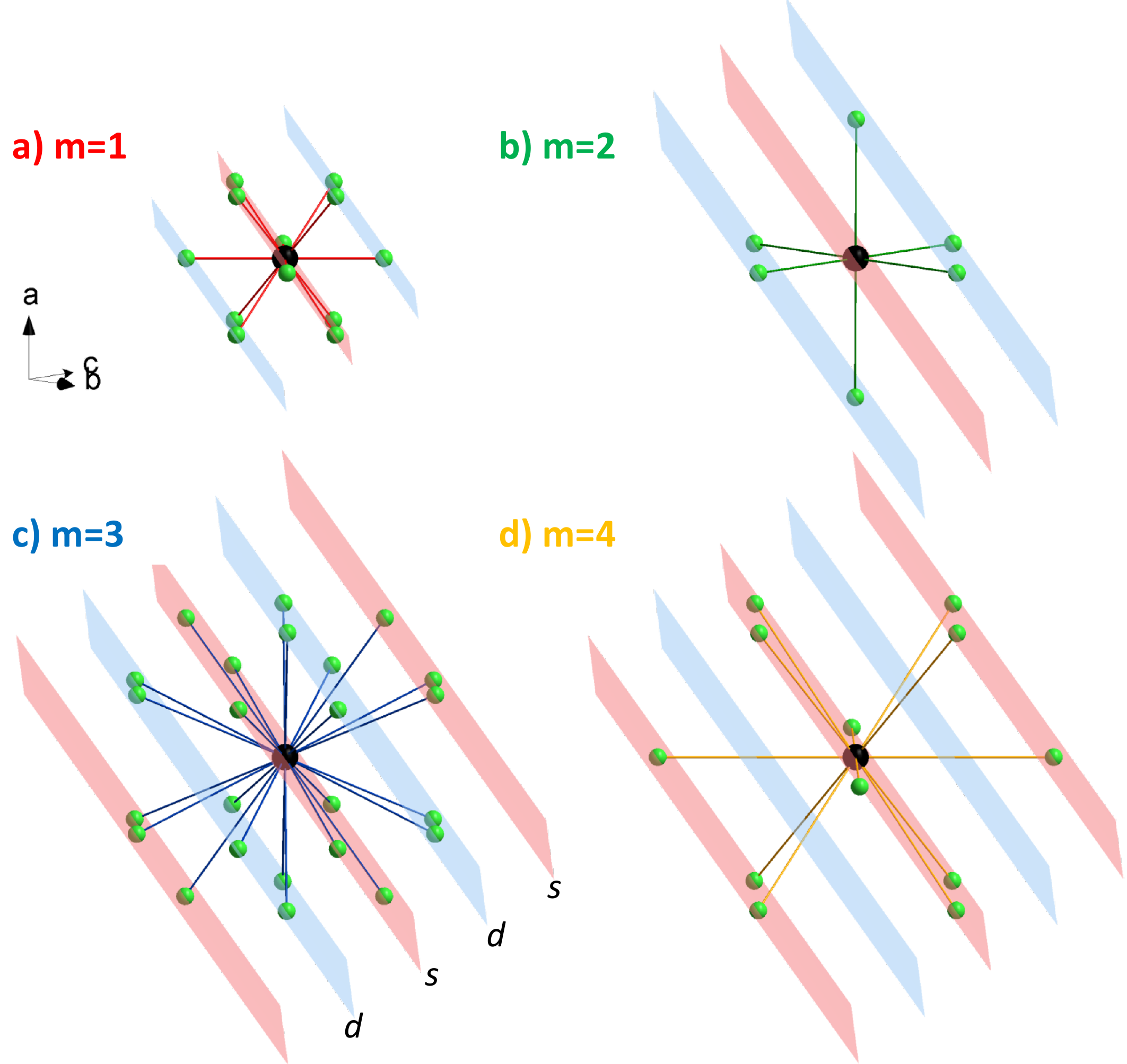}
	\caption{Isometric view of all Co$^{2+}$ cations located in the (a) first ($m=1$), (b) second ($m=2$), (c) third ($m=3$), and (d) fourth ($m=4$) coordination shells of
		the CoO rocksalt structure. For the purposes of reference, all (111) planes are labeled as either $s$ and $d$ planes with respect to the reference Co$^{2+}$ (central black site). All displacement vectors $\mathbf{d}_{m,ij}$ are listed in Tab.~SI in the \emph{Supplementary Information}~\cite{suppl} and used to calculate $J_{s}(\mathbf{Q})$ and $J_{d}(\mathbf{Q})$ discussed in the text.}
	\label{fig:figurejacky}
\end{figure}

In contrast to the single-site response function, $J(\mathbf{Q})$ does explicitly depend on the particular magnetic sublattice under consideration, stemming from the presence of the position indices $i$ and $j$ in its definition given by Eq.~\ref{eq:j_q}. The Co$^{2+}$ sites in the $d$- and $s$-sublattices for a particular coordination shell $m$ was determined by first selecting a reference Co$^{2+}$ cation, thus defining a reference (111) plane as illustrated in Fig. \ref{fig:figurejacky}. By definition of the type-II antiferromagnetic structure, Co$^{2+}$ cations located on odd integer number of $(111)$ planes away from the reference plane are defined as belonging to the $d$ sublattice, while Co$^{2+}$ cations located in the same or an even integer number of $(111)$ planes away are on the $s$ sublattice. 

The program VESTA~\cite{VESTA} was used to determine the displacement vectors to calculate $J_{s,d}$ from Eq.~\ref{eq:j_q} giving: 
 
 \begin{widetext}
 \begin{align}
 \label{eq:def1}
&J_{s}(\mathbf{Q})= \\
&  2J_{1}\{\cos(\pi(H-K)) + \cos(\pi(K-L)) + \cos(\pi(L-H))\} + ... \tag{$m=1$} \\
& 2J_{3} \left\{\cos\left(\pi\left(2H-K-L\right)\right) + \cos\left(\pi\left(2H+K+L\right)\right)+ \cos\left(\pi\left(H-2K+L\right)\right) \right.+... \tag{$m=3$} \\
& \left.{} \ \ \ \ \ \cos\left(\pi\left(-H-2K-L\right)\right)  + \cos\left(\pi\left(H-K+2L\right)\right) + \cos\left(\pi\left(-H-K+2L\right)\right)\right\}+ ...  \nonumber \\
& 2J_{4}\left\{\cos(2\pi(H-L)) + \cos(2\pi(H-K)) + \cos(\pi(L-K)) \right. + ... \tag{$m=4$} \\
& \left.{} \ \ \ \ \ \cos(\pi(H+L))  +\cos(\pi(H+K)) + \cos(\pi(K+L)) \right\} \nonumber
 \end{align}
 
 \noindent and
 
 \begin{align}
 \label{eq:def2}
&J_{d}(\mathbf{Q})=  \\
& 2J_{1}\{\cos(\pi(H+K)) + \cos(\pi(K+L)) + \cos(\pi(L+H))\} +... \tag{$m=1$} \\
&  2J_{2}\left\{\cos(2\pi H) + \cos(2\pi K) + \cos(2\pi L) \right\} +... \tag{$m=2$} \\
&  2J_{3}\left\{\cos\left(\pi\left(2H-K+L\right)\right) + \cos\left(\pi\left(2H+K-L\right)\right)  + \cos\left(\pi\left(-H-2K+L\right)\right) \right.+...\tag{$m=3$}\\
& \left.{} \ \ \ \ \ \cos\left(\pi\left(H-2K-L\right)\right)+ \cos\left(\pi\left(-H+K+2L\right)\right) +  \cos\left(\pi\left(H-K+2L\right)\right)\right\}, \nonumber
 \end{align}
 
\end{widetext}
		
\noindent where the contributions from each coordination shell $m$ have been labeled explicitly. 

By employing the definitions of the single-site response function $g^{\alpha\beta}$ (Eq.~\ref{eq:small_g}), and the Fourier transform of the exchange interaction $J_{s,d}(\mathbf{Q})$ (Eqs.~\ref{eq:def1} and \ref{eq:def2}), the total response function $G(\mathbf{Q},E)$ given by  

\begin{equation}
\begin{split}
& G(\mathbf{Q},E)  \equiv \sum\limits_{\alpha\beta}G^{\alpha\beta}(\mathbf{Q},E) = \\
& G^{+-}(\mathbf{Q},E)+G^{-+}(\mathbf{Q},E)+G^{zz}(\mathbf{Q},E),
\end{split}
\label{eq:total_green}
\end{equation} 

\noindent can be calculated numerically. In the T $\rightarrow$ 0~K limit, the imaginary part of $G(\mathbf{Q},E)$ is proportional to the dynamical structure factor (Eq.~\ref{eq:spin_structure}), and thus Eq.~\ref{dynamic_ss} may be reduced to  

\begin{equation}
S(\mathbf{Q},E) \appropto  -f^{2}(\mathbf{Q})\Im G(\mathbf{Q},E),
\nonumber
\end{equation} 

\noindent demonstrating that the imaginary component of the total response function given by Eq.~\ref{eq:total_green}, with the inclusion of the magnetic form factor which here has been approximated by the isotropic magnetic form factor $f(Q)$, is directly proportional to the neutron magnetic cross section.

\FloatBarrier 

\subsection{Parameters: Initial Values \& Orbital Configurations} \label{sec:parameters}

Having presented our model, we now discuss the parameterization of the spin-orbit excitations in CoO.  Since our model approximates CoO as a tetragonally distorted type-II antiferromagnet, the single-site response $g^{\alpha\beta}$, itself being a function of the single-ion Hamiltonian $\hat{\mathcal{H}}_{1}$ (Eq.~\ref{single_ion}), is defined by three parameters: $\lambda$, $\Gamma$, and $H_{MF}$. The spin-orbit coupling parameter $\lambda$ was taken to be $-16$~meV, corresponding to its value reported by Cowley \emph{et al.}~\cite{cowley13:88}.  An initial estimate for the mean molecular field $H_{MF}$ was determined by first extracting the value for $\sum\limits_{i>j}z_{ij}J_{ij}$ from the reported~\cite{kanamori57:17,singer56:104,nagamiya55:4} Curie-Weiss temperature $\theta{\rm{_{CW}}}=-330$~K ($-28.4$~meV) \emph{via} its mean field definition

\begin{equation}
\theta{\rm{_{CW}}} = -\frac{2}{3\zeta}S(S + 1)\sum\limits_{i>j}z_{ij}J_{ij},
\label{eq:CW} 
\end{equation}     

\noindent where $\zeta$ is a scale factor of 1.9 calculated by Kanamori~\cite{kanamori57:17} accounting for mixing between the $^{4}F$ and $^{4}P$ free-ion states. Inserting the value of $\sum\limits_{i>j}z_{ij}J_{ij}$ into the definition of $H_{MF}$ given by Eq.~\ref{eq:molecular_field_2}, yields an initial estimate of 64.8~meV. An initial estimate of the tetragonal distortion parameter $\Gamma$=$-8.76$~meV was determined by scaling the value of $-1.49$~meV that was reported for KCoF$_{3}$~\cite{buyers71:4} by an empirical factor of 0.0116/0.00197=5.89 corresponding to the ratio of their respective tetragonal distortions $\delta a/a$.       

To define $J(\mathbf{Q})$, we have taken the values for the exchange constants for the dilute monoxide Mg$_{0.97}$Co$_{0.03}$O~\cite{sarte18:98} as estimates for pure undiluted CoO since these exchange constants correspond to a Curie-Weiss temperature (Eq.~\ref{eq:CW}) in close agreement with the value reported for CoO. However, our investigation on Mg$_{0.97}$Co$_{0.03}$O also revealed that each coordination shell possessed the possibility for both antiferromagnetic and ferromagnetic coupling, with the exception of the second nearest neighbor which is fixed to be antiferromagnetic by the 180$^{\rm{\circ}}$ Co$^{2+}$-O$^{2-}$-Co$^{2+}$ superexchange pathway. As illustrated in Fig.~\ref{fig:fig3}(c,inset), such dual behavior is a direct consequence of the $t_{2g}$ degeneracy of the high spin $d^{7}$ configuration of Co$^{2+}$, and thus a particular choice of $J$, be it antiferromagnetic or ferromagnetic, corresponds to a specific local orbital configuration. By incorporating this dual behavior for coordination shells 1, 3, and 4, we must consider all $2^{3}$=8 sets of exchange constants of the form $xAxx$, where $x$ can be either antiferromagnetic (A) or ferromagnetic (F). Furthermore, since our model incorporates the effects of a tetragonal (uniaxial) distortion, we must also distinguish the involvement of a distorted or undistorted bonding configuration for each of the 8 $xAxx$ orbital configurations. Thus, with these 2 additional degrees of freedom, our model must consider 16 different orbital configurations of the form $xAxx\gamma$, where the index $\gamma$ =1 or 2, distinguishing the presence or absence of distorted bonding configurations. 

Each of these 16 $xAxx\gamma$ combinations may be interpreted as a unique orbital configuration, physically corresponding to a unique type of ``domain" in the bulk CoO single crystal, each of which is subject to a different mean molecular field $H_{MF}$. In contrast, since all cations under consideration are assumed to be Co$^{2+}$ in octahedral coordination subject to a cooperative Jahn-Teller tetragonal distortion, $\lambda$, $\Gamma$, and the individual $J_{n}$ values (where $n$ denotes a particular type of coupling in a coordination shell $m$) are fixed to be equal for each of the 16 $xAxx\gamma$ orbital configurations. By noting that our neutron spectroscopic measurements were performed with a large experimental beam that irradiated a macroscopic number of domains in the single crystal of CoO, our model considers the mean contribution from all 16 equally weighted $xxAx\gamma$ orbital configurations.  The initial parameters considered in the model and their initial values are summarized in Tab.~\ref{tab:1}.      

\begin{table}[htb!]
	\caption{Initial values (in meV) for the parameters of the spin-orbit exciton model.}
	{\renewcommand{\arraystretch}{2.5}
		\begin{tabular}{|c|c|c|}
			\hline
			~~~~~Parameter~~~~~&~~~~~Initial Value~~~~~&~~~~~Reference~~~~~\\ 
			\hline
			$\lambda$ 	&  $-16$	& \onlinecite{cowley13:88}		\\ \hline
			$\Gamma$   	& $-8.76$ 	 & \onlinecite{buyers71:4},\onlinecite{gladney66:146} 		\\ \hline
			$J_{1F}$   	& $-0.918$  	&   \multirow{8}{*}{\onlinecite{sarte18:98}}  		\\
			$J_{1AF}$ 	& 1.000  	&  		\\  
			$J_{2}$    		& 3.09 	& 	\\   
			$J_{3F}$    	& $-0.182$ 	 &	\\ 
			$J_{3AF}$    	& 0.262 	& 		\\   
			$J_{4F}$    	& $-0.0504$ 	& \\ 
			$J_{4AF}$    	& 0.0759  	& 		\\  \hline
			$H_{MF}$		& 64.8  & \onlinecite{kanamori57:17},\onlinecite{singer56:104},\onlinecite{nagamiya55:4} 					\\ \hline
	\end{tabular}}
	\label{tab:1}
\end{table}

\section{Experiment}

Having discussed both the underlying theory of Co$^{2+}$ magnetism and the corresponding physical parameters that constitute our spin-orbit exciton model, we will now address the experimental results from neutron scattering experiments on a single crystal of CoO. This section begins with a description of the experimental techniques, followed by a summary of
the neutron spectroscopic data. We conclude this section with a description of how our model was used to interpret the low-energy fluctuations of CoO deep within the N\'{e}el regime.

\subsection{Experimental Details}

\textit{Sample Preparation:} Polycrystalline samples of CoO were synthesized by annealing high purity Co$_{3}$O$_{4}$ ($>$~99.99\%) under flowing Ar at 1200$\rm{^{o}}$C for 36~h with intermittent grinding until laboratory x-ray diffraction confirmed the absence of the Co$_{3}$O$_{4}$ precursor.~The phase pure CoO powder was compressed into cylindrical rods using a hydraulic press and subsequently annealed under flowing Ar at 1275$\rm{^{o}}$C for 24~h in a horizontal annealing furnace.~Crystal growth was performed using the floating zone technique with a four-mirror optical floating zone halogen furnace (CSI system Inc.), yielding a 10~g single crystal of CoO ($l$~=~50~mm, $\phi$~=~8~mm).~The feed and seed rods were counter-rotated at 35~rpm with a vertical translation of 2 to 4~mm~hr$^{-1}$ in a pure Ar atmosphere.~The initial polycrystalline seed rod was replaced for subsequent runs by single crystal seeds from earlier growths. Previous~\cite{cowley13:88} optical and scanning electron microscopy, x-ray diffraction and DC susceptibility measurements on the single crystal confirmed the presence of a single growth domain with a mosaic spread of approximately 0.1$\rm{^{o}}$ and the absence of both multiple magnetic or strain domains on the crystal surface and Co$_{3}$O$_{4}$ impurities, respectively.     

\textit{Neutron Inelastic Scattering:} Neutron inelastic scattering measurements were performed on the MERLIN direct geometry chopper spectrometer~\cite{bewley06:385} at the ISIS neutron spallation source (Didcot, U.K.).~The $t_{0}$ chopper was spun at 50~Hz in parallel with the ``sloppy" Fermi chopper package to fix the incident energy with the energy transfer defined as $E=E_{i}-E_{f}$.~To access a large dynamic range, three fixed incident energies $E_{i}$ of 110~meV, 75~meV, and 45~meV were selected with Fermi chopper frequencies of 350, 300, and 250~Hz, providing a resolution at the elastic line (E=0) of 7.3, 4.8, and 2.7~meV, respectively.~A 5~g portion of the CoO single crystal was mounted in a top loading closed cycle refrigerator such that the [110] and [001] crystallographic axes lay within the horizontal plane. A tomographic reconstruction in momentum-space was accomplished by rotating the crystal about the [0$\overline{1}$0] axis over 120$\rm{^{o}}$ in 0.5$\rm{^{o}}$ steps. 

The four dimensional $(\mathbf{Q},E)$ experimental data at each angle $\Psi$ and $E_{i}$ was collected at 5~K for an accumulated charge of 30 $\rm{\mu A\cdot hr}$ on the spallation target.~The raw experimental data was normalized by accumulated proton charge, corrected for detector-efficiency using a vanadium reference sample, and reduced by the Mantid data analysis software.\cite{arnold14:764,taylor12:57}~Visualization and manipulation of reduced experimental data including rebinning and projections were performed using the HORACE software package distributed by ISIS~\cite{ewings16:834}.

\subsection{Experimental Data}

\begin{figure}[t!]
	\centering
	\includegraphics[width=0.89\linewidth]{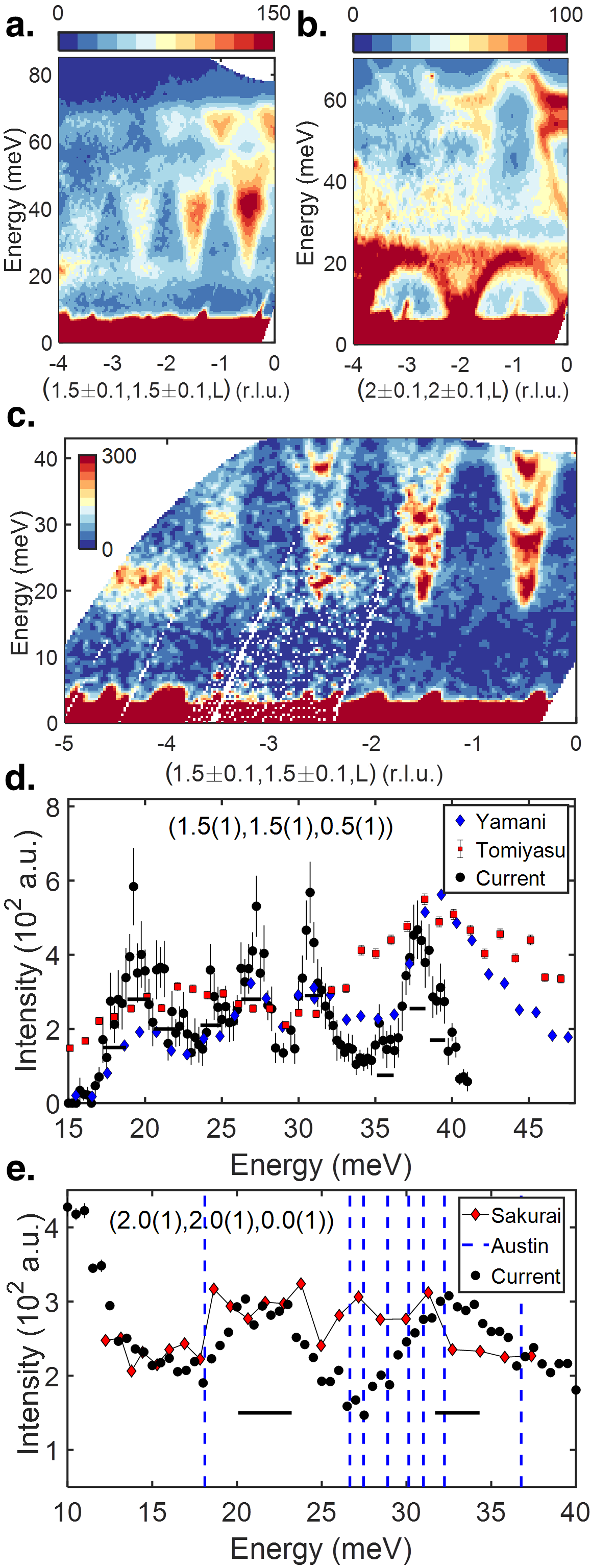}
	\caption{$(\mathbf{Q},E)$ slices of CoO measured on MERLIN at 5 K with an $E_{i}$ of (a) 110 meV, (b) 75 meV, and (c) 45 meV. All $(\mathbf{Q},E)$ slices have been folded along [001]. A comparison of $\mathbf{Q}$-integrated cuts of (c) and (b) with previous measurements in the literature at the (d) magnetic zone centers, and (e) magnetic zone boundaries, respectively. Solid lines in (e) indicate the location of excitations previously determined by IR spectroscopy~\cite{Austin70:33}. Horizontal bars indicate instrumental resolution.}
	\label{fig:figure3}
\end{figure}

\begin{figure*}[htb!]
\centering
\includegraphics[width=0.8\linewidth]{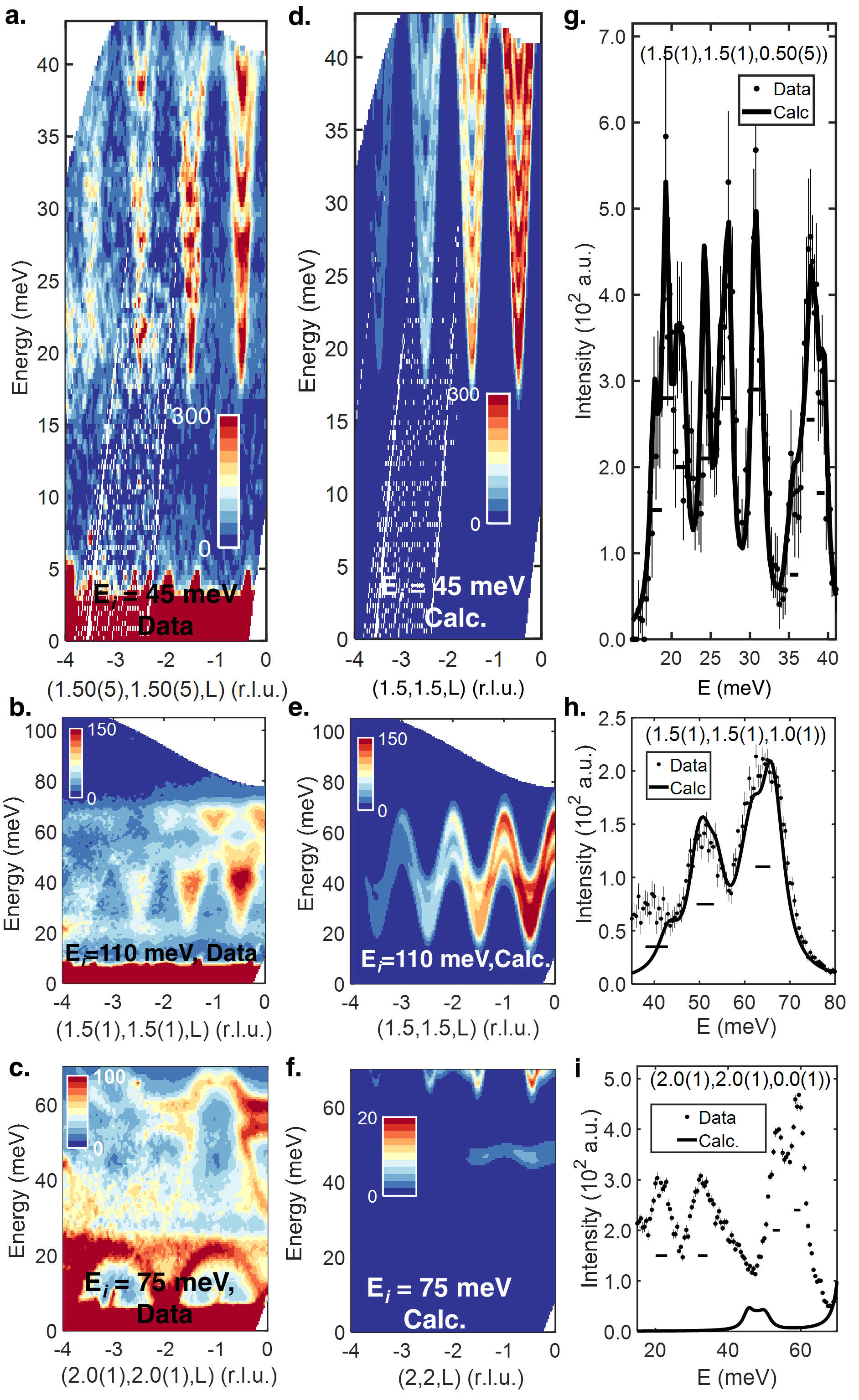}
\caption{Comparison of $(\mathbf{Q},E)$ slices and corresponding $\mathbf{Q}$-integrated cuts for CoO measured on MERLIN at 5~K and calculated with a mean-field multi-level spin-orbit exciton model employing the refined parameters listed in Tabs.~\ref{tab:2} and~\ref{tab:3} for an $E_{i}$ of (a,d,g) 45~meV, (b,e,h) 110~meV, and (c,f,i) 75~meV. Horizontal bars in $\mathbf{Q}$-integrated cuts indicate experimental resolution. All $(\mathbf{Q},E)$ slices have been folded along [001]. Individual contributions for each $xAxx\gamma$ orbital configuration to the $\mathbf{Q}$-integrated cuts presented in (g,h) are illustrated in Fig.~S1.}
\label{fig:modelA}
\end{figure*}

We begin by first presenting a summary of the experimental data from single crystal neutron spectroscopic measurements allowing for a direct comparison with previous work on CoO to establish consistency.  The experimental data from the MERLIN chopper spectrometer at 5~K is presented in Fig. \ref{fig:figure3} in the form of $(\mathbf{Q},E)$ slices along both $(1.5\pm0.1,1.5\pm0.1,L)$ (Figs. \ref{fig:figure3}(a,c)) and $(2.0\pm0.1, 2.0\pm0.1, L)$ (Fig. \ref{fig:figure3}(b)) capturing both the magnetic zone center and boundary, where both are compared to previously published work in the form of $\mathbf{Q}$-integrated cuts presented in Figs. \ref{fig:figure3}(d) and (e), respectively.~As illustrated in Fig. \ref{fig:figure3}(a), a $(\mathbf{Q},E)$ slice with an incident energy $E_{i}$=110~meV exhibits a band of excitations extending from $\sim$ 20~meV up to $\sim$ 60~meV energy transfer, corresponding to a similar range in energy transfer reported by previous THz~\cite{Satoh12:116} and Raman~\cite{Hayes74:14} spectroscopic measurements. ~These excitations decrease in intensity with increasing $L$, as is expected for the Co$^{2+}$ magnetic form factor, thus indicating these excitations are possibly magnetic.~A higher resolution slice with an $E_{i}$= 45~meV (Fig.~\ref{fig:figure3}(c)) reveals that the band of excitations corresponds to a fine structure consisting of a series of modes that are unevenly spaced in energy, in agreement with previous triple-axis,~\cite{yamani10:88,sakurai68:167} time-of-flight data,~\cite{tomiyasu06:75} and Raman spectroscopy,~\cite{Struzhkin93:A168} with the exception of a broader peak reported for triple-axis measurements at $\sim40$~meV. However, it is important to note that the previously reported triple-axis measurements employed final energies $E_{f}$=14.6~meV and 30.5~meV, both of which potentially produce spurious signals near 40~meV as a result of weak elastic scattering corresponding to $E_{i}$ $\rightarrow$ 4 $E_{f}$ ($\lambda_{f}/2$) and 4 $E_{i}$ ($\lambda_{i}/2$) $\rightarrow$ 9 $E_{f}$ ($\lambda_{f}/3)$~\cite{Shirane:book}.~This weak elastic process may have contributed to the extra scattering intensity that was observed in the previously reported triple-axis data, yet is absent in our current time-of-flight data.  We also note the presence of nearly dispersionless optical phonon branches near this energy may also contribute to the overall neutron cross section~\cite{Wdowik07:75,goyal77:79}.~We will later discuss how this fine structure near the magnetic zone center can be understood in terms of the spin-orbit excitations.

Slices and $\mathbf{Q}$-integrated cuts through the magnetic zone boundary presented in Figs.~\ref{fig:figure3}(b) and (e) shows additional complexity present in the neutron response.  At low energy transfers below $\sim$20 meV, strong acoustic phonons can be seen in  Fig.~\ref{fig:figure3}(b) to disperse from the even integer positions in momentum transfers.~The phonon nature of these excitations is confirmed by the fact that the cross section grows with increasing momentum transfer $L$, in contrast to the magnetic cross section that is subject to the Co$^{2+}$ form factor.  In addition to the acoustic phonons, a flat band is observed at $E \sim$ 20~meV.~This band also exhibits higher intensity at large momentum transfers, indicating lattice fluctuations as the origin. However, it should be noted that a peak in the aluminum phonon density of states exists near this energy transfer, thus suggesting that this particular band likely corresponds to scattering from the sample can and walls of the cryostat.  At higher energy transfers, two distinct bands of excitations are present over a large range of energy transfers spanning from $\sim$30 to $\sim$70~meV.  The intensity of these bands weaken with momentum transfer $L$; however, as illustrated in Fig.~\ref{fig:figure3}(e), these excitations overlap with lattice vibrational modes previously identified by infrared spectroscopy~\cite{Austin70:33}.  We will later discuss the origin of these excitations in terms of magneto-vibrational scattering~\cite{Egelstaff:book} by comparing the dispersion of these two high energy bands at the magnetic zone boundary to those of phonons measured at large momentum transfers.

\subsection{Comparison between Experimental Data and the Spin-Orbit Exciton Calculation}~\label{sec:C}

Having summarized our experimental data, we now present a direct comparison to our calculated parameterization based on the Green's function approach that has been discussed above.   Here, we first present the final model used to describe the experimental data. This is followed by a discussion concerning how such a model and its refined parameters were obtained.  

\begin{figure}[htb!] 
\centering
\includegraphics[width=1.0\linewidth]{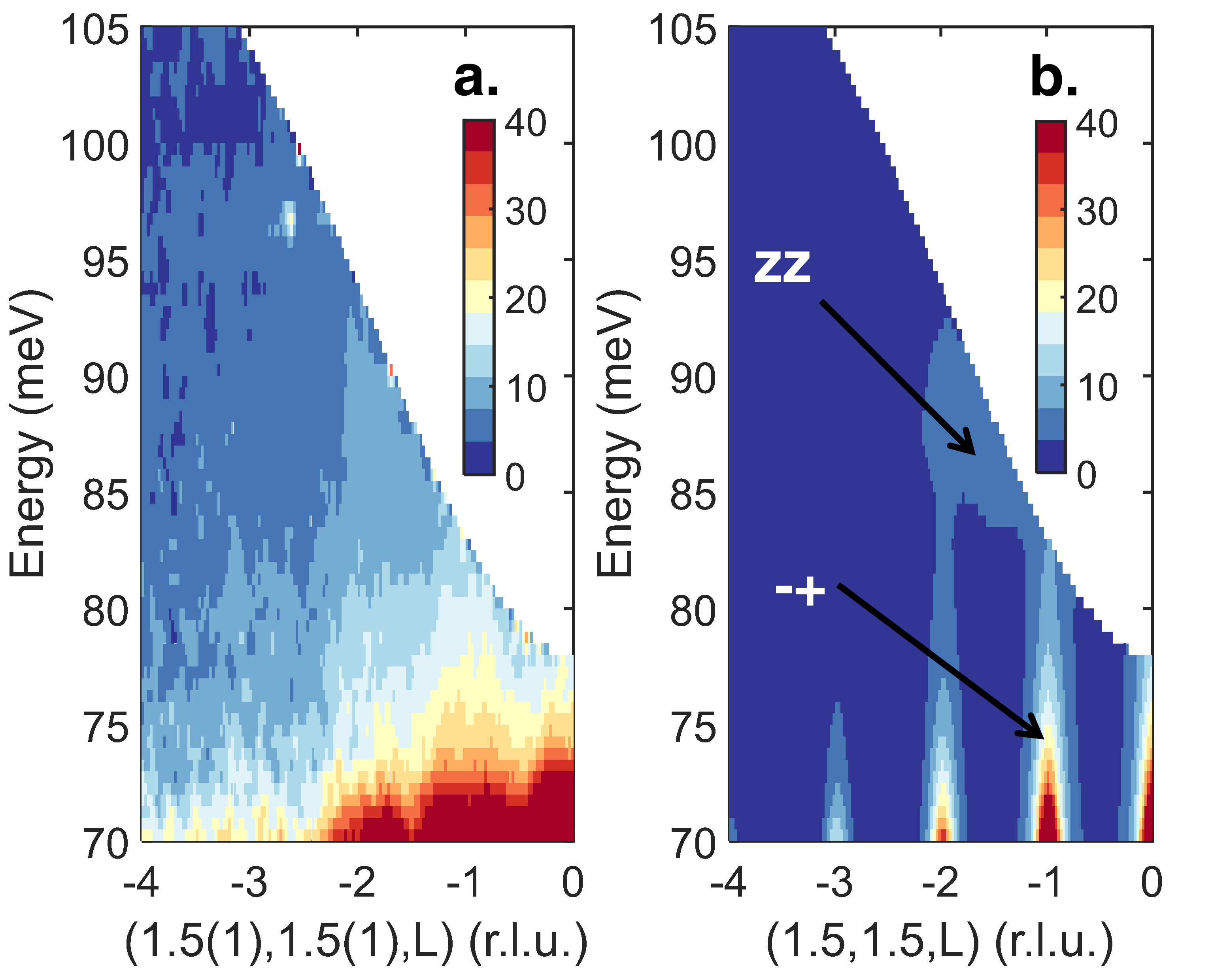}
\caption{Comparison of (a) measured ($E_{i}=110$~meV, 5~K) and (b) calculated $(\mathbf{Q},E)$ slices at high energy transfers illustrating the individual contributions of the $G^{-+}$ and $G^{zz}$ components to the total neutron response. The $(\mathbf{Q},E)$ slice presented in (a) has been folded along [001].}
\label{fig:highe_comparison}
\end{figure}

As illustrated in Fig.~\ref{fig:modelA}, by allowing the value of mean molecular field $H_{MF}$ to refine independently for each of the equally weighted 16 $xAxx\gamma$ domains, each possessing identical refined values of $\lambda$, $J$, and $\Gamma$, our mean-field multi-level spin-orbit exciton model successfully reproduces both the fine structure at the magnetic zone center and the broad excitations at the $(1.5,1.5,-1)$ zone boundary, while capturing the steeply dispersive columns of scattering observed at higher energy transfers (Fig.~\ref{fig:highe_comparison}). The refined values of the mean molecular field $H_{MF}$ are listed in Tab.~\ref{tab:3} for the 16 $xAxx\gamma$  orbital configurations, each with refined values of $\lambda$, $\Gamma$, and $J_{m,\xi}$ summarized in Tab.~\ref{tab:2}.

\begin{table}[htb!]
\caption{Refined values (in meV) of the mean molecular field parameter $H_{MF}$ for all 16 $xAxx\gamma$ orbital configurations considered in a mean-field multi-level spin-orbit exciton model, each with refined values of $\lambda$, $\Gamma$, and $J_{m,\xi}$ listed in Tab.~\ref{tab:2}. Numbers in parentheses indicate statistical errors.}
{\renewcommand{\arraystretch}{1.8}
\begin{tabular}{|c|c|}
\hline
~~~~Orbital Configuration~~~~&~~~~Refined Value~~~~\\ 
\hline\hline
AAAA1 & 62.4(2) \\ \hline
AAAA2 &46.2(1) \\ \hline\hline
AAAF1 & 55.3(2) \\ \hline 
AAAF2 & 53.9(1) \\ \hline \hline
AAFA1 & 46.2(1)\\ \hline
AAFA2 & 49.9(1)\\ \hline \hline
AAFF1 & 56.2(2)\\ \hline
AAFF2 & 56.0(2)\\ \hline \hline
FAAA1 & 47.3(1)\\ \hline
FAAA2 & 47.9(1)\\ \hline \hline
FAAF1 & 58.9(3)\\ \hline
FAAF2 & 58.8(2)\\ \hline \hline
FAFA1 & 61.6(3) \\ \hline
FAFA2 & 60.9(3)\\ \hline \hline
FAFF1 & 48.9(1)\\ \hline
FAFF2 & 59.5(2)\\ \hline \hline
Average & 54.4(4)\\ \hline 
\end{tabular}}
\label{tab:3}
\end{table}

\begin{table}[htb!]
\caption{Summary of the initial values, parameter spaces, and refined values for the parameters of the mean-field multi-level spin-orbit exciton model. All values are reported in meV and numbers in parentheses indicate statistical errors.}
{\renewcommand{\arraystretch}{1.8}
\begin{tabular}{|c|c|c|c|}
\hline
Parameter&Initial Value&~~Range~~&Refined Value\\ 
\hline
$\lambda$ & $-16$ & [$-19$,$-13$] & $-19.00(1)$ \\ \hline
$\Gamma$ & $-8.76$ & [$-8.76$,$-6.16$] & $-6.16(1)$ \\ \hline
$J_{1F}$ & $-0.918$ & [$-1.134$,$-0.730$]& $-0.780(1)$ \\ \hline
$J_{1AF}$ & 1.000 & [0.798,1.24] & 0.848(1) \\ \hline
$J_{2}$ & 3.09 & [2.29,4.55]& 2.43(1) \\ \hline
$J_{3F}$ & $-0.182$  & [$-0.220$,$-0.145$]& $-0.154(1)$ \\ \hline 
$J_{3AF}$ & 0.262 & [0.209,0.316]& 0.223(1) \\ \hline
$J_{4F}$ & $-0.0504$ & [$-0.0581$,$-0.0402$]& $-0.0428(1)$ \\ \hline
$J_{4AF}$ & 0.0759 & [0.0606,0.0874] & 0.0645(1) \\ \hline
$H_{MF}$ & 64.8 & [0,100] & Tab.~\ref{tab:3} \\ \hline
\end{tabular}}
\label{tab:2}
\end{table}

\begin{figure*}[!htbp]
\centering
\includegraphics[width=1.0\linewidth]{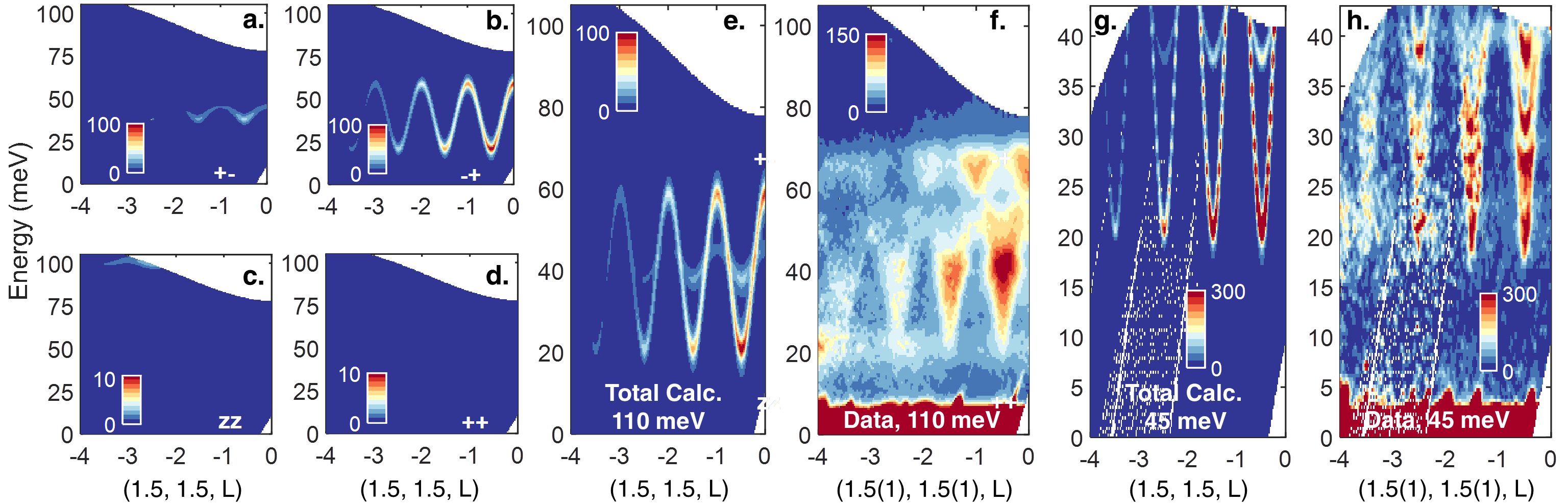}
\caption{A comparison of the calculated $(\mathbf{Q},E)$ slices along $(1.5,1.5,L)$ for (a) $G^{+-}$, (b) $G^{-+}$, (c) $G^{zz}$, and (d) $G^{++}$ components of (e) the total response function $G$ and (f) the corresponding slice measured on MERLIN at T=5~K with an $E_{i}$=110~meV, with (g,h) the same comparison for an $E_{i}$=45~meV. The calculated model presented here only includes the $AAAA$ orbital configuration. The model's parameters' values were fixed to those initial values listed in Tab.~\ref{tab:1}. The $(\mathbf{Q},E)$ slices presented in (f) and (h) have been folded along [001]. The intensity modulation observed in the calculated response for high resolution measurements presented in (g) is an artifact of the steep disperson of the excitation, manifesting itself as singularities in $g^{\alpha\beta}$ and creating numerical difficulties with sampling the $G^{+-}$ mode.}
\label{fig:Ei110meV_comp}
\end{figure*}

\begin{figure}[htb!]
\centering
\includegraphics[width=1.0\linewidth]{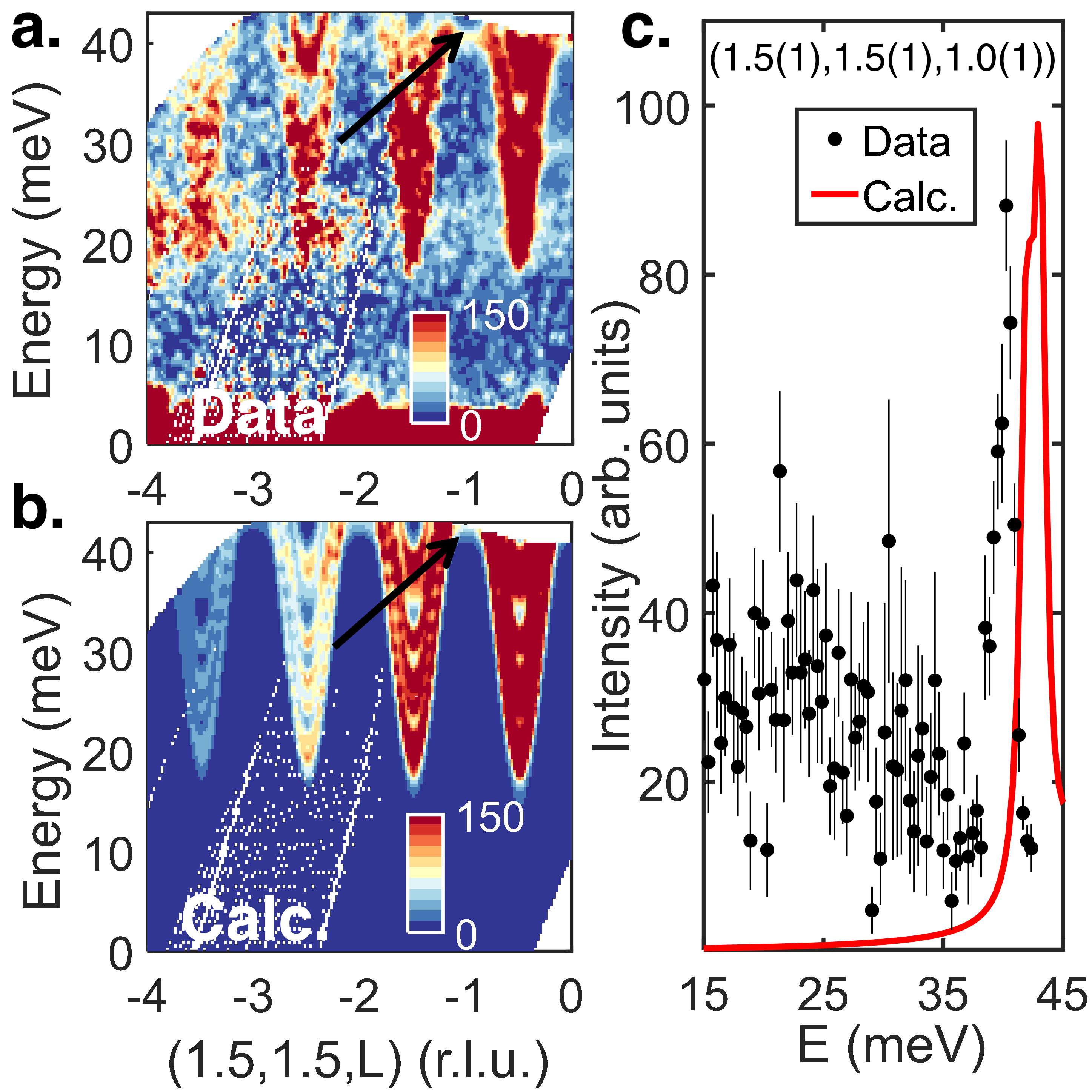}
\caption{A comparison of $(\mathbf{Q},E)$ slices (a) measured at 5~K on MERLIN and (b) calculated using spin-orbit exciton model for an $E_{i}$=110~meV. (c) Comparison between (a) and (b) in a corresponding $\mathbf{Q}$-integrated cut. The success of the spin-orbit exciton model to reproduce the $G^{+-}$ component is emphasized. $(\mathbf{Q},E)$ slices presented in (a) and (b) have been folded along [001].}
\label{fig:pm_mode}
\end{figure}

The need for all 16 $xAxx\gamma$ domains is illustrated in Figs.~\ref{fig:Ei110meV_comp}(e) and (f), where despite the success of the spin-orbit exciton model in reproducing the bandwidth of the excitations between $\sim$20~meV up to $\sim$60~meV with the initial parameters listed in Tab.~\ref{tab:1}, the use of a single $xAxx\gamma$ orbital configuration fails to reproduce the low energy fine structure at the zone center and the steeply dispersive columns of high energy scattering at the zone boundary, instead predicting the presence of a single dominant highly dispersive $G^{-+}$ mode (Figs.~\ref{fig:Ei110meV_comp}(b) and (g)). While both $G^{++}$ and $G^{--}$ modes exhibit negligible intensity (Fig.~\ref{fig:Ei110meV_comp}(d)), a direct consequence of the definition of $G^{\alpha\beta}$ given by Eq.~\ref{eq:alphabeta} above, the spin-orbit exciton model does predict two additional gapped modes corresponding to the longitudinal $G^{zz}$ mode and the transverse $G^{+-}$ mode.  As summarized in  Figs.~\ref{fig:Ei110meV_comp}(a,c), it is clear that both the weakly dispersive $G^{+-}$ and G$^{zz}$ modes cannot account for the missing spectral weight in the fine structure, as both modes are significantly weaker in intensity relative to the dominant $G^{+-}$ mode, while contributing solely at higher energy transfers ($\gtrsim$40~meV). As illustrated in Figs.~\ref{fig:highe_comparison} and~\ref{fig:pm_mode}, while the longitudinal $G^{zz}$ corresponds to a band of scattering that is broad in both momentum and energy while being located above the dominant $G^{-+}$ component, the transverse $G^{+-}$ corresponds to a flat band of scattering centered about $\sim$40~meV.

\begin{figure*}[htb!]
\centering
\includegraphics[width=0.875\linewidth]{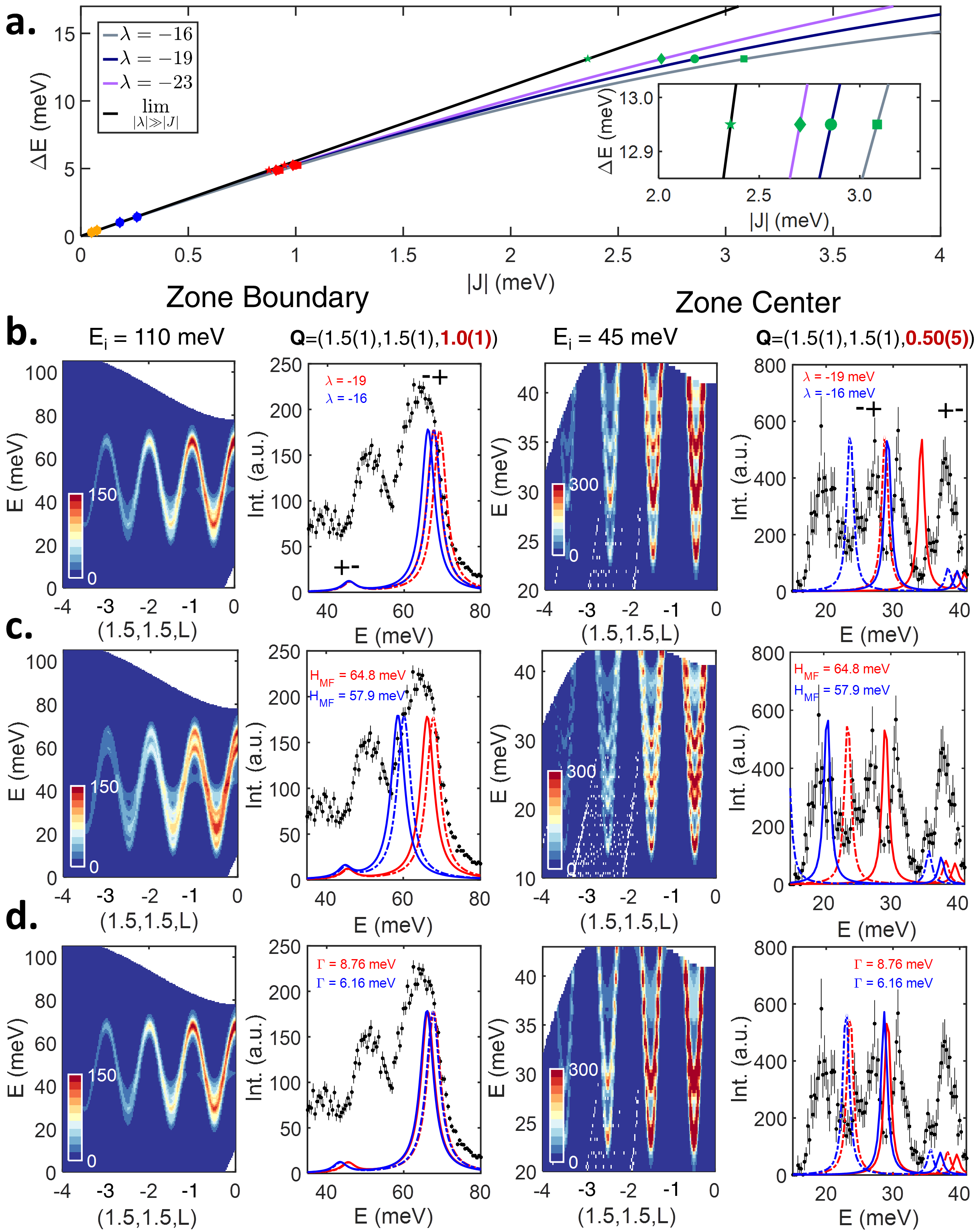}
\caption{(a) Influence of the spin-orbit exchange constant on the splitting of the $j\rm{_{eff}}=\frac{1}{2}$ manifold as a function of $|J|$ for the pair Hamiltonian $\hat{\mathcal{H}}_{pair}$ (Eq.~\ref{pair_ham}). Co$^{2+}$ pair excitations previously measured on MARI are shown explicitly with the corresponding value of $J_{2}$ (inset) particularly influenced by the non-linearity at high energy transfers. Calculated $(\mathbf{Q},E)$ slices for $E_{i}=$110 and 45~meV with corresponding $\mathbf{Q}$-integrated cuts at the zone boundary and center, respectively, illustrating the influences of (b) the spin-orbit coupling constant $\lambda$, (c) the mean molecular field $H_{MF}$, and (d) the tetragonal distortion $\Gamma$ parameters. Solid and dashed lines denote AAAA and AAAF orbital configurations, respectively. The prominent $G^{-+}$ and much weaker $G^{+-}$ have been labeled explicitly for reference. Unless otherwise stated, the values of $\lambda$, $H_{MF}$, $\Gamma$, and magnetic exchange constants $J$ were set to their initial values described in the main text and listed in Tab.~\ref{tab:1}.}
\label{fig:influences}
\end{figure*}

In addition to the necessity for 16 $xAxx\gamma$ domains, the discrepancy between experiment and the calculated dispersion for each of the $G^{\alpha\beta}$ components that were presented in Fig.~\ref{fig:Ei110meV_comp} confirmed the need for the optimization of the spin-orbit exciton model's parameters.  A summary of the influences of each of the model's parameters for a fixed orbital configuration: $J$ \emph{via} $\lambda$, $H_{MF}$, and $\Gamma$ on the calculated spectra is presented in Fig.~\ref{fig:influences}. Calculations are shown by false colormaps and compared against constant momentum cuts at the $\bf{Q}$=$(1.5, 1.5, -1.0)$ magnetic zone boundary and $\bf{Q}$=$(1.5,1.5,-0.5)$ zone center. For illustrative purposes, two different orbital configurations $AAAA$ and $AAAF$ are shown being denoted by solid and dashed lines, respectively. 

As was previously noted in the discussion of the inter-ion Hamiltonian $\hat{\mathcal{H}}_{2}$, the conversion factor $\tilde{\alpha}$ (Fig.~\ref{fig:fig3}) between the energy transfers measured by neutron spectroscopy and the magnitude of the corresponding magnetic exchange constants were determined by diagonalizing the pair Hamiltonian $\hat{\mathcal{H}}_{pair}$ (Eq.~\ref{pair_ham}). Since deviations away from the linear dependence predicted by the projection theorem occurs when the value of $|J| \rightarrow |\lambda|$, the value of larger magnetic exchange constants, such as the strong 180$^{\circ}$ antiferromagnetic superexchange $J_{2}$, is particularly sensitive to $|\lambda|$. As illustrated in Fig.~\ref{fig:influences}(a), the strong sensitivity of $|J_{2}|$ on the value of $|\lambda|$ is of particular concern for CoO since the experimentally determined value of $\lambda=-16(3)$~meV possesses a significant relative error of almost 20\%~\cite{cowley13:88}, corresponding to a large range of possible exchange values. 

The influence of the large relative error for $\lambda$ is summarized in Fig.~\ref{fig:influences}(b), illustrating that an increase of $|\lambda|$ by 20\% to $-19$~meV from its initial value of $-16$~meV for a fixed value of a given molecular field $H_{MF}$ results in a significant shift to higher energy transfers for the dominant $G^{-+}$ component at both the magnetic zone center and boundary. This increase in energy transfer is much more prominent for the $G^{-+}$ component compared to the less intense $G^{+-}$, with the greatest increase for both components occuring at the magnetic zone center. Similar behavior in the calculated response is observed for a change in the mean molecular field $H_{MF}$, as illustrated in Fig.~\ref{fig:influences}(c), with shifts in energy transfers significantly larger compared to the same relative change in the value of $|\lambda|$. In contrast with the three parameters presented so far, the influence of the tetragonal distortion parameter $\Gamma$ is most pronounced on the less intense weakly dispersive $G^{+-}$ mode. As illustrated in Fig.~\ref{fig:influences}(d), $\Gamma$ provides a mechanism to shift the energy transfers of the $G^{+-}$ mode without inducing a comparable shift for the dominant $G^{-+}$ mode for a fixed set of values for $H_{MF}$ and $J$ \emph{via} a fixed value for $\lambda$. 

With such a large number of domains under consideration, constraints on the parameter space for $J$ \emph{via} $\lambda$, $\Gamma$, and $H_{MF}$ were required to ensure convergence for a least squares optimization. As summarized in Tab.~\ref{tab:2}, the model presented in Fig.~\ref{fig:modelA} allowed $\lambda$ to vary from $-13$ to $-19$~meV, corresponding to the experimental error associated with its reported value~\cite{cowley13:88}, while the value of $\Gamma$ varied from $-8.76$ to $-6.16$~meV, corresponding to the range of its values reported in the literature~\cite{cowley13:88}. The exchange constants $J_{m,\xi}$ for each coordination shell $m$ = 1 \dots 4, and type $\xi$=$A$ and $F$ were allowed to vary $\pm$20\% from their starting values obtained from the excitation energies derived from Mg$_{0.97}$Co$_{0.03}$O~\cite{sarte18:98} with a Hamiltonian $\hat{\mathcal{H}}_{pair}$ employing the value of $\lambda$ under consideration. Such a deviation of the exchange constants was rationalized by noting that the estimates of the exchange constants listed in Tab.~\ref{tab:1} are based on the superexchange pathways present in Mg$_{0.97}$Co$_{0.03}$O. Due to the prevalence of the non-magnetic Mg$^{2+}$ in the dilute monoxide, these pathways and their superexchange constants most likely differ compared to those present in pure CoO. As a first approximation, the relative deviation of all magnetic exchange constants $J$ from their respective values in Mg$_{0.97}$Co$_{0.03}$O were set to be equal for all coordination shells $m$ and type $\xi$. As previously discussed in Section~\ref{sec:parameters}, the values of $\lambda$, $J$, and $\Gamma$ were constrained to be equal for each of the 16 domains, while no such restriction was applied to the mean molecular field $H_{MF}$ which was itself allowed to vary independently for each domain from a value of 0 to an arbitrarily large upper limit. In our model, this limit was set to 100~meV, corresponding to $\sim$1.5 times the value of the initial value of 64.8~meV. 

Among all the spin-orbit exciton model's parameters, the expansion of the parameter space for the exchange constants $J_{m,\xi}$ proved to be of particular importance for the model's success in reproducing both the fine structure at the magnetic zone center and the broad excitations at the $(1.5,1.5,-1)$ zone boundary simultaneously. As summarized in Section VI of the \emph{Supplementary Information}~\cite{suppl}, the restriction of the exchange constants $J_{m,\xi}$ to be equal to their respective values originally reported for Mg$_{0.97}$Co$_{0.03}$O~\cite{sarte18:98} for a given value of $\lambda$, resulted in failure of the spin-orbit exciton model to reproduce both broad excitations at the $(1.5,1.5,-1)$ zone boundary  (Fig.~S2). As summarized by Tab.~SIII, such a restriction placed on the values of exchange constants resulted in the value of $\lambda$ being refined to its most negative permissible value. Such behavior is a reflection of the model's attempts to minimize the average value for $H_{MF}$ in order to capture the intensity at lower energy transfers at the $(1.5,1.5,L)$ zone boundary. As illustrated in Fig.~S3, the clear failure of the model, even by expanding the parameter space of $\lambda$ to include all values down to $-23$~meV, confirmed the inability of the model to reproduce the data at the zone boundary while employing the exact exchange constants measured in Mg$_{0.97}$Co$_{0.03}$O.  

\begin{figure*}[htb!]
\centering
\includegraphics[width=0.975\linewidth]{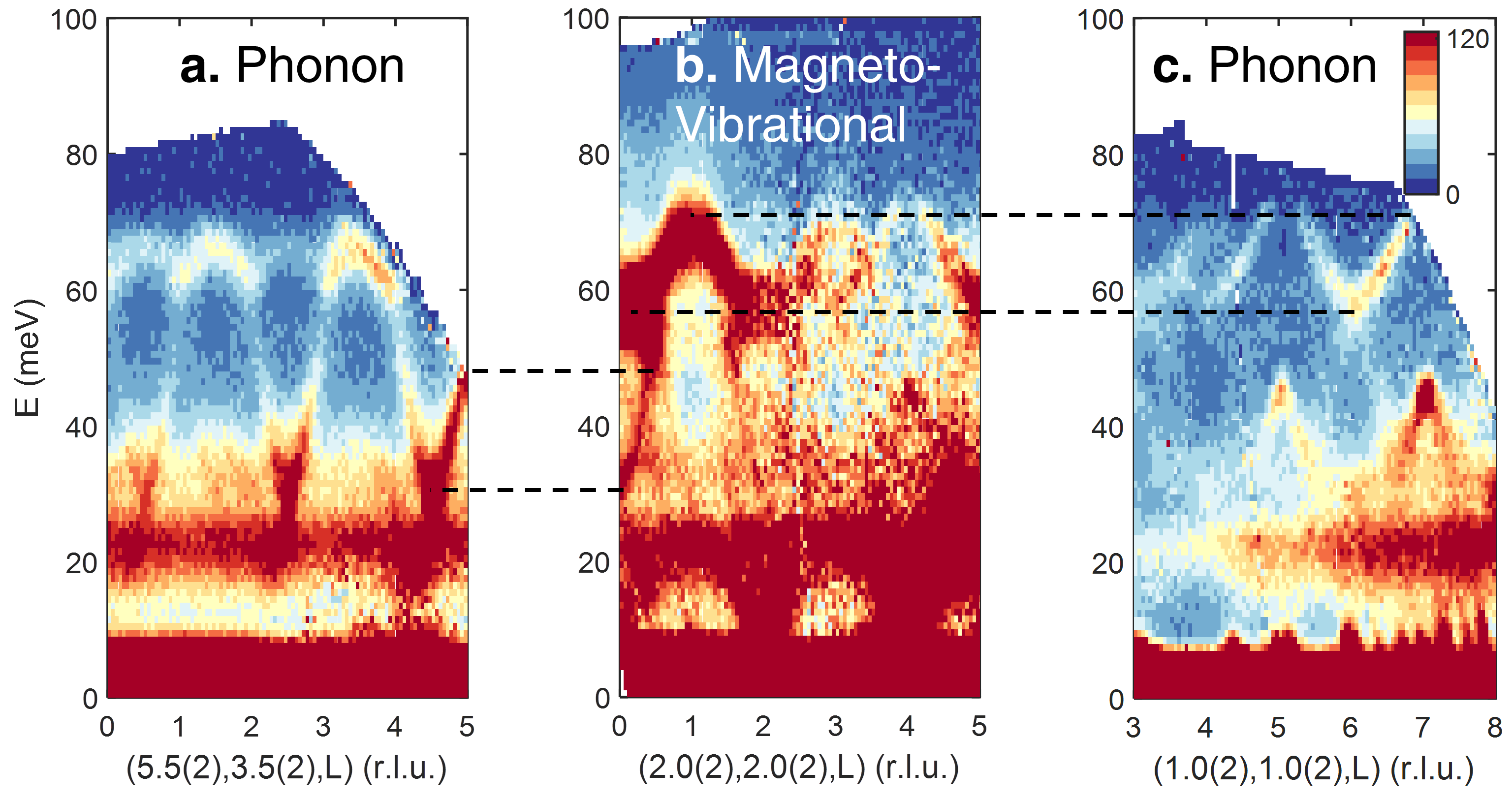}
\caption{A comparison between (a,c) the phonon scattering at large momentum transfers centered about the nuclear zone boundaries and (b) the modes located near the magnetic zone boundary along the $(2,2,0)$ direction that are unaccounted by our spin-orbit exciton model. The overlap of the energy transfers range between the (b) low-$\mathbf{Q}$ scattering near the $(2,2,0)$ magnetic zone boundary and both gapped optical phonon modes centered near the (a) $(5.5,3.5,4.5)$ and (c) $(1,1,6)$ nuclear zone boundaries with energy transfers above $\sim$30~meV and $\sim$60~meV, respectively, are emphasized with dashed lines. All three $(\mathbf{Q},E)$ slices have been folded along [001] and have been renormalized to share a common relative intensity scale.}
\label{fig:magneto_vib}
\end{figure*}

\subsection{Magneto-Vibrational Scattering and the Magnetic Zone Boundary} \label{sec:D}

Despite the success of the spin-orbit exciton parameterization of the experimental data along $(1.5,1.5,L)$, the model still fails at the $(2, 2, L)$ zone boundary, suggesting the presence of additional physics that is beyond our localized model. The presence of such prominent magnetic scattering near $(2, 2, L)$ is particularly unusual due to the predictions from both our spin-orbit exciton model and linear spin-wave theory for nearly zero intensity for magnetic fluctuations at these points in reciprocal space for an antiferromagnetic structure.

By comparing the scattering along $(2,2, L)$ to phonon branches near different nuclear zone boundaries, we note that there is a distinct overlap in dispersion.  As is summarized in Fig.~\ref{fig:magneto_vib}(a), a strong and steeply dispersing optical phonon mode near ${\bf{Q}}=(5.5,3.5,4.5)$ exhibits an identical dispersion to that of the mode centered about $(2,2,0)$, a mode that is not accounted for by our localized model.  The phonon nature of the gapped mode in Fig.~\ref{fig:magneto_vib}(a) is confirmed by the fact that the intensity increases with ${\bf{Q}}$. ~A similar observation is illustrated in Fig.~\ref{fig:magneto_vib}(c) for energy transfers greater than 60~meV, where a gapped phonon mode, identified by both its ${\bf{Q}}$ dependence and previously reported first principles calculations~\cite{Wdowik07:75}, exhibits a dispersion that is identical to the magnetic scattering around $(2,2,0)$. 

The apparent similarity between the dispersion of phonons at high $\mathbf{Q}$ and the magnetic scattering that was originally unaccounted for in our model may suggest that our data at the zone boundaries includes a magneto-vibrational contribution to the neutron cross section. Corresponding to an indirect energy exchange between the neutron and a vibrating nucleus \emph{via} the electro-magnetic interaction between the neutron and the unpaired electrons of the nucleus, the magneto-vibrational neutron cross section~\cite{Egelstaff:book} is identical to the more commonly employed direct one-phonon cross section employing the nuclear force, with the exception that the nuclear scattering length $b$ is replaced by the magnetic scattering length given by 

\begin{equation}
b_{mag}=\frac{e^{2}\gamma}{m c^{2}}f(Q)|{\bf{\mu}}| |\sin \alpha|,
\nonumber
\end{equation}

\noindent where $\alpha$ is the angle between the momentum transfer ${\bf{Q}}$ and the ordered magnetic moment direction ${\bf{\mu}}$.  The presence of the isotropic magnetic form factor $f(Q)$ guarantees that the cross section will ultimately decay with momentum transfer, regardless of the underlying phonon origin of the scattering.  This particular cross section requires an ordered magnetic structure and has been used previously to characterize the dynamic magnetic form factor~\cite{brown88:49}.  Similar cross sections have been reported in the rare earth magnetic pyrochlores~\cite{fennell14:112} and doped manganites~\cite{Hennion02:312}.~We note that in the case of CoO, the magnetic ions are strongly correlated and coupled in all three dimensions, implying that that the correlations are maintained at large energy transfers in the range of 30-60~meV where prominent phonon modes exist.   Based on the comparison presented above, we suggest that both the excitations along $(2,2,L)$, and other excitations that were unaccounted for by our model such as the weak magnetic scattering at $\sim$40~meV at the $(1.5,1.5,L)$ zone boundary all originate from phonon modes, rather than the underlying magnetic Hamiltonian $\hat{\mathcal{H}}$.  

While we have cast this discussion in terms of the magneto-vibrational cross section, which itself does not provide any new information on the underlying Hamiltonian, these excitations may be indicative of a coupling between the lattice and magnetic degrees of freedom. Such magneto-elastic coupling in the underlying Hamiltonian has been investigated recently in CeAuAl$_{3}$~\cite{Cermak19:116}. It is particularly compelling to consider such a claim given that these modes are located near the expected energy scale for the single-ion spin-orbit transitions from $|j{\rm{_{eff}}}=\frac{1}{2} \rangle \rightarrow |j{\rm{_{eff}}}= \frac{3}{2}\rangle$.  We will not investigate this coupling further given the need for first principles calculations to reconcile past reported phonon data and the modes identified in our current study.

\subsection{The High-Energy Response and the Co$^{2+}$ Form Factor}

\begin{figure}[htb!]
	\centering
	\includegraphics[width=1.0\linewidth]{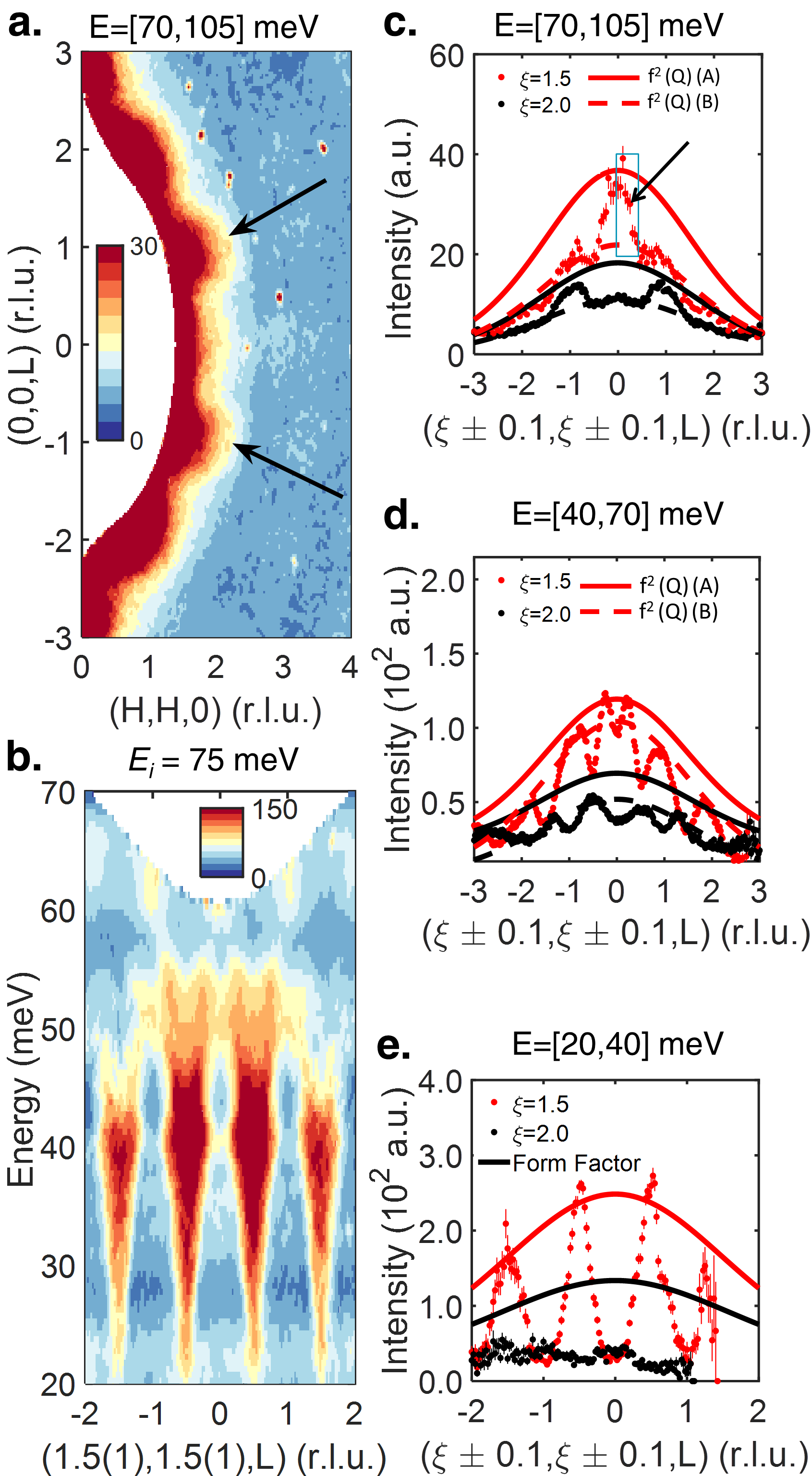}
	\caption{(a) Energy-integrated ($E=[70,105]$~meV) slice measured on MERLIN at 5~K with an $E_{i}$ of 110~meV. (b) $(\mathbf{Q},E)$ slice folded along [001] measured on MERLIN at 5 K with an $E_{i}$ of 75~meV. $\mathbf{Q}$-integrated ($\xi$,$\xi$,$L$) cuts measured on MERLIN at 5~K with an $E_{i}$ of (c) 110~meV, (d) 75~meV, and (e) 45~meV. Solid and dashed lines in $\mathbf{Q}$-integrated cuts correspond to the Co$^{2+}$ magnetic form factor $f^{2}(Q)$ that includes and excludes the intensity at (1.5,1.5,$\pm$0.5), respectively.  Both $(\mathbf{Q},E)$ slices presented in (a) and (b) have been folded along [001]. Arrows in panels (a), (b), and (c) indicate fluctuations exhibiting potential itinerant-like behavior. For the purposes of comparison, the specific region in $ (\mathbf{Q},E)$ space (blue-green) identified in (c) as possibly containing itinerant-like fluctuations has been labeled explicitly in Fig.~\ref{fig:manila}(b).}
	\label{fig:chris}
\end{figure}

We have demonstrated that the experimental data at low energy transfers can be succesfully parameterized in terms of spin-orbit excitons with crystal field and exchange parameters based on our previous work reported on MgO substituted with Co$^{2+}$. In Section~\ref{sec:C}, such a mean-field multi-level spin-orbit exciton model was shown to successfully reproduce the data in pure CoO near the zone center for energy transfers below 40~meV, while the failure of the model near the magnetic zone boundaries up to energy transfers of $\sim$60~meV was attributed in Section~\ref{sec:D} to either magneto-vibrational scattering or a coupling to underlying phonon excitations that were identified at large momentum transfers.  In this final subsection, we will address the magnetic excitations at large energy transfers above $\sim$ 65~meV,  where no phonon scattering is expected, exceeding the dynamic range predicted by first principle calculations~\cite{Wdowik07:75} and measured by both our time-of-flight experiment and previous experiments~\cite{kant08:78,sakurai68:167,goyal77:79}.   

In Fig. \ref{fig:chris} we present a comparison between the scattering for high energy transfers ([70, 105]~meV) with the lower energy magnetic fluctuations that were discussed above in terms of our spin-orbit exciton parameterization.  As summarized in Fig. \ref{fig:chris}(a), excitations located at low momentum transfers extend up to high energies. Previously identified in Fig.~\ref{fig:highe_comparison} as being magnetic in origin, it appeared that these excitations were successfully reproduced by the spin-orbit exciton model. These ``columns'' of scattering were interpreted as a result of the overlap at the magnetic zone boundary of multiple $G^{-+}$ components from different local orbital arrangements that were required to capture the fine structure of the strongly dispersing fluctuations (Fig.~\ref{fig:chris}(b)) centered about the zone centers and found at lower energy transfers. While the analysis presented so far suggest that the $\mathbf{Q}$-dependence for all magnetic excitations appear to follow the Co$^{2+}$ form factor, the energy-integrated slice in Fig.~\ref{fig:chris}(a) demonstrates this is not the case for high energy transfers, where the intensity decays more rapidly than $f(Q)$. Such an observation can be confirmed by comparing momentum cuts of the magnetic fluctuations at small momentum transfers with the magnetic form factor for Co$^{2+}$.  In contrast with the lowest energy transfers ([20-40]~meV, Fig.~\ref{fig:chris}(e)) where the magnetic fluctuations follow the form factor, deviations from such localized magnetic behavior begin to appear at $(1.5,1.5,\pm0.5)$ for intermediate energy transfers ([40,70]~meV, Fig.~\ref{fig:chris}(d)), with the deviations being particularly prominent around $(1.5,1.5,0)$ at the highest energy transfers ([70,105]~meV, Fig.~\ref{fig:chris}(e)).  The rapid decay of intensity with momentum transfer maybe indicative of a real space object that is extended spatially, and not due to localized magnetism. 

\begin{figure}[t!]
	\centering
	\includegraphics[width=1.0\linewidth]{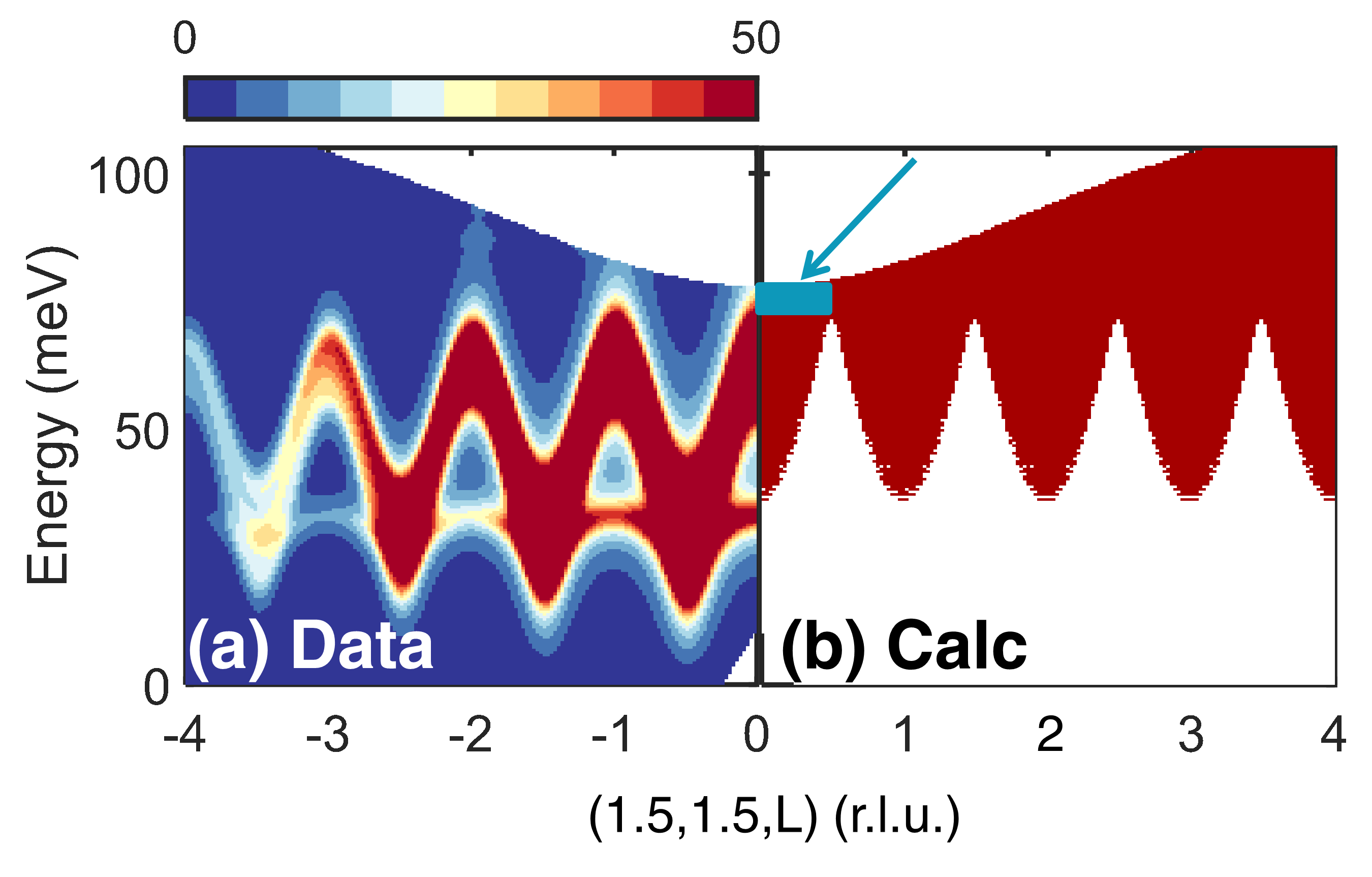}
	\caption{Comparison of (a) $(\mathbf{Q},E)$ slice measured on MERLIN with $E_{i}=110$~meV and the corresponding (b) calculated kinematically permissable $(\mathbf{Q},E)$ region for the 2-magnon continuum. The specific region in $(\mathbf{Q},E)$ space that was previously identified in Fig.~\ref{fig:chris}(c) being labeled explicitly. The $(\mathbf{Q},E)$ slice presented in (a) has been folded along [001].}
	\label{fig:manila}
\end{figure}

We now speculate as to the possible origin for such delocalized magnetism.  In the case of the highest energy transfers in CoO, where the relative deviation from the form factor is greatest, the scattering is steeply ``dispersive" in energy, indicative of a large underlying energy scale.  These fluctuations are highly reminiscent of the magnetic response found in itinerant magnets such as CeRhIn$_{5}$~\cite{stock15:114}, the cuprates~\cite{Stock07:75,Stock10:82}, and also iron-based systems~\cite{stock14:90,plumb18:97}.  In the case of CeRhIn$_{5}$, the high-energy steeply dispersive excitations were found to be longitudinally polarized and occupied a region in $(\mathbf{Q},E)$ phase space where two magnon processes were kinematically allowed.  Termed the ``1+2'' model, multi-magnon decay processes were used to provide a heuristic description of the data in the case of CeRhIn$_{5}$.  Motivated by the qualitative similarities in the excitation spectrum between CeRhIn$_{5}$ and CoO, we investigated the possibility that the steeply dispersing excitations observed in CoO overlap a similarly allowed region of $(\mathbf{Q},E)$ phase space.  The phase space permitted for such a decay of the low energy magnetic fluctuations was calculated using a simple model based on energy and momentum conservation that is given by~\cite{huberman05:72,stock18:2,songvilay18:121}

\begin{equation}
G(\mathbf{Q},E) = \sum\limits_{\mathbf{Q}_{1},\mathbf{Q}_{2}}\delta(\mathbf{Q}-\mathbf{Q}_{1}-\mathbf{Q}_{2})\delta(E-E_{\mathbf{Q}_{1}}-E_{\mathbf{Q}_{2}}),
\nonumber
\end{equation}

\noindent  where $E_{\mathbf{Q}_{1,2}}$ are the energies of transverse excitations at a given momentum transfer.  As shown in Fig.~\ref{fig:manila}, the kinematically allowed region overlaps in both momentum and energy with the steeply dispersing excitations at the highest energy transfers; a region where the spin-orbit exciton model predicts the presence of a longitudinally polarized excitations ($i.e.$ $G^{zz}$ mode), as was previously illustrated in Fig. \ref{fig:highe_comparison}(b).  Employing the observed overlap that is summarized in Fig.~\ref{fig:manila}, we speculate that the spectral weight for these steeply dispersive high energy excitations that are localized in momentum draws from this longitudinally polarized mode, analogous to what was observed in previously investigated itinerant magnets.  Magnetism spatially extending beyond the Co$^{2+}$ site has been suggested theoretically with some moment expected to be presented on the oxygen atom~\cite{Elp91:44}.  Given the extended nature of the magnetism in such a scenario, the magnetism would be expected to decay faster in momentum transfer than the isotropic Co$^{2+}$ form factor.   An intermediate example has been reported in the case of Sr$_{2}$CuO$_{3}$~\cite{walters09:5} which found excellent agreement between the magnetic form factor and a model including strong covalent bonding and hybridization of the $3d$ orbitals.  It could be that the higher energy excitations are more sensitive to such a situation in CoO.  
 
\section{Concluding Remarks}

We have presented a neutron spectroscopic study of the magnetic fluctuations in the Mott insulator CoO.  We have parameterized the low energy magnetic excitations near the magnetic zone center in terms of a mean-field multi-level spin-orbit exciton model incorporating multiple structural domains owing to multiple local orbital arrangements, as well as a prominent tetragonal Jahn-Teller distortion. Dispersive excitations at the zone boundaries mimicking magnetic scattering at low $\mathbf{Q}$ that were originally unaccounted for by the spin-orbit exciton model were found to exhibit similar dispersions as phonons measured at larger momentum transfers, suggesting that the model's failures at the zone boundaries may be attributed to magneto-vibrational scattering or possibly coupling to lattice degrees of freedom.  Finally, we report a discrepancy between the $\mathbf{Q}$-dependence of excitations at high energy transfers and the behavior predicted by the Co$^{2+}$ form factor. Despite the strong insulating nature of CoO, we speculate that such a discrepancy corresponds to a breakdown of spin-orbit excitons that may be accompanied by a crossover from localized to spatially-extended magnetism, reminisicent of an itinerant-like response or strong covalency.     
 
\begin{acknowledgements} 

\indent We would like to convey our sincere gratitude to both W.~J.~L. Buyers and the late R.~A.~Cowley for their assistance throughout the preliminary stages of this investigation.  We acknowledge fruitful conversations with C.~R.~Wiebe,~G.~M.~McNally,~J.~A.~M.~Paddison,~G.~Perversi, S.~E.~Maytham, and K.~J.~Camacho. We are grateful to the Royal Society,~the STFC,~the ERC, the EPSRC, and the Carnegie Trust for the Universities of Scotland for financial support.~P.~M.~S. acknowledges additional financial support from the University of California, Santa Barbara through the Elings Fellowship, the CCSF,~RSC,~ERC, and the University of Edinburgh through the GRS and PCDS. A portion of this work was supported by the DOE, Office of Science, Basic Energy Sciences under Award DE-SC0017752. 

\end{acknowledgements}

%

\end{document}


\title{Supplementary Information: ``Spin-Orbit Excitons in CoO"}

\author{P.~M.~Sarte}
\affiliation{California NanoSystems Institute, University of California, Santa Barbara, California 93106-6105, USA}
\affiliation{Materials Department, University of California, Santa Barbara, California 93106-5050, USA} 
\affiliation{School of Chemistry, University of Edinburgh, Edinburgh EH9 3FJ, United Kingdom}
\affiliation{Centre for Science at Extreme Conditions, University of Edinburgh, Edinburgh EH9 3FD, United Kingdom}
\author{M.~Songvilay}
\affiliation{Centre for Science at Extreme Conditions, University of Edinburgh, Edinburgh EH9 3FD, United Kingdom}
\affiliation{School of Physics and Astronomy, University of Edinburgh, Edinburgh EH9 3FD, United Kingdom}
\author{E.~Pachoud}
\affiliation{School of Chemistry, University of Edinburgh, Edinburgh EH9 3FJ, United Kingdom}
\affiliation{Centre for Science at Extreme Conditions, University of Edinburgh, Edinburgh EH9 3FD, United Kingdom}
\author{R.~A.~Ewings}
\affiliation{ISIS Pulsed Neutron and Muon Source, STFC Rutherford Appleton Laboratory, Harwell Campus, Didcot, Oxon, OX11 OQX, United Kingdom}
\author{C. D.~Frost}
\affiliation{ISIS Pulsed Neutron and Muon Source, STFC Rutherford Appleton Laboratory, Harwell Campus, Didcot, Oxon, OX11 OQX, United Kingdom}
\author{D.~Prabhakaran} 
\affiliation{Department of Physics, Clarendon Laboratory, University of Oxford, Park Road, Oxford, OX1 3PU, United Kingdom}
\author{K.~H.~Hong}
\affiliation{School of Chemistry, University of Edinburgh, Edinburgh EH9 3FJ, United Kingdom}
\affiliation{Centre for Science at Extreme Conditions, University of Edinburgh, Edinburgh EH9 3FD, United Kingdom}
\author{A.~J.~Browne}
\affiliation{School of Chemistry, University of Edinburgh, Edinburgh EH9 3FJ, United Kingdom}
\affiliation{Centre for Science at Extreme Conditions, University of Edinburgh, Edinburgh EH9 3FD, United Kingdom}
\author{Z.~Yamani}
\affiliation{National Research Council, Chalk River, Ontario K0J 1JO, Canada }
\author{J.~P.~Attfield}
\affiliation{School of Chemistry, University of Edinburgh, Edinburgh EH9 3FJ, United Kingdom}
\affiliation{Centre for Science at Extreme Conditions, University of Edinburgh, Edinburgh EH9 3FD, United Kingdom}
\author{E.~E.~Rodriguez} 
\affiliation{Department of Chemistry and Biochemistry, University of Maryland, College Park, Maryland 20742, USA}
\author{S.~D.~Wilson} 
\affiliation{California NanoSystems Institute, University of California, Santa Barbara, California 93106-6105, USA}
\affiliation{Materials Department, University of California, Santa Barbara, California 93106-5050, USA} 
\author{C.~Stock}
\affiliation{Centre for Science at Extreme Conditions, University of Edinburgh, Edinburgh EH9 3FD, United Kingdom}
\affiliation{School of Physics and Astronomy, University of Edinburgh, Edinburgh EH9 3FD, United Kingdom}

\date{\today}

\begin{abstract}

\end{abstract}

\maketitle
\renewcommand{\figurename}{Figure S\hspace*{-1.00mm}}
\renewcommand{\tablename}{Table S\hspace*{-1.00mm}}

\onecolumngrid

\section{D\lowercase{erivation of the} E\lowercase{quation-of-}M\lowercase{otion for the} R\lowercase{esponse} F\lowercase{unction}} \label{sec:equation-of-motion}

\hspace*{5.00mm} The derviation begins by first taking the time derivative of the definition of the retarded Green's function (Eq.~3 in the main text) which yields   

\begin{equation}
\begin{split}
\frac{d}{d t}G(\hat{S}_{\alpha}(i,t),\hat{S}_{\beta}(j,0)) & = -i\delta(t)\langle[\hat{S}_{\alpha}(i,t),\hat{S}_{\beta}(j,0)]\rangle -i\Theta(t)\langle[\frac{d}{d t}\hat{S}_{\alpha}(i,t),\hat{S}_{\beta}(j,0)]\rangle. 
\end{split}
\label{eq:eofm1} 
\end{equation} 

Since none of the spin operators have an explicit time dependence, the second term of the equation-of-motion in the Heisenberg picture of quantum mechanics given by 

\begin{equation}
\frac{d\hat{A}(t)}{d t} = -i[\hat{A}(t),\hat{\mathcal{H}}] + \left(\frac{\partial \hat{A}}{\partial t}\right),
\label{eq:eofm2} 
\end{equation}	

\noindent must be zero, yielding
 
\begin{equation}
i\frac{d\hat{A}(t)}{d t}= [\hat{A}(t),\hat{\mathcal{H}}],
\label{eq:eofm3} 
\end{equation}

\noindent where $\hat{\mathcal{H}}$ is the Hamiltonian of interest.

The insertion of Eq.~\ref{eq:eofm3} into Eq.~\ref{eq:eofm1}, and the subsequent application of the inverse Fourier transform yields   

\begin{equation}
\begin{split}
& \int_{-\infty}^{\infty}dte^{i\omega t}\frac{d}{d t}G(\hat{S}_{\alpha}(i,t),\hat{S}_{\beta}(j,0))  = -i\int_{-\infty}^{\infty}dte^{i\omega t} \delta(t)\langle[\hat{S}_{\alpha}(i,t),\hat{S}_{\beta}(j,0)]\rangle - i\int_{-\infty}^{\infty}dte^{i\omega t}\{-i\Theta(t)\langle[[\hat{S}_{\alpha}(i,t),\hat{\mathcal{H}}],\hat{S}_{\beta}(j,0)]\rangle\}.
\end{split}
\label{eq:eofm4} 
\end{equation}

\noindent By noting that the first term on the RHS may be immediately simplified due to the presence of the delta function, while the second term on the RHS contains the definition of the Green's function (Eq.~3 in the main text) but with $[\hat{S}_{\alpha}(i,t),\hat{\mathcal{H}}]$ instead of $\hat{S}_{\alpha}$ and thus can be relabeled, Eq.~\ref{eq:eofm4} can be rewritten as 

 \begin{align}
\begin{split}
 \int_{-\infty}^{\infty}dte^{i\omega t}\frac{d}{d t}G(&\hat{S}_{\alpha}(i,t),\hat{S}_{\beta}(j,0)) =  -i\langle[\hat{S}_{\alpha}(i,0),\hat{S}_{\beta}(j,0)]\rangle  - i\int_{-\infty}^{\infty}dte^{i\omega t}G([\hat{S}_{\alpha}(i,t),\hat{\mathcal{H}}], \hat{S}_{\beta}(j,0)).
\end{split}
\label{eq:eofm5} 
\end{align}

\noindent By noting that 
\begin{equation}
\begin{split}
\frac{d}{d t}\{e^{i\omega t}G(\hat{S}_{\alpha}(i,t),\hat{S}_{\beta}(j,0)) \} &= i\omega e^{i\omega t}G(\hat{S}_{\alpha}(i,t),\hat{S}_{\beta}(j,0))  + e^{i\omega t}\frac{d}{d t}G(\hat{S}_{\alpha}(i,t),\hat{S}_{\beta}(j,0)), 
\end{split}
\end{equation}

\noindent or equivalently  

\begin{equation}
\begin{split}
e^{i\omega t}\frac{d}{d t}G(\hat{S}_{\alpha}(i,t),\hat{S}_{\beta}(j,0)) &= \frac{d}{d t}\{e^{i\omega t}G(\hat{S}_{\alpha}(i,t),\hat{S}_{\beta}(j,0))\} - i\omega e^{i\omega t}G(\hat{S}_{\alpha}(i,t),\hat{S}_{\beta}(j,0)) 
\end{split}
\end{equation}

\noindent must hold \emph{via} the product rule, then Eq.~\ref{eq:eofm5} can be rewritten as 

\begin{equation}
\begin{split}
& \int_{-\infty}^{\infty}dt\left\{\frac{d}{d t}e^{i\omega t}G(\hat{S}_{\alpha}(i,t),\hat{S}_{\beta}(j,0)) - i\omega e^{i\omega t}G(\hat{S}_{\alpha}(i,t),\hat{S}_{\beta}(j,0)) \right\} \\
& = -i\langle[\hat{S}_{\alpha}(i,0),\hat{S}_{\beta}(j,0)]\rangle - i\int_{-\infty}^{\infty}dte^{i\omega t}G([\hat{S}_{\alpha}(i,t),\hat{\mathcal{H}}], \hat{S}_{\beta}(j,0)).
\end{split}
\label{eq:eofm6} 
\end{equation} 

\noindent By employing the boundary conditions of Green's functions, the first term on the LHS of Eq.~\ref{eq:eofm6} must be zero. Such an observation follows directly from the fact that Green's functions are well-behaved (normalizable), and thus must approach zero in the limit of $\pm\infty$.~Eq.~\ref{eq:eofm6} can be further simplified by noting that the second terms on both the LHS and the RHS are simply the definition of the inverse Fourier transform from $t$ to $\omega$. Finally, by dividing both sides by the $-i$ prefactor and noting that the expression above must hold for generic $\alpha,\beta$ components of the spin operators, without the loss of generality, Eq.~\ref{eq:eofm6} can be rewritten as 

\begin{equation}
\omega G(\hat{A},\hat{B},\omega) = \langle [\hat{A},\hat{B}] \rangle + G([\hat{A},\hat{\mathcal{H}}],\hat{B},\omega),
\label{eq:10}
\end{equation}

\noindent corresponding to the equation-of-motion of the response function given by Eq.~4 in the main text, where the specific components of the spin operators have been replaced by generic labels $\hat{A}$ and $\hat{B}$.  

\section{C\lowercase{rystalline} E\lowercase{lectric} F\lowercase{ield} \& S\lowercase{teven} O\lowercase{perators}} \label{sec:stevens} 

The crystalline electric field describes the electrostatic interaction between the metal center's electrons and its surrounding charged ligands~\cite{Abragam:book}. The CEF Hamiltonian for a single electron of charge $-e$ located at position $\mathbf{r}$ is given by 

\begin{equation}
\hat{\mathcal{H}}_{CEF} = -e\hat{V}(\mathbf{r}),
\label{eq:HCF}
\end{equation}

\noindent or in the case of a multi-electron atom, 

\begin{equation}
\hat{\mathcal{H}}_{CEF} = -e\sum\limits_{i}\hat{V}(\mathbf{r}_{i}),
\label{eq:HCF2}
\end{equation} 

\noindent where the $i^{th}$ electron has position vector $\mathbf{r}_{i}$. The potential $\hat{V}(\mathbf{r})$ corresponds to the Coulomb potential at a position $\mathbf{r}$ due to the presence of ligands of charge $q_{j}$ at position $\mathbf{R}_{j}$. According to classical electromagnetism~\cite{Long1991}, the classical form for such a potential would be given by   

\begin{equation} 
V(\mathbf{r}) = \sum\limits_{j}\frac{q_{j}}{|\mathbf{R}_{j}-\mathbf{r}|},
\end{equation}  

\noindent where $|\dots|$ denotes the modulus. If one considers a position $\mathbf{r}\ll\mathbf{R}_{j}$~$\forall~j$, then one can expand the potential in terms of Legendre polynomials such that

\begin{equation}
V(\mathbf{r}) = \sum\limits_{j}\frac{q_{j}}{R_{j}}\sum_{n=0}^{\infty}\left(\frac{r}{R_{j}}\right)^{n}P_{n}(\cos(\theta)),
\label{eq:CF1}
\end{equation} 

\noindent where $P_{n}(\cos(\theta))$ are the Legendre polynomials and the angle $\theta$ is defined such that $|\mathbf{r}-\mathbf{R}_{j}|=r^{2}+R^{2}_{j}-2rR_{j}\cos(\theta)$. Utilizing the mathematical relationship~\cite{walter84:45} between the Legendre polynomials and the tesseral harmonics $Z(\theta,\phi)$ given by 

\begin{equation}
P_{n}(\cos(\theta)) = \frac{4\pi}{2n+1}\sum\limits_{\alpha}Z_{n\alpha}(\theta,\phi)Z_{n\alpha}(\theta_{j},\phi_{j}),
\end{equation}

\noindent then Eq.~\ref{eq:CF1} can be rewritten as 

\begin{equation}
V(\mathbf{r}) = V(r,\theta,\phi) = \sum\limits_{j}q_{j}\sum_{n=0}^{\infty}\frac{r^{n}}{R^{n+1}_{j}}\sum\limits_{\alpha}\left\{\frac{4\pi}{2n+1}Z_{n\alpha}(\theta,\phi)Z_{n\alpha}(\theta_{j},\phi_{j})\right\},
\end{equation}

\noindent which is often condensed to 

\begin{equation}
V(\mathbf{r}) = V(r,\theta,\phi) = \sum_{n=0}^{\infty}\sum\limits_{\alpha}A_{n\alpha}r^{n}Z_{n\alpha}(\theta,\phi),
\label{eq:CF2}
\end{equation}

\noindent where 

\begin{equation}
A_{n\alpha} = \sum\limits_{j}\frac{4\pi}{2n+1}q_{j}\frac{Z_{n\alpha}(\theta_{j},\phi_{j})}{R^{n+1}_{j}}.
\label{eq:Aalpha} 
\end{equation}

By rewriting Eq.~\ref{eq:CF1} in terms of the tesseral harmonics, the potential can now be written as a function of $(x,y,z)$ explicitly. By noting the spin-orbit coupling is considerably weaker than the crystal field contribution for the $3d$ Co$^{2+}$, whose energy scale is significantly weaker than the energy differences between the free-ion terms, the complete set of commuting observables are $\hat{L}^{2}$, $\hat{L}_{z}$, $\hat{S}^{2}$ and $\hat{S}_{z}$ with corresponding good quantum numbers $L$, $m_{L}$, $s$ and $m_{s}$ in the Russell-Saunders $L$-$S$ coupling scheme~\cite{cowley13:88}. Consequently, for the case of Co$^{2+}$ in CoO, $\hat{\mathcal{H}}_{CEF}$ may be considered as a perturbative potential to the eigenstates $|L,S,m_{L},m_{s}\rangle$ for fixed values of $L=3$ and $S=\frac{3}{2}$ for the $^{4}F$ free-ion ground state term that is defined by Hund's rules incorporating the effects of electron-electron repulsion and the Pauli Exclusion Principle. Since the position operator transforms in an identical manner as the angular momentum operator~\cite{Hutchings64:16}, one may employ the Wigner-Eckhart theorem, which yields

\begin{equation}
\hat{\mathbf{R}} \propto \hat{\mathbf{L}}.
\end{equation}     
 
By employing both the correspondence principle and the Wigner-Eckhart theorem to Eq.~\ref{eq:CF2}, the CEF potential and thus the CEF Hamiltonian can be rewritten in terms of the operators $\hat{L}^{2}$, $\hat{L}_{z}$ and their corresponding raising and lowering operators. In the Stevens approach, Eq.~\ref{eq:CF2} is further condensed into the form 

\begin{equation}
\hat{V} = \sum\limits_{l,m}B^{m}_{l}\hat{O}^{m}_{l},
\label{eq:CF3} 
\end{equation}

\noindent where $B^{m}_{l}$ and $\hat{O}^{m}_{l}$ are the Stevens coefficients and operators, respectively. The particular Stevens parameters that are present in the sum of Eq.~\ref{eq:CF3} are determined by symmetry and have been tabulated by Walter~\cite{walter84:45}, while the expressions of the Stevens operators in terms of angular momentum operators have been tabulated by Hutchings~\cite{Hutchings64:16}. The two relevant Stevens operators $\hat{\mathcal{O}}^{0}_{4}$ and $\hat{\mathcal{O}}^{4}_{4}$ employed in this particular investigation correspond to those for ideal octahedral coordination and are given by

\begin{equation}
\hat{\mathcal{O}}^{0}_{4}=35\hat{L}_{z}^{4}-30\hat{L}^{2}\hat{L}_{z}^{2}+25\hat{L}_{z}^{2}-6\hat{L}^{2}+3\hat{L}^{4},
\nonumber
\end{equation} 

\noindent and 

\begin{equation}
\hat{\mathcal{O}}^{4}_{4}=\frac{1}{2}\left[\hat{L}_{+}^{4} + \hat{L}_{-}^{4} \right].
\nonumber
\end{equation} 

\section{C\lowercase{ommutators}} \label{sec:commutators}

\subsection{Derivation of the \emph{Diagonal} Commutator} \label{sec:diagonal} 

The \emph{diagonal} commutator given by $\sum\limits_{k}\sum\limits_{r}[\hat{C}^{\dagger}_{m}(i)\hat{C}_{n}(i),\hat{C}^{\dagger}_{r}(k)\hat{C}_{r}(k)]\omega_{r}$ stems from the single-ion term $\hat{\mathcal{H}}_{1}$.  In this derivation, we will prove that the \emph{diagonal} commutator can be simplified to $\delta_{ij}(\omega_{n}-\omega_{m})\hat{C}^{\dagger}_{m}(i)\hat{C}_{n}(j)$.

The simplification of the \emph{diagonal} commutator begins by employing the four term commutator identity given by

\begin{equation}
[\hat{\mathcal{A}}\hat{\mathcal{B}}, \hat{\mathcal{C}}\hat{\mathcal{D}}] = \hat{\mathcal{A}}[\hat{\mathcal{B}},\hat{\mathcal{C}}]\hat{\mathcal{D}} + [\hat{\mathcal{A}},\hat{\mathcal{C}}]\hat{\mathcal{B}}\hat{\mathcal{D}} + \hat{\mathcal{C}}\hat{\mathcal{A}}[\hat{\mathcal{B}},\hat{\mathcal{D}}] + \hat{\mathcal{C}}[\hat{\mathcal{A}},\hat{\mathcal{D}}]\hat{\mathcal{B}}. 
\label{eq:identity_four}
\end{equation}

\noindent When applied to the original \emph{diagonal} commutator,  the identity yields 

\begin{align}
& \sum\limits_{k}\sum\limits_{r}[\hat{C}^{\dagger}_{m}(i)\hat{C}_{n}(i),\hat{C}^{\dagger}_{r}(k)\hat{C}_{r}(k)]\omega_{r} = \nonumber \\ & \sum\limits_{k}\sum\limits_{r} \left\{ \hat{C}^{\dagger}_{m}(i)[\hat{C}_{n}(j),\hat{C}^{\dagger}_{r}(k)]\hat{C}_{r}(k)\omega_{r} + [\hat{C}^{\dagger}_{m}(i),\hat{C}^{\dagger}_{r}(k)]\hat{C}_{n}(j)\hat{C}_{r}(k)\omega_{r} \right. \nonumber \\ &~~~~~~~\left. +  \hat{C}^{\dagger}_{r}(k)\hat{C}^{\dagger}_{m}(i)[\hat{C}_{n}(j),\hat{C}_{r}(k)]\omega_{r} + \hat{C}^{\dagger}_{r}(k)[\hat{C}^{\dagger}_{m}(i),\hat{C}_{r}(k)]\hat{C}_{n}(j)\omega_{r} \right\}. 
\label{eq:followup}
\end{align}

\noindent By utilizing the commutator identity for creation/annihilation operators given by 
	\begin{equation}
	[\hat{C}_{m},\hat{C}^{\dagger}_{n}]_{\pm} = \delta_{mn}, 
	\label{eq:commutation}
	\end{equation}
		
	\noindent where $\pm$ denotes the anticommutator$(+)$ and commutator$(-)$, respectively, for both indices pairs for eigenstates $m,n$ and sites $i,j$, Eq.~\ref{eq:followup} may be simplified to 
	
\begin{align}
\sum\limits_{k}\sum\limits_{r}[\hat{C}^{\dagger}_{m}(i)\hat{C}_{n}(i),\hat{C}^{\dagger}_{r}(k)\hat{C}_{r}(k)]\omega_{r} =  \sum\limits_{k}\sum\limits_{r} \left\{ \hat{C}^{\dagger}_{m}(i)\delta_{nr}\delta_{jk}\hat{C}_{r}(k)\omega_{r}  + \hat{C}^{\dagger}_{r}(k)\delta_{rm}\delta_{ik}\hat{C}_{n}(j)\omega_{r} \right\}, 
\label{eq:followup2}
\end{align}

\noindent where the second and third terms have been removed because $[\hat{C}_{m}^{\dagger}(i)\hat{C}^{\dagger}_{n}(j)] = 0$ and $[\hat{C}_{m}(i)\hat{C}_{n}(j)] = 0~~\forall~ \{m,n,i,j\}$.~By both expanding the sum and utilizing the property that the Kronecker delta will remove both indices $k$ and $r$, Eq.~\ref{eq:followup2} becomes 

\begin{equation}
\sum\limits_{k}\sum\limits_{r}[\hat{C}^{\dagger}_{m}(i)\hat{C}_{n}(i),\hat{C}^{\dagger}_{r}(k)\hat{C}_{r}(k)]\omega_{r} = (\omega_{n}-\omega_{m})\hat{C}^{\dagger}_{m}(i)\hat{C}_{n}(j).
\label{eq:diagonal_final}
\end{equation}

\noindent Finally, it is important to note that the expression above will be non-zero only if the raising and lowering operators operate on the same site ($i.e.$ $i$ = $j$).~This physical interpretation may be summarized by the final expression 

\begin{equation}
\sum\limits_{k}\sum\limits_{r}[\hat{C}^{\dagger}_{m}(i)\hat{C}_{n}(i),\hat{C}^{\dagger}_{r}(k)\hat{C}_{r}(k)]\omega_{r} = \delta_{ij}(\omega_{n}-\omega_{m})\hat{C}^{\dagger}_{m}(i)\hat{C}_{n}(i),
\label{eq:diagonal_final2}
\end{equation}

\noindent where the removal of the $j$ index in $\hat{C}_{n}$ is a direct consequence of the requirement that all indices must be the same. 

\subsection{Derivation of the \emph{Transverse} Commutator \emph{via} the \emph{Random Phase Decoupling Method}} \label{sec:transverse} 

The \emph{transverse} commutator given by $\sum\limits_{kl}J(kl)[\hat{C}^{\dagger}_{m}(i)\hat{C}_{n}(i),\hat{S}_{+}(k)\hat{S}_{-}(l)]$ is a result of the presence of the $S_{+}S_{-}$ terms in the inter-ion Hamiltonian $\hat{\mathcal{H}}_{2}$ (Eq.~7 in he main text). In this derivation, it will be shown that the application of mean field theory through the random phase decoupling method~\cite{wolff60:120,cooke73:7,yamada67:22,yamada66:21} given by

\begin{equation}
\begin{split}
\hat{C}^{\dagger}_{m}(i)\hat{C}_{s}(i)\hat{C}^{\dagger}_{q}(l)\hat{C}_{p}(l) &\simeq f_{m}(i)\delta_{ms}\hat{C}^{\dagger}_{q}(l)\hat{C}_{p}(l) \\
& + f_{q}(l)\delta_{pq}\hat{C}^{\dagger}_{m}(i)\hat{C}_{s}(i), 
\end{split}
\label{eq:RPA}  
\end{equation}

\noindent which decouples individual sites $i$ and $l$ (Section~\ref{sec:RPA}), results in the simplification of the \emph{transverse} commutator to  

\begin{equation} 
\sum\limits_{l}J(il) \left\{\hat{S}_{+nm}(f_{m}-f_{n})\sum_{pq}\hat{S}_{-qp}\hat{C}^{\dagger}_{q}(l)\hat{C}_{p}(l)
+ \hat{S}_{-nm}(f_{m}-f_{n})\sum_{pq}\hat{S}_{+qp}\hat{C}^{\dagger}_{q}(l)\hat{C}_{p}(l) \right\}.
\nonumber 
\end{equation}

The simplification of the \emph{transverse} commutator begins  by switching the positions of the spin and raising/lowering operarators, where a negative sign is introduced by the anticommutivity property yielding

\begin{align}
\sum\limits_{kl}J(kl)[\hat{C}^{\dagger}_{m}(i)\hat{C}_{n}(i),\hat{S}_{+}(k)\hat{S}_{-}(l)] =  \sum\limits_{kl}J(kl)\{-[\hat{S}_{+}(k)\hat{S}_{-}(l),\hat{C}^{\dagger}_{m}(i)\hat{C}_{n}(i)]\} .
\label{eq:derivation_1}
\end{align}

\noindent Utilizing the four term commutation identity given by Eq.~\ref{eq:identity_four},  Eq.~\ref{eq:derivation_1} becomes

\begin{align}
\sum\limits_{kl}J(kl)[\hat{C}^{\dagger}_{m}(i)\hat{C}_{n}(i),\hat{S}_{+}(k)\hat{S}_{-}(l)] = \sum\limits_{kl}\left\{-\hat{S}_{+}(k)[\hat{S}_{-}(l),\hat{C}^{\dagger}_{m}(i)\hat{C}_{n}(i)] - [\hat{S}_{+}(k),\hat{C}^{\dagger}_{m}(i)\hat{C}_{n}(i)]\hat{S}_{-}(l) \right\}.
\label{eq:derivation_2}
\end{align}  

\noindent By employing the definition of the spin operators $\hat{S}_{\alpha} = \sum\limits_{pq}\hat{S}_{\alpha pq} \hat{C}^{\dagger}_{p}\hat{C}_{q}$ given by Eq.~12 in the main text, Eq.~\ref{eq:derivation_2} becomes 

\begin{align}
& \sum\limits_{kl}J(kl)[\hat{C}^{\dagger}_{m}(i)\hat{C}_{n}(i),\hat{S}_{+}(k)\hat{S}_{-}(l)] = \nonumber \\ & \sum\limits_{kl}J(kl)\left\{-\hat{S}_{+}(k)\sum\limits_{pq}\hat{S}_{-pq}(l)[\hat{C}^{\dagger}_{p}(l)\hat{C}_{q}(l),\hat{C}^{\dagger}_{m}(i)\hat{C}_{n}(i)] - \sum\limits_{pq}\hat{S}_{+pq}(k)[\hat{C}^{\dagger}_{p}(k)\hat{C}_{q}(k),\hat{C}^{\dagger}_{m}(i)\hat{C}_{n}(i)]\hat{S}_{-}(l)  \right\}.
\label{eq:derivation_3}
\end{align} 

\noindent Reversing both commutators to remove the negative sign \emph{via} anticommutivity yields 

\begin{align}
& \sum\limits_{kl}J(kl)[\hat{C}^{\dagger}_{m}(i)\hat{C}_{n}(i),\hat{S}_{+}(k)\hat{S}_{-}(l)] =   \nonumber \\
& \sum\limits_{kl}\left\{\hat{S}_{+}(k)\sum\limits_{pq}\hat{S}_{-pq}(l)[\hat{C}^{\dagger}_{m}(i)\hat{C}_{n}(i),\hat{C}^{\dagger}_{p}(l)\hat{C}_{q}(l)] + \sum\limits_{pq}\hat{S}_{+pq}(k)[\hat{C}^{\dagger}_{m}(i)\hat{C}_{n}(i),\hat{C}^{\dagger}_{p}(k)\hat{C}_{q}(k)]\hat{S}_{-}(l)  \right\}.
\label{eq:derivation_4}
\end{align} 

\noindent Before proceeding, it will prove useful to derive two key commutation relations given below:

\begin{subequations} 
	\begin{align}
	[\hat{C}^{\dagger}_{m}\hat{C}_{n},\hat{C}^{\dagger}_{r}\hat{C}_{s}]  &= \delta_{nr}\hat{C}^{\dagger}_{m}\hat{C}_{s} - \delta_{sm}\hat{C}^{\dagger}_{r}\hat{C}_{n}   \label{eq:der_14} \\ 
	\sum\limits_{rs}[\hat{C}^{\dagger}_{m}\hat{C}_{n},\hat{C}^{\dagger}_{r}\hat{C}_{s}]\hat{S}_{\alpha rs} &= \sum\limits_{s}\left(\hat{C}^{\dagger}_{m}\hat{C}_{s}\hat{S}_{\alpha ns} - \hat{C}^{\dagger}_{s}\hat{C}_{n}\hat{S}_{\alpha sm}\right).  \label{eq:der_15}
	\end{align}
\end{subequations}

\noindent By combining the four term commutation identity (Eq.~\ref{eq:identity_four}) and the commutator identity (Eq.~\ref{eq:commutation}), the LHS of Eq.~\ref{eq:der_14} becomes 

\begin{subequations} 
	\begin{align}
	[\hat{C}^{\dagger}_{m}\hat{C}_{n},\hat{C}^{\dagger}_{r}\hat{C}_{s}]  &= \hat{C}^{\dagger}_{m}[\hat{C}_{n},\hat{C}^{\dagger}_{r}]\hat{C}_{s} + [\hat{C}^{\dagger}_{m},\hat{C}^{\dagger}_{r}]\hat{C}_{n}\hat{C}_{s} \nonumber \\ &  + \hat{C}^{\dagger}_{r}\hat{C}^{\dagger}_{m}[\hat{C}_{n},\hat{C}_{s}] + \hat{C}^{\dagger}_{r}[\hat{C}^{\dagger}_{m}, \hat{C}_{s}]\hat{C}_{n}  \\
	&= \hat{C}^{\dagger}_{m}[\hat{C}_{n},\hat{C}^{\dagger}_{r}]\hat{C}_{s} + \hat{C}^{\dagger}_{r}[\hat{C}^{\dagger}_{m},\hat{C}_{s}]\hat{C}_{n} \label{eq:try} \\
	&= \hat{C}^{\dagger}_{m}\delta_{nr}\hat{C}_{s} - \hat{C}^{\dagger}_{r}\delta_{sm}\hat{C}_{n}, 
	\end{align}
\end{subequations}

\noindent yielding the desired expression. Note that in Eq.~\ref{eq:try} we have employed the observation that the commutators $[\hat{C}_{n},\hat{C}_{m}]$ and $[\hat{C}^{\dagger}_{n},\hat{C}^{\dagger}_{m}]$ are zero $\forall~m,n$. \\

\hspace*{5.00mm} The derivation of the second identity begins with the observation that the identity that was just derived above (Eq.~\ref{eq:der_14}) implies that the LHS of  Eq.~\ref{eq:der_15} can be expressed as 

\begin{equation}
\sum\limits_{rs}[\hat{C}^{\dagger}_{m}\hat{C}_{n},\hat{C}^{\dagger}_{r}\hat{C}_{s}]\hat{S}_{\alpha rs} = \sum\limits_{rs}\{\delta_{nr}\hat{C}^{\dagger}_{m}\hat{C}_{s} - \delta_{sm}\hat{C}^{\dagger}_{r}\hat{C}_{n}\}\hat{S}_{\alpha rs}.  
\end{equation}

\noindent The Kronecker delta eliminates the indices $r$ and $s$ in the first and second series, respectively, yielding 

\begin{equation}
\sum\limits_{s}\hat{C}^{\dagger}_{m}\hat{C}_{s}\hat{S}_{\alpha ns} -\sum\limits_{r}\hat{C}^{\dagger}_{r}\hat{C}_{n}\hat{S}_{\alpha rm}.
\label{eq:filler2}
\end{equation} 

\noindent By replacing the index $r$ by a dummy index $s$ in the second series, and combining both series into one, Eq.~\ref{eq:filler2} becomes 

\begin{equation}
\sum\limits_{s}\left(\hat{C}^{\dagger}_{m}\hat{C}_{s}\hat{S}_{\alpha ns} - \hat{C}^{\dagger}_{s}\hat{C}_{n}\hat{S}_{\alpha sm}\right), 
\end{equation} 

\noindent corresponding to the final desired expression. 

 \hspace*{5.00mm} A comparison to the commutation relation that was just derived (Eq.~\ref{eq:der_15}) and Eq.~\ref{eq:derivation_4} suggests that both equations will have the same form if $r = p$ and $s = q$.~It is also important to note that the terms $\hat{S}_{\alpha mn}$ are scalars, and thus will commute with the commutator itself. Applying the definition of the spin operator $\hat{S}_{\alpha}$ (Eq.~12 in the main text), and relabeling $q$ as a dummy variable $s$, Eq.~\ref{eq:derivation_4} becomes

\begin{align}
& \sum\limits_{kl}J(kl)[\hat{C}^{\dagger}_{m}(i)\hat{C}_{n}(i),\hat{S}_{+}(k)\hat{S}_{-}(l)] =  \nonumber \\
& \sum\limits_{kl}J(kl)\left\{\hat{S}_{+}(k)\sum\limits_{s}\left(\hat{C}^{\dagger}_{m}(i)\hat{C}_{s}(k)\hat{S}_{-ns}(l) - \hat{C}^{\dagger}_{s}(k)\hat{C}_{n}(i)\hat{S}_{-sm}(l) \right)  \right.\nonumber \\ &~~~~~~~~~\left. + \sum\limits_{s}\left(\hat{C}^{\dagger}_{m}(i)\hat{C}_{s}(k)\hat{S}_{+ns}(k) - \hat{C}^{\dagger}_{s}(k)\hat{C}_{n}(i)\hat{S}_{+sm}(k) \right)\hat{S}_{-}(l)\right\}.
\label{eq:derivation_5}
\end{align} 

\noindent It is worthwhile to note that in order for the sum over $s$ to be physically relevant, all indices in the three product terms must be equal. This observation suggests that  in the case of the first sum, $k$~=~$l$~=~$i$, while in the case of the second term, $k$~=~$i$, and as a result Eq.~\ref{eq:derivation_5} may be rewritten as   

\begin{align}
& \sum\limits_{kl}J(kl)[\hat{C}^{\dagger}_{m}(i)\hat{C}_{n}(i),\hat{S}_{+}(k)\hat{S}_{-}(l)] = \nonumber \\
& \sum\limits_{kl}J(kl)\left\{\hat{S}_{+}(k)\sum\limits_{s}\left(\hat{C}^{\dagger}_{m}(i)\hat{C}_{s}(k)\hat{S}_{-ns}(l) - \hat{C}^{\dagger}_{s}(k)\hat{C}_{n}(i)\hat{S}_{-sm}(l) \right)\delta_{kl}\delta_{ik}  \right.\nonumber \\ &~~~~~~~~~ \left. + \sum\limits_{s}\left(\hat{C}^{\dagger}_{m}(i)\hat{C}_{s}(k)\hat{S}_{+ns}(k) - \hat{C}^{\dagger}_{s}(k)\hat{C}_{n}(i)\hat{S}_{+sm}(k) \right) \hat{S}_{-}(l)\delta_{ik} \right \}.
\label{eq:derivation_6}
\end{align}

\noindent In the presence of the Kronecker delta terms, by applying the sum over $k$, Eq.~\ref{eq:derivation_6} becomes  

\begin{align}
& \sum\limits_{kl}J(kl)[\hat{C}^{\dagger}_{m}(i)\hat{C}_{n}(i),\hat{S}_{+}(k)\hat{S}_{-}(l)] =    \nonumber \\
& \sum\limits_{l}J(il)\left\{\hat{S}_{+}(l)\sum\limits_{s}\left(\hat{C}^{\dagger}_{m}(i)\hat{C}_{s}(i)\hat{S}_{-ns}(i) - \hat{C}^{\dagger}_{s}(i)\hat{C}_{n}(i)\hat{S}_{-sm}(i) \right) \right.\nonumber \\ &~~~~~~~~~\left.  + \sum\limits_{s}\left(\hat{C}^{\dagger}_{m}(i)\hat{C}_{s}(i)\hat{S}_{+ns}(i) - \hat{C}^{\dagger}_{s}(i)\hat{C}_{n}(i)\hat{S}_{+sm}(i) \right) \hat{S}_{-}(l) \right \}.
\label{eq:derivation_7}
\end{align}

\noindent By combining the definition of $\hat{S}_{\alpha}$ with dummy indices $q$ and $p$, with the observation that the $\hat{S}_{\alpha}$ operators in Eq.~\ref{eq:derivation_7} refer to site $l$ and thus commute with site $i$, Eq.~\ref{eq:derivation_7} can be rewritten as 

\begin{align}
& \sum\limits_{kl}J(kl)[\hat{C}^{\dagger}_{m}(i)\hat{C}_{n}(i),\hat{S}_{+}(k)\hat{S}_{-}(l)] =\nonumber \\ &\sum\limits_{l}J(il)\sum\limits_{s} [\hat{S}_{+ns}(i)\hat{C}^{\dagger}_{m}(i)\hat{C}_{s}(i) -
\hat{S}_{+sm}(i)\hat{C}^{\dagger}_{s}(i)\hat{C}_{n}(i)] 
\sum\limits_{pq}\hat{S}_{-qp}(l)\hat{C}^{\dagger}(l)\hat{C}_{p}(l)  \nonumber \\  & + \sum\limits_{l}J(il)\sum\limits_{s}[\hat{S}_{-ns}(i)\hat{C}^{\dagger}_{m}(i)\hat{C}_{s}(i) - \hat{S}_{-sm}(i)\hat{C}^{\dagger}_{s}(i)\hat{C}_{n}(i)] \sum\limits_{pq}\hat{S}_{+qp}(l)\hat{C}^{\dagger}(l)\hat{C}_{p}(l). ~\label{eq:filler}
\end{align}

\noindent  By multiplying the sum over $ls$ by the sum over $pq$, Eq.~\ref{eq:filler} becomes 

\begin{align}
& \sum\limits_{kl}J(kl)[\hat{C}^{\dagger}_{m}(i)\hat{C}_{n}(i),\hat{S}_{+}(k)\hat{S}_{-}(l)] =  \nonumber \\
& \sum\limits_{l}\sum\limits_{s}\sum\limits_{pq}J(il)\left\{\hat{S}_{+ns}(i)\hat{S}_{-qp}(l)\hat{C}^{\dagger}_{m}(i)\hat{C}_{s}(i)\hat{C}^{\dagger}_{q}(l)\hat{C}_{p}(l)  - \hat{S}_{+sm}(i)\hat{S}_{-qp}(l)\hat{C}^{\dagger}_{s}(i)\hat{C}_{n}(i)\hat{C}^{\dagger}_{q}(l)\hat{C}_{p}(l)  \right.\nonumber \\ &~~~~~~~~~~~~~~~~~~\left. +\hat{S}_{-ns}(i)\hat{S}_{+qp}(l)\hat{C}^{\dagger}_{m}(i)\hat{C}_{s}(i)\hat{C}^{\dagger}_{q}(l)\hat{C}_{p}(l) - \hat{S}_{-sm}(i)\hat{S}_{+qp}(l)\hat{C}^{\dagger}_{s}(i)\hat{C}_{n}(i)\hat{C}^{\dagger}_{q}(l)\hat{C}_{p}(l) \right \}.
\label{eq:derivation_8}
\end{align}

\noindent It is important to note that Eq.~\ref{eq:derivation_8} produces four product terms of the form $\hat{C}^{\dagger}_{m}(i)\hat{C}_{n}\hat{C}^{\dagger}_{q}(l)\hat{C}_{p}(l)$, representing simultaneous $mn$ and $qp$ transitions at sites $i$ and $l$, respectively.~These sites can be decoupled by the application of mean field theory through the \emph{random phase decoupling method}~\cite{cooke73:7,buyers75:11,wolff60:120,yamada67:22,yamada66:21}. As described later in Section~\ref{sec:RPA} and summarized by Eq.~\ref{eq:RPA}, such decoupling results in the replacement of terms of the form $\hat{C}^{\dagger}_{m}(i)\hat{C}_{s}(i)\hat{C}^{\dagger}_{q}(l)\hat{C}_{p}(l)$ by $f_{m}(i)\delta_{ms}\hat{C}^{\dagger}_{q}(l)\hat{C}_{p}(l) + f_{q}(l)\delta_{pq}\hat{C}^{\dagger}_{m}(i)\hat{C}_{s}(i)$, thus allowing Eq.~\ref{eq:derivation_8} to be rewritten as 

\begin{align}
& \sum\limits_{kl}J(kl)[\hat{C}^{\dagger}_{m}(i)\hat{C}_{n}(i),\hat{S}_{+}(k)\hat{S}_{-}(l)] =  \nonumber \\
\sum\limits_{l}\sum\limits_{s}\sum\limits_{pq}J(il)&\left\{\hat{S}_{+ns}(i)\hat{S}_{-qp}(l)(f_{m}(i)\delta_{ms}\hat{C}^{\dagger}_{q}(l)\hat{C}_{p}(l) + f_{q}(l)\delta_{pq}\hat{C}^{\dagger}_{m}(i)\hat{C}_{s}(i)) \right.\nonumber \\ & \left. - \hat{S}_{+sm}(i)\hat{S}_{-qp}(l)(f_{s}(i)\delta_{sn}\hat{C}^{\dagger}_{q}(l)\hat{C}_{p}(l) + f_{q}(l)\delta_{pq}\hat{C}^{\dagger}_{s}(i)\hat{C}_{n}(i))  \right.\nonumber \\ &\left. + \hat{S}_{-ns}(i)\hat{S}_{+qp}(l)(f_{m}(i)\delta_{ms}\hat{C}^{\dagger}_{q}(l)\hat{C}_{p}(l) + f_{q}(l)\delta_{pq}\hat{C}^{\dagger}_{m}(i)\hat{C}_{s}(i)) \right.\nonumber \\ 
&\left. - \hat{S}_{-sm}(i)\hat{S}_{+qp}(l)(f_{s}(i)\delta_{sn}\hat{C}^{\dagger}_{q}(l)\hat{C}_{p}(l) + f_{q}(l)\delta_{pq}\hat{C}^{\dagger}_{s}(i)\hat{C}_{n}(i))
\right \}.
\label{eq:derivation_9}
\end{align}

\noindent The equation above may be simplified by first observing that there is a $\delta_{pq}$ prefactor in four out of the eight terms.~This observation reduces the sum over $p,q$ to $p=q$. Next, it is important to recall that $\hat{S}_{\pm nn}$ must be zero $\forall n$ since the elements of the transverse spin operators $\hat{S}_{\pm}$ must be entirely off-diagonal. Consequently, any term with $\delta_{pq}$ can be neglected. Therefore, Eq.~\ref{eq:derivation_9} may be reduced to a sum of four remaining terms given by  

\begin{align}
& \sum\limits_{kl}J(kl)[\hat{C}^{\dagger}_{m}(i)\hat{C}_{n}(i),\hat{S}_{+}(k)\hat{S}_{-}(l)] = ~~~~~  \nonumber \\
& \sum\limits_{l}\sum\limits_{s}\sum\limits_{pq}J(il)\left\{\hat{S}_{+ns}(i)\hat{S}_{-qp}(l)(f_{m}(i)\delta_{ms}\hat{C}^{\dagger}_{q}(l)\hat{C}_{p}(l)) - \hat{S}_{+sm}(i)\hat{S}_{-qp}(l)(f_{s}(i)\delta_{sn}\hat{C}^{\dagger}_{q}(l)\hat{C}_{p}(l))  \right.\nonumber \\ &~~~~~~~~~~~~~~ \left. \hat{S}_{-ns}(i)\hat{S}_{+qp}(l)(f_{m}(i)\delta_{ms}\hat{C}^{\dagger}_{q}(l)\hat{C}_{p}(l)) - \hat{S}_{-sm}(i)\hat{S}_{+qp}(l)(f_{s}(i)\delta_{sn}\hat{C}^{\dagger}_{q}(l)\hat{C}_{p}(l))
\right \}.
\label{eq:derivation_10}
\end{align}

\noindent Due to the presence of the Kronecker delta terms, the index $s$ is replaced by $m$ or $n$ for terms 2 and 4 or 1 and 3, respectively.~Therefore, the final expression for the \emph{transverse} commutator is 

\begin{align}
& \sum\limits_{kl}J(kl)[\hat{C}^{\dagger}_{m}(i)\hat{C}_{n}(i),\hat{S}_{+}(k)\hat{S}_{-}(l)] = \\
&  \sum\limits_{l}J(il) \left\{\hat{S}_{+nm}(f_{m}-f_{n})\sum_{pq}\hat{S}_{-qp}\hat{C}^{\dagger}_{q}(l)\hat{C}_{p}(l) + \hat{S}_{-nm}(f_{m}-f_{n})\sum_{pq}\hat{S}_{+qp}\hat{C}^{\dagger}_{q}(l)\hat{C}_{p}(l) \right\}.
\label{eq:long6_2}
\end{align}

\noindent It should be noted that the site indices have been dropped for the thermal weighting factor $f$ since it assumed that in the case of a perfect crystal, the values of the single-ion energy levels are identical for all sites.

\subsection{Derivation of the \emph{Longitudinal} Commutator} \label{sec:longitudinal} 

The \textit{longitudinal} commutator given by $\sum\limits_{lk}J(kl)[\hat{C}^{\dagger}_{m}(i)\hat{C}_{n}(i),\hat{S}_{z}(k)\{\hat{S}_{z}(l)-2\langle \hat{S}_{z}(l)\rangle\}]$ stems from the $\hat{S}_{z}(i)[\hat{S}_{z}(j)-2\langle \hat{S}_{z}(j)\rangle]$ term in $\hat{\mathcal{H}}_{2}$ (Eq.~7 in the main text). In this derivation, it will be shown that the \emph{longitudinal} commutator may be simplified to  $2\sum\limits_{l}J(il)\hat{S}_{znm}(f_{m}-f_{n})\sum\limits_{pq}\hat{S}_{zpq}\hat{C}^{\dagger}_{q}(l)\hat{C}_{p}(l)$.

\hspace*{5.00mm} The simplification of the longitudinal commutator begins with the observation that the first component of the \emph{longitudinal} commutator is similar to the commutator that was previously solved for the \emph{transverse} terms above.~One difference in particular is that terms with $\delta_{pq}$ are not immediately discarded due to the spin matrices elements since $\hat{S}_{zpp}$ is not necessarily zero. Instead, the cancellation of these terms results from the presence of the additional $-2\langle \hat{S}_{z}(l)\rangle$ contribution. Therefore, the expression for the \emph{longitudinal} commutator is nearly identical to the one that was previously observed for the case of the \emph{transverse commutator}, with an additional factor of 2 since $\hat{S}_{+nm}$ and $\hat{S}_{-nm}$ in the previous derivation are now both $\hat{S}_{znm}$. \\
\hspace*{5.00mm} This cancellation may be demonstrated explicitly by first focusing on the commutator with the $\langle \hat{S}_{z}(l)\rangle$ pre-factor in the \emph{longitudinal} commuator that is given by

\begin{equation}
\sum\limits_{lk}J(kl)\left\{-2\langle \hat{S}_{z}(l)\rangle [\hat{C}^{\dagger}_{m}(i)\hat{C}_{n}(i),\sum\limits_{qp}\hat{S}_{zqp}(k)\hat{C}^{\dagger}(k)\hat{C}_{p}(k)] \right\}.
\label{eq:filler3}
\end{equation}

\noindent Pulling out the sum over $qp$,~\ref{eq:filler3} becomes

\begin{equation}
\sum\limits_{lk}\sum\limits_{qp}J(kl)\left\{-2\langle \hat{S}_{z}(l)\rangle \hat{S}_{zqp}(k) [\hat{C}^{\dagger}_{m}(i)\hat{C}_{n}(i),\hat{C}^{\dagger}(k)\hat{C}_{p}(k)] \right\}.
\label{eq:filler4}
\end{equation}

\noindent Employing the commutator identity from Eq.~\ref{eq:der_15},~\ref{eq:filler4} becomes

\begin{equation}
\sum\limits_{lk}\sum\limits_{p}J(kl)\left\{-2\langle \hat{S}_{z}(l)\rangle( \hat{C}^{\dagger}_{m}(i)\hat{C}_{p}(k)\hat{S}_{znp}(k) - \hat{C}^{\dagger}_{p}(k)\hat{C}_{n}(i)\hat{S}_{zpm}(k))  \right\},
\end{equation} 

\noindent which becomes  

\begin{equation}
\sum\limits_{lk}\sum\limits_{p}\sum\limits_{a}J(kl)\left\{-2(f_{a}(l)\hat{S}_{zaa}(l))(\hat{C}^{\dagger}_{m}(i)\hat{C}_{p}(k)\hat{S}_{znp}(k) - \hat{C}^{\dagger}_{p}(k)\hat{C}_{n}(i)\hat{S}_{zpm}(k)) \right\}
\end{equation} 

\noindent after the replacement $\langle \hat{S}_{z} \rangle$ by its definition given by Eq.~8 in the main text. Finally, by relabeling the indices $a$ as $p$ and $p$ as $s$, these terms are identical to the terms containing $\delta_{pq}$ in Eq.~\ref{eq:derivation_9}.   

\subsection{Derivation of $\left\langle [\hat{A},\hat{B}] \right\rangle$} \label{sec:AB}

The final commutator given by $\langle [\hat{A},\hat{B}] \rangle$ corresponds to the first term in the equation-of-motion of the response function discussed in Section~\ref{sec:equation-of-motion}. In this derivation, it will be shown that the $\langle [\hat{A},\hat{B}] \rangle$ commutator may be simplified to $(f_{m}(i)-f_{n}(i))\hat{S}_{\beta n m}(j)\delta_{ij}$.

\hspace*{5.00mm} The derivation begins by noting that the definition of the inter-level susceptibility (Eq.~13 in the main text) allows for the expectation value $\langle [\hat{A},\hat{B}] \rangle$, where $\hat{A}= \hat{C}_{m}(i)\hat{C}_{n}(i)$ and $\hat{B}=\hat{S}_{\beta}$, to be expressed as 

\begin{equation}
\left\langle \left[\hat{C}_{m}(i)\hat{C}_{n}(i),\sum\limits_{pq}\hat{S}_{\beta pq}(j)\hat{C}^{\dagger}_{p}(j)\hat{C}_{q}(j)\right]\right\rangle.  
\label{eq:ab1}
\end{equation} 

\noindent Pulling out the sum over $pq$ and $\hat{S}_{\beta pq}$, expression~\ref{eq:ab1} becomes 

\begin{equation}
\left\langle \sum\limits_{pq} [\hat{C}_{m}(i)\hat{C}_{n}(i),\hat{C}^{\dagger}_{p}(j)\hat{C}_{q}(j)]\hat{S}_{\beta pq}(j) \right\rangle,
\end{equation}

\noindent which becomes 

\begin{equation}
\left\langle \sum\limits_{q}\left\{\hat{C}^{\dagger}_{m}(i)\hat{C}_{q}(i)\hat{S}_{\beta nq}(j) - \hat{C}^{\dagger}_{q}(i)\hat{C}_{n}(i)\hat{S}_{\beta qm}(j) \right\} \right\rangle,
\label{eq:ab2}
\end{equation}

\noindent after employing the commutation identity (Eq.~\ref{eq:der_15}). In order for the expression above to be non-zero, all the indices must be equal to one another, and thus a $\delta_{ij}$ is introduced, allowing expression~\ref{eq:ab2} to rewritten as

\begin{equation}
\left\langle \sum\limits_{q}\left\{\hat{C}^{\dagger}_{m}(i)\hat{C}_{q}(i)\hat{S}_{\beta nq}(j)\delta_{ij} - \hat{C}^{\dagger}_{q}(i)\hat{C}_{n}(i)\hat{S}_{\beta qm}(j)\delta_{ij} \right\} \right\rangle.
\label{eq:ab3}
\end{equation}

\noindent The first term in expression~\ref{eq:ab3} will now be addressed.~By pulling both the sum and $\hat{S}_{\beta nq}$ out of the expectation value, the first term may be rewritten as 

\begin{equation}
\sum\limits_{q}\hat{S}_{\beta nq}(j)\delta_{ij} \left\langle  \hat{C}^{\dagger}_{m}(i)\hat{C}_{q}(i) \right\rangle.
\label{eq:ab4}
\end{equation}

\noindent By noting that the term $\hat{C}^{\dagger}_{m}(i)\hat{C}_{q}(i)$ is an operator and employing the definition of the expectation value, the expression above is equivalent to 

\begin{equation}
\sum\limits_{q}\hat{S}_{\beta n q}(j)\delta_{ij}\sum_{m}f_{m}(i)\left\langle m|\hat{C}^{\dagger}_{m}(i)\hat{C}_{q}(i)|m\right\rangle. 
\end{equation} 

\noindent Since this inner product will only be non-zero if $q = m$, the expression above can be rewritten as  

\begin{equation}
\sum\limits_{q}\sum_{m}\hat{S}_{\beta n q}(j)\delta_{ij}f_{m}(i)\left\langle m|\hat{C}^{\dagger}_{m}(i)\hat{C}_{q}(i)\delta_{qm}|m\right\rangle, 
\end{equation}

\noindent which reduces the sum over $q$ to 

\begin{equation}
\sum_{m}\hat{S}_{\beta n m}(j)\delta_{ij}f_{m}(i)\left\langle m|\hat{C}^{\dagger}_{m}(i)\hat{C}_{m}(i)|m\right\rangle. 
\label{eq:ab5}
\end{equation}

\noindent Since the definition of the single-ion Hamiltonian requires that $\sum\limits_{m}\left\langle m|\hat{C}^{\dagger}_{m}(i)\hat{C}_{m}(i)|m\right\rangle$ must be unity, the first term may be expressed as  

\begin{equation}
\hat{S}_{\beta n m}(j)\delta_{ij}f_{m}(i),
\end{equation} 

\noindent where the sum over $m$ is implied based on the definition of $G^{\beta}$. The repetition of the process described above (expressions~\ref{eq:ab4}-\ref{eq:ab5}) for the second term yields   

\begin{equation}
\hat{S}_{\beta n m}(j)\delta_{ij}f_{n}(i).
\end{equation} 

\noindent Finally, a sum of both contributions in expression~\ref{eq:ab3} yields 

\begin{equation}
\left\langle [\hat{A},\hat{B}] \right\rangle = (f_{m}(i)-f_{n}(i))\hat{S}_{\beta n m}(j)\delta_{ij},
\label{eq:final_inner_2}   
\end{equation} 

\noindent corresponding to the desired simplified expression.

\section{\emph{R\lowercase{andom} P\lowercase{hase} D\lowercase{ecoupling} M\lowercase{ethod}}} ~\label{sec:RPA} 

Similar to the decoupling method used in the band theory of magnetism, the \emph{random phase decoupling method}~\cite{cooke73:7,buyers75:11,wolff60:120,yamada67:22,yamada66:21} begins by employing a mean-field approach where the excitations for each of the sites $i$ and $k$ are approximated as a variation away $\delta_{n}$ from a ``mean'' value $n$. In the case of the four operator product $\hat{C}^{\dagger}_{m}(i)\hat{C}_{s}(i)\hat{C}^{\dagger}_{q}(l)\hat{C}_{p}(l)$, such an approximation corresponds to   

\begin{equation}
 \left\{n_{m}(i)\delta_{ms} + \delta_{n_{ms}(i)} \right\}\left\{n_{q}(l)\delta_{qp} + \delta_{n_{qp}(l)} \right\},
\label{eq:RPA_1}
\end{equation}

\noindent where $n_{m}(i)$ correspond to the occupation number at a site $i$ for the eigenstate $|m\rangle$ if indices $m$=$s$, while $\delta_{n_{ms}(i)}$ corresponds to deviations away from this occupation number at the site $i$ due to excitations between $|m\rangle$ and $|s\rangle$. By expanding the product, ignoring both the constant $n_{m}(i)\delta_{ms}n_{q}(l)\delta_{qp}$ and $O(\delta^{2})$ terms, expression~\ref{eq:RPA_1} may be simplified to 

\begin{equation}
n_{m}(i)\delta_{ms}\delta_{n_{qp}(l)} + n_{q}(l)\delta_{qp}\delta_{n_{ms}(i)}.
\label{eq:RPA_2}
\end{equation}
 
\noindent Since the occupation number of an eigenstate $|m\rangle$ at a site $i$ is given by the thermal weighting factor $f_{m}(i)$ that was defined in Section~IIA in the main text, while deviations from $n_{m}(i)$ are due to excitations between $|m\rangle$ and $|s\rangle$ eigenstates given by $\hat{C}^{\dagger}_{m}(i)\hat{C}_{s}(i)$ in the second quantization formalism, expression~\ref{eq:RPA_2} can be alternatively expressed as 

\begin{equation}
f_{m}(i)\delta_{ms}\hat{C}^{\dagger}_{q}(l)\hat{C}_{p}(l) + f_{q}(l)\delta_{qp}\hat{C}^{\dagger}_{m}(i)\hat{C}_{s}(i),
\label{eq:RPA_2}
\end{equation}

\noindent corresponding to the RHS of Eq.~\ref{eq:RPA}. As discussed in Section~IIC in the main text, the importance of the \emph{random phase decoupling method} and its conversion of the four operator product $\hat{C}^{\dagger}_{m}(i)\hat{C}_{s}(i)\hat{C}^{\dagger}_{q}(l)\hat{C}_{p}(l)$ into expression~\ref{eq:RPA_2} is that its use results in the simplification of the equation-of-motion of the response function (Eq.~\ref{eq:10}) to a set of coupled linear equations given by Eq.~16 in the main text, thus faciliating a direct comparison to the experimental data that can be readily calculated numerically.   

\clearpage
\newpage

\section{D\lowercase{isplacement} V\lowercase{ectors in $s$- and $d$-}S\lowercase{ublattices of} C\lowercase{o}O} \label{sec:tables} 
\begin{table*}[h!]
	\caption{Summary of displacement vectors $\mathbf{d}_{m,ij}$ for each coordination shell $m$ in $s$- and $d$-sublattices of CoO assuming a type-II collinear antiferromagnetic structure. All vectors were calculated using the VESTA visualization software package~\cite{VESTA}, employing the cubic unit cell parameters at room temperature~\cite{jauch01:64}. Numbers in parentheses indicate statistical errors.} 	
	\centering
	{\renewcommand{\arraystretch}{1.25}
		\begin{tabular}{|c|c|c|c|c|}
			\hline
			~~~$m$~~~&~~~$|\mathbf{d}_{m,ij}|$ (\AA)~~~&~~~Number of Neighbors~~~&~~~$\mathbf{d}_{m,ij}$ in $s$-Sublattice ($a$)~~~&~~~$\mathbf{d}_{m,ij}$ in $d$-Sublattice ($a$)~~~\\  
			\hline 
			\multirow{6}{*}{1} 	&   \multirow{6}{*}{3.0144(8)} & \multirow{6}{*}{12} & $\left(~0-\frac{1}{2}~\frac{1}{2}\right)$ & $\left(-\frac{1}{2}-\frac{1}{2}~0\right)$ \\  
			&   &  & $\left(~\frac{1}{2}-\frac{1}{2}~0\right)$ & $\left(-\frac{1}{2}0-\frac{1}{2}\right)$   \\ 
			&   &  &$\left(~\frac{1}{2}~0-\frac{1}{2}\right)$ & $\left(~0-\frac{1}{2}-\frac{1}{2}\right)$   \\ 
			&   & &$\left(~0~\frac{1}{2}-\frac{1}{2}\right)$  & $\left(~0~\frac{1}{2}~\frac{1}{2}\right)$    \\ 
			&   &  & $\left(-\frac{1}{2}~\frac{1}{2}~0 \right)$ & $\left(~\frac{1}{2}~0~\frac{1}{2}\right)$   \\ 
			&   &  & $\left(-\frac{1}{2}~0~\frac{1}{2}\right)$ &  $\left(~\frac{1}{2}~\frac{1}{2}~0\right)$  \\
			\hline
			\multirow{6}{*}{2} 	&   \multirow{6}{*}{4.2630(11) } & \multirow{6}{*}{6} &  & $(-1~0~0)$ \\  
			&   &  &   & $(~0~0-1)$ \\ 
			&   &  &   & $(~0-1~0)$ \\ 
			&   &  &  &  $(~1~0~0)$ \\
			&   &  &   & $(~0~0~1)$ \\  
			&   &  & &  $(~0~1~0)$  \\ \hline
			\multirow{12}{*}{3} 	&   \multirow{12}{*}{5.2211(13)} & \multirow{12}{*}{24}  & $\left(~1-\frac{1}{2}-\frac{1}{2}\right)$ & $\left(~1-\frac{1}{2}~\frac{1}{2}\right)$ \\  
			&   &  & $\left(~1~\frac{1}{2}~\frac{1}{2}\right)$ & $\left(~1~\frac{1}{2}-\frac{1}{2}\right)$ \\ 
			&   &  & $\left(-1~\frac{1}{2}~\frac{1}{2}\right)$ & $\left(-1-\frac{1}{2}~\frac{1}{2}\right)$  \\ 
			&   &  & $\left(-1-\frac{1}{2}-\frac{1}{2} \right)$ & $\left(-1~\frac{1}{2}-\frac{1}{2}\right)$   \\ 
			&   &  & $\left(~\frac{1}{2}-1~\frac{1}{2}\right)$ & $\left(-\frac{1}{2}-1~\frac{1}{2}\right)$   \\
			&   &  & $\left(-\frac{1}{2}-1-\frac{1}{2}\right)$ & $\left(~\frac{1}{2}-1-\frac{1}{2}\right)$   \\  
			&   & & $\left(-\frac{1}{2}~1-\frac{1}{2}\right)$  & $\left(-\frac{1}{2}~1~\frac{1}{2}\right)$   \\ 
			&   &  & $\left(~\frac{1}{2}~1~\frac{1}{2}\right)$ & $\left(~\frac{1}{2}~1-\frac{1}{2}\right)$   \\ 
			&   &  & $\left(-\frac{1}{2}-\frac{1}{2}~1\right)$ & $\left(-\frac{1}{2}~\frac{1}{2}~1\right)$   \\ 
			&   &  & $\left(~\frac{1}{2}~\frac{1}{2}~1\right)$ & $\left(~\frac{1}{2}-\frac{1}{2}~1\right)$   \\
			&   &  & $\left(~\frac{1}{2}~\frac{1}{2}-1\right)$ & $\left(-\frac{1}{2}~\frac{1}{2}-1\right)$   \\ 
			&   &  & $\left(-\frac{1}{2}-\frac{1}{2}-1\right)$ &  $\left(~\frac{1}{2}-\frac{1}{2}-1\right)$   \\ \hline
			\multirow{12}{*}{4} 	&   \multirow{12}{*}{6.0288(15)} & \multirow{12}{*}{12} &   $(~1~0~1)$ & \\  
			&   &  &    $(~1~0-1)$ & \\ 
			&   &  &    $(~1-1~0)$ & \\ 
			&   &  &   $(~0-1~1)$ & \\ 
			&   &  &   $(~0~1-1)$ & \\
			&   &  &   $(-1~0~1)$ & \\  
			&   &  &   $(-1~1~0)$ &\\ 
			&   &  &   $(~1~1~0)$ &\\ 
			&   &  &   $(~0~1~1)$ &\\ 
			&   &  &   $(~0-1-1)$ &\\
			&   &  &   $(-1-1~0)$ &\\ 
			&   &  &   $(-1~0-1)$ & \\ \hline
	\end{tabular}}
\label{tab:long}
\end{table*}

\FloatBarrier
\clearpage

\section{Supplementary Analysis} ~\label{sec:figures}

\begin{figure}[H]
\centering
\includegraphics[width=1.0\linewidth]{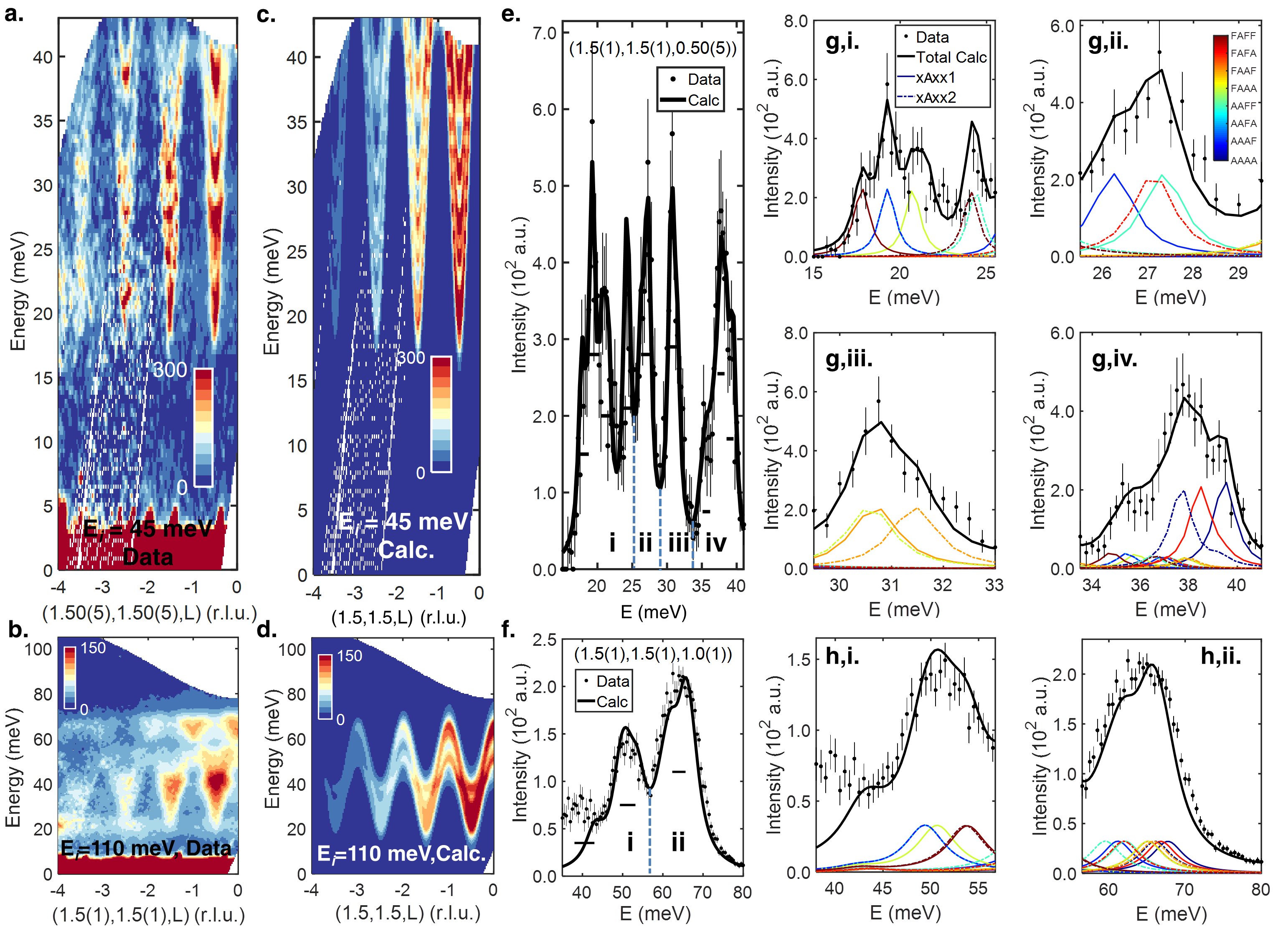}
\caption{Comparison of $(\mathbf{Q},E)$ slices and corresponding $\mathbf{Q}$-integrated cuts for CoO measured on MERLIN at 5~K and calculated with a mean-field multi-level spin-orbit exciton model employing the refined parameters listed in Tabs.~2 and~3 for an $E_{i}$ of (a,c,e) 45~meV, and (b,d,f) 110~meV, where the individual contributions for each $xAxx\gamma$ orbital configuration to the $\mathbf{Q}$-integrated cuts have been explicitly labeled in (g,h). Horizontal bars in $\mathbf{Q}$-integrated cuts indicate experimental resolution.}
\label{fig:modelA}
\end{figure}

\begin{figure}[H]
	\centering
	\includegraphics[width=0.55\linewidth]{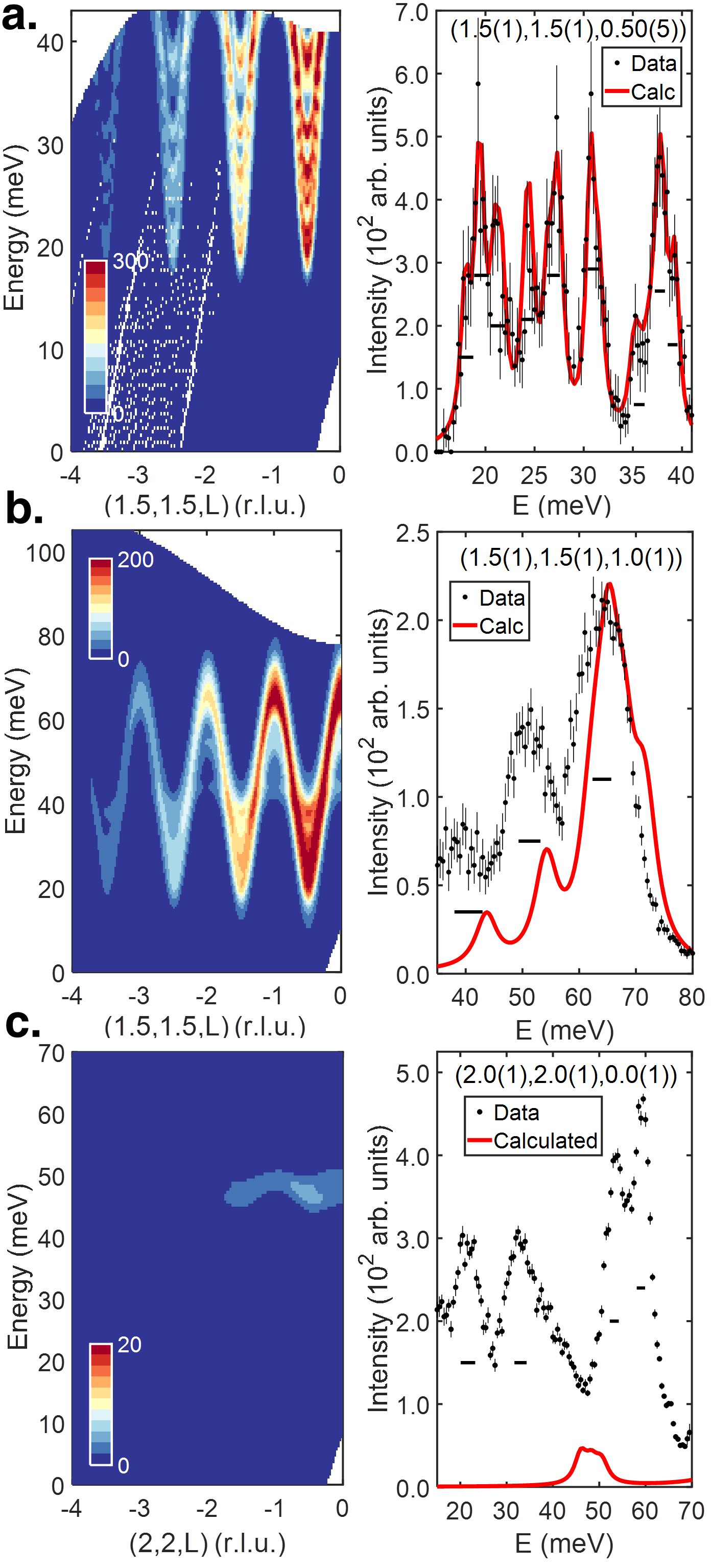}
	\caption{Calculated ($\mathbf{Q},E$) slices for an $E_{i}$ of (a) 45 meV, (b) 110 meV and (c) 70 meV for the spin-orbit exciton model with the values of $J_{m,\xi}$ restricted to be equal to their respective values originally reported~\cite{sarte18:98} for Mg$_{0.97}$Co$_{0.03}$O for a given refined value of $\lambda$. (d-f) Comparison of measured (5~K) and calculated $\mathbf{Q}$-integrated cuts of (a-c). Horizontal bars indicate instrumental resolution. The refined values of the mean molecular field $H_{MF}$ are listed in Tab.~S\ref{tab:5} for the 16 $xAxx\gamma$  orbital configurations, each with refined values of $\lambda$, $\Gamma$, and $J_{m,\xi}$ summarized in Tab.~S\ref{tab:4}.}
	\label{fig:modelC}
\end{figure}

\begin{figure}[H]
	\centering
	\includegraphics[width=0.55\linewidth]{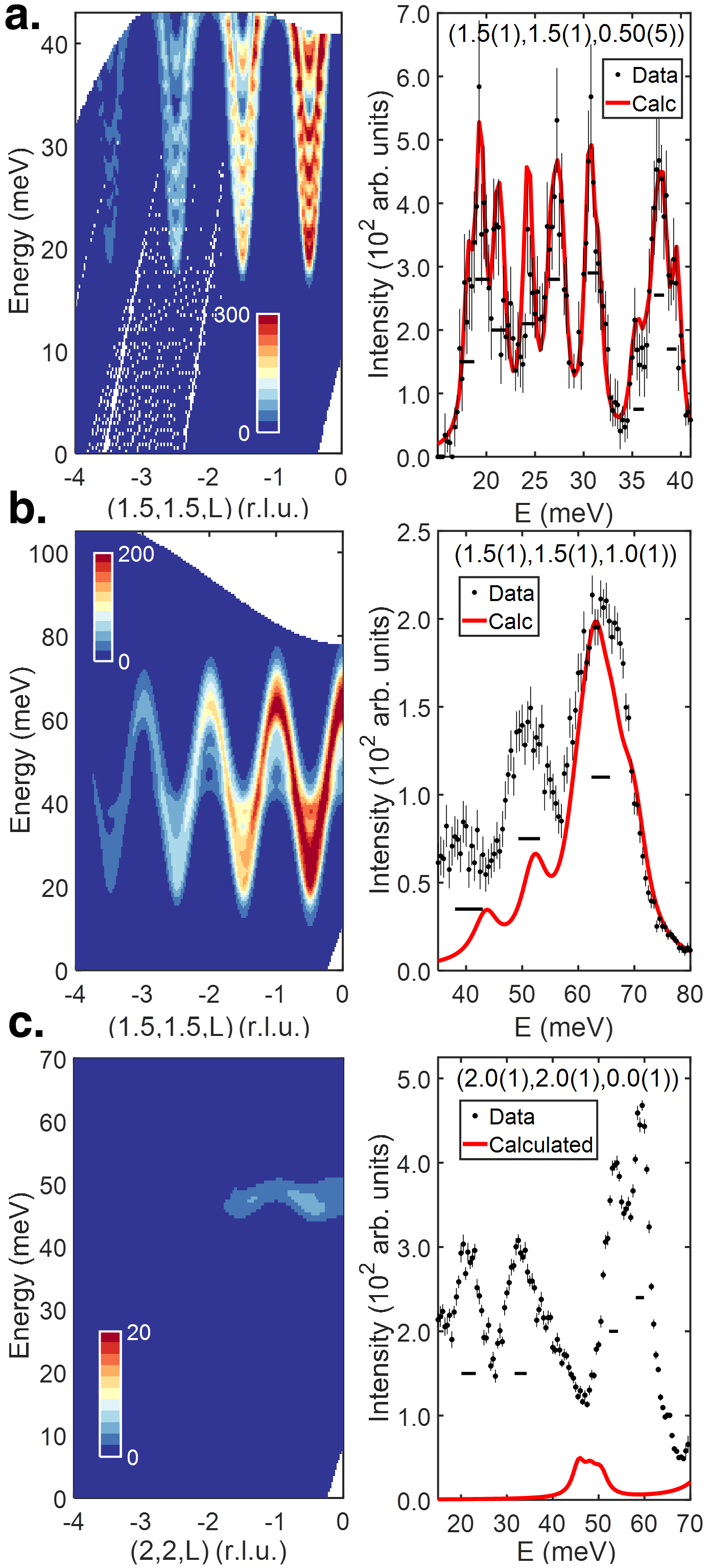}
	\caption{Calculated ($\mathbf{Q},E$) slices for an $E_{i}$ of (a) 45 meV, (b) 110 meV and (c) 70 meV. (d-f) Comparison of measured (5~K) and calculated $\mathbf{Q}$-integrated cuts of (a-c). The calculated model employed the same assumptions as the previous model presented in Fig.~S\ref{fig:modelC} with a spin-orbit coupling constant $\lambda$ allowed to vary down to $-23$~meV \emph{in lieu} of $-19$~meV.  Horizontal bars indicate instrumental resolution. The refined values of the mean molecular field $H_{MF}$ are listed in Tab.~S\ref{tab:5} for the 16 $xAxx\gamma$  orbital configurations, each with refined values of $\lambda$, $\Gamma$, and $J_{m,\xi}$ summarized in Tab.~S\ref{tab:4}.}
	\label{fig:modelD}
\end{figure}

In this final section, a summary of the supplementary analysis performed for this investigation is presented. Figure~S\ref{fig:modelA} illustrates the contributions to the total calculated scattering intensity from the 16 individual $xAxx\gamma$ orbital configurations at both the $(1.5,1.5,0.5)$ zone center and the $(1.5,1.5,1.0)$ zone boundary. Figures~S\ref{fig:modelC} and~S\ref{fig:modelD} both summarize the neutron scattering intensity calculated using the spin-orbit exciton model employing the refined values of the mean molecular field $H_{MF}$ that are listed in Tab.~S\ref{tab:5} for the 16 $xAxx\gamma$  orbital configurations, each with refined values of $\lambda$, $\Gamma$, and $J_{m,\xi}$ that are summarized in Tab.~S\ref{tab:4}. In both figures, the values for the exchange constants $J_{n}$ were fixed to be equal to the values reported~\cite{sarte18:98} for the dilute monoixde Mg$_{0.97}$Co$_{0.03}$O for a particular value of $\lambda$, which was itself allowed to vary in a range of $[-19,-13]$~meV for the former, while the latter employed an extended range of $[-23,-13]$~meV.  

\begin{table}[htb!]
	\caption{Refined values (in meV) of the mean molecular field parameter $H_{MF}$ for all 16 $xAxx\gamma$ orbital configurations that are considered in a spin-orbit exciton model, each with refined values of $\lambda$, $\Gamma$, $J_{m,\xi}$ listed in Tab.~S\ref{tab:4}, and where the values of $J_{m,\xi}$ are restricted to be equal to their respective values originally reported~\cite{sarte18:98} for Mg$_{0.97}$Co$_{0.03}$O for a given refined value of $\lambda$ that is allowed to vary down to $-19$ (Model A) and $-23$~meV (Model B).  Numbers in parentheses indicate statistical errors.}
	{\renewcommand{\arraystretch}{1.8}
		\begin{tabular}{|c|c|c|}
			\hline
			~~Orbital Configuration~~&~~Value (Model A)~~&~~Value (Model B)~~\\ 
			\hline\hline
			AAAA1   & 68.0(3) & 65.2(2) \\ \hline
			AAAA2   &69.4(4)  & 66.9(2)  \\ \hline\hline
			AAAF1   & 63.2(2) & 60.7(2)  \\ \hline 
			AAAF2   & 63.9(3) & 61.2(2)  \\ \hline \hline
			AAFA1   & 52.4(2) & 49.7(1)\\ \hline
			AAFA2   & 53.4(2) & 50.7(2)\\ \hline \hline
			AAFF1   & 61.8(3)& 58.9(3)\\ \hline
			AAFF2   & 64.2(3) & 61.6(3)\\ \hline \hline
			FAAA1   & 63.0(2) & 60.2(3)\\ \hline
			FAAA2   & 68.3(3) & 65.7(3)\\ \hline \hline
			FAAF1   & 59.6(2) & 56.8(2)\\ \hline
			FAAF2   & 61.5(2) & 58.8(3)\\ \hline \hline
			FAFA1   & 62.3(3) & 59.7(2) \\ \hline
			FAFA2   & 62.3(3) & 59.4(2)\\ \hline \hline
			FAFF1   & 58.0(2) &55.1(2)\\ \hline
			FAFF2   & 59.0(2) & 56.3(2)\\ \hline  \hline
			Average & 61.9(3) & 59.2(4)\\ \hline  
	\end{tabular}}
	\label{tab:5}
\end{table}

\begin{table*}[htb!]
	\caption{Summary of the initial values, parameter spaces, and refined values for the parameters of the spin-orbit exciton model where the values of $J_{m,\xi}$ are restricted to be equal to their respective values originally reported~\cite{sarte18:98} for Mg$_{0.97}$Co$_{0.03}$O for a given refined value of $\lambda$ that is allowed to vary down to $-19$ (Model A) and $-23$~meV (Model B). All values are reported in meV and numbers in parentheses indicate statistical errors.}
	{\renewcommand{\arraystretch}{1.8}
		\begin{tabular}{|c|c|c|c|c|c|}
			\hline
			$~~$Parameter$~~$&   Initial Value &Range (Model A)&Range  (Model B)&Refined Value (Model A)&Refined Value (Model B)\\ 
			\hline
			$\lambda$ 	&  $-16$	& [$-19$,$-13$] & [$-23$,$-13$]  	&  $-19.00(1)$ & -23.00(1)	\\ \hline
			$\Gamma$   	& $-8.76$ 	& \multicolumn{2}{c|}{[$-8.76$,$-6.16$]} 	&   \multicolumn{2}{c|}{$-6.16(1)$}    		\\ \hline
			$J_{1F}$   	& $-0.918$  	&  [$-0.945$,$-0.912$]& [$-1.134$,$-0.730$]&  $-0.912(1)$ & $-0.908(1) $ 	\\ \hline
			$J_{1AF}$ 	& 1.000  	& [0.998,1.031]& [0.798,1.24] &  0.998(1) & 	0.986(1)\\  \hline
			$J_{2}$    		& 3.09 	& [2.86,3.79]& [2.29,4.55]&  2.86(1) & 2.702(1) 		\\   \hline
			$J_{3F}$    	& $-0.182$ 	& [$-0.183$,$-0.181$] 	& [$-0.220$,$-0.145$]&   \multicolumn{2}{c|}{$-0.181(1)$} 	\\ \hline 
			$J_{3AF}$    	& 0.262 	& [0.261,0.263] & [0.209,0.316]&   \multicolumn{2}{c|}{0.261(1)}  	\\   \hline
			$J_{4F}$    	& $-0.0504$ 	& [$-0.0606$,$-0.0503$] 	& [$-0.0581$,$-0.0402$]&    \multicolumn{2}{c|}{$-0.0503(1)$}  \\ \hline
			$J_{4AF}$    	& 0.0759  	& [0.0758,0.0760] &  [0.0606,0.0874]  &   \multicolumn{2}{c|}{0.0758(1)} 	\\  \hline
			$H_{MF}$		& 64.8	& \multicolumn{2}{c|}{[0,100]} & 		 \multicolumn{2}{c|}{Tab.~S\ref{tab:5}}		\\ \hline
	\end{tabular}}
	\label{tab:4}
\end{table*}


%